\documentclass[12pt]{article}
\usepackage{graphicx} 
\usepackage{feynmf}
\usepackage{amssymb}
\usepackage{amsmath}
\usepackage{xcolor}
\usepackage{cite}
\usepackage{xr}
\usepackage{dsfont}
\usepackage{setspace}
\usepackage{tikz}
\usetikzlibrary{arrows,shapes,positioning,shadows,trees}
\newcommand{\cals}{\text{$\cal S$}}

\newcommand{\dilog}{\mbox{Li}_2}

\newcommand{\tx}{\tilde{x}}

\newcommand{\bbar}{\overline{b}}
\newcommand{\detg}{\det{(G)}}
\newcommand{\dets}{\det{(\cals)}}
\newcommand{\sign}{\mbox{sign}}

\newcommand{\baru}{\bar{u}}

\newcommand{\tD}{\widetilde{D}}
\newcommand{\bbj}[2]{\overline{b}_{#1}^{\{#2\}}}

\newcommand{\detgj}[1]{\det{(G^{\{#1\}})}}
\newcommand{\detsj}[1]{\det{(\cals^{\{#1\}})}}

\renewcommand\Re{\operatorname{Re}}
\renewcommand\Im{\operatorname{Im}}

\newsavebox{\Gammap}
\newsavebox{\Gammam}
\numberwithin{equation}{section}
\tikzset{
  basic/.style  = {draw, drop shadow, rectangle, inner sep=2pt, minimum size=6mm},
  root/.style   = {basic, rounded corners=2pt, thin, align=center,
                   fill=green!30},
  level 2/.style = {basic, rounded corners=6pt, thin,align=center, fill=green!60
                   },
  level 3/.style = {basic, thin, align=left, fill=pink!60}
}

\begin{document}

\setlength{\unitlength}{1mm}
\begin{fmffile}{samplepics}

\begin{titlepage}

\vspace{1.cm}

\long\def\symbolfootnote[#1]#2{\begingroup%
\def\thefootnote{\fnsymbol{footnote}}\footnote[#1]{#2}\endgroup} 

\begin{center}

{\large \bf A novel approach to \\
the computation of one-loop three- and four-point functions. \\
\vspace{0.1cm}
III - The infrared divergent case}\\[2cm]

{\large  J.~Ph.~Guillet$^{a}$, E.~Pilon$^{a}$, 
Y.~Shimizu$^{b}$ and M. S. Zidi$^{c}$ } \\[.5cm]

\normalsize
{$^{a}$ Univ. Grenoble Alpes, Univ. Savoie Mont Blanc, CNRS, LAPTH, F-74000 Annecy, France}\\
{$^{b}$ KEK, Oho 1-1, Tsukuba, Ibaraki 305-0801, Japan\symbolfootnote[2]{Y. Shimizu passed away during the completion of this series of articles.}}\\
{$^{c}$ LPTh, Universit\'e de Jijel, B.P. 98 Ouled-Aissa, 18000 Jijel, Alg\'erie}\\
      
\today
\end{center}

\vspace{2cm}

\begin{abstract} 
\noindent
This article is the third and last of a series presenting an alternative 
method to compute the one-loop scalar integrals.
It extends the results of first two articles to the infrared divergent case.
This novel method enjoys a couple of
interesting features as compared with
the methods found in the literature.
It directly proceeds in terms of the quantities driving algebraic 
reduction methods. 
It yields a simple decision tree based on the vanishing 
of internal masses and one-pinched kinematic matrices which avoids a profusion of cases.
Lastly, it extends to kinematics more general than the one of
physical e.g. collider processes relevant at one loop. This last feature may be
useful when considering the application of this method beyond one loop using 
generalised one-loop integrals as building blocks. 
\end{abstract}

\vspace{1cm}

\begin{flushright}
LAPTH-45/18\\
\end{flushright}

\vspace{2cm}

\end{titlepage}
\savebox{\Gammap}[7mm][r]{%
  \begin{fmfgraph*}(7,7)
  \fmfset{arrow_len}{3mm}
  \fmfset{arrow_ang}{10}
  \fmfipair{o,xm,xp,ym,yp}
  \fmfipath{c[]}
  \fmfipair{r[]}
  \fmfiequ{o}{(.1w,-.2h)}
  \fmfiequ{xm}{(0,-.2h)}
  \fmfiequ{ym}{(.1w,0)}
  \fmfiequ{r3}{(.35w,-.2h)}
  \fmfiset{c1}{fullcircle scaled 1.5w shifted o}
  \fmfi{fermion}{subpath (length(c1)/4,0) of c1}
  \fmfiequ{r1}{point 0 of c1}
  \fmfiequ{r2}{point length(c1)/4 of c1}
  \fmfi{fermion}{r1--r3}
  \fmfi{fermion}{o--r2}
  \fmfiv{l={\tiny 0},l.a=-120,l.d=0.05w}{o}
  \fmfiv{l={\tiny 1},l.a=-70,l.d=0.06w}{r3}
\end{fmfgraph*}}

\savebox{\Gammam}[7mm][r]{%
  \begin{fmfgraph*}(7,7)
  \fmfset{arrow_len}{3mm}
  \fmfset{arrow_ang}{10}
  \fmfipair{o,xm,xp,ym,yp}
  \fmfipath{c[]}
  \fmfipair{r[]}
  \fmfiequ{o}{(.1w,.5h)}
  \fmfiequ{xm}{(0,.5h)}
  \fmfiequ{ym}{(.1w,0)}
  \fmfiequ{r3}{(.35w,.5h)}
  \fmfiset{c1}{fullcircle scaled 1.5w shifted o}
  \fmfiequ{r1}{point 0 of c1}
  \fmfiequ{r2}{point 3length(c1)/4 of c1}
  \fmfi{fermion}{subpath (3length(c1)/4,length(c1)) of c1}
  \fmfi{fermion}{r1--r3}
  \fmfi{fermion}{o--r2}
  \fmfiv{l={\tiny 0},l.a=90,l.d=0.05w}{o}
  \fmfiv{l={\tiny 1},l.a=70,l.d=0.06w}{r3}
\end{fmfgraph*}}

\newpage

\section{Introduction}\label{intro}

This article is the third of a triptych. The first one \cite{paper1} presented 
a method exploiting a Stokes-type identity to compute ``generalised'' (in the sense of the underlying kinematics) one-loop three- and 
four-point scalar integrals for the real mass case. The second article \cite{paper2} 
extended the results of the first paper to the case of general complex masses.
The present article widens the results of \cite{paper1} and \cite{paper2} to 
the case where some internal masses are vanishing leading to infrared divergences. 
We refer the reader to ref. \cite{letter} for more details on the motivation of this work.

\vspace{0.3cm}

\noindent
The scalar Feynman integrals for one-loop three- and four-point functions are 
all known and have been compiled in a useful article \cite{Ellis:2007qk}. 
This article relies mainly on the results of other publications, especially the important 
work of Beenakker and Denner \cite{Beenakker:1988jr}. 
Let us mention also ref. \cite{Denner:2010tr}
which provides a complete set of results for soft and/or collinear divergent four-point functions using different kind of IR regulators.
The purpose of the present 
article is to extend these results for more general kinematics beyond those relevant 
for collider processes at the one-loop order. Note that despite the fact that some 
internal masses may vanish, the others can be real or complex and we treat both 
cases in this article. The soft and collinear divergences are dealt with using 
dimensional regularisation, $n = 4 - 2 \, \varepsilon$, and doing an $\varepsilon$ expansion.



\vspace{0.3cm}

\noindent
The outline of this article follows closely the one of our preceding articles \cite{paper1} and \cite{paper2}.
We start by considering the three-point function $I_{3}^{n}$ in a space-time dimension 
shifted by a small amount from $4$ to $n$. The kinematics leading to infrared divergences is discussed.
It is considered as a warm-up for sec. \ref{sectfourpointir}.
We successively present two variants of the method. The simplest variant, 
labelled ``direct way'', is presented in subsec. \ref{dirway}. It is 
well suited for the three-point function, but cannot be extended to the case of the four-point function.
Then, in subsec.~\ref{exp_exemp_ir}, practical implementation of the
results of the preceding subsection is discussed and some explicit examples are computed and compared to \cite{Ellis:2007qk}.
In subsec. \ref{indirway} we present an alternative 
coined ``indirect way'' easily applicable to the four-point case which is the subject of sec.~\ref{sectfourpointir}.
We first explain, in subsec.~\ref{compI4n} how to extend the calculation of $I_4^4$ developed 
in \cite{paper1} to the case where the infrared divergences are regulated in $n$-dimension. 
The net result is that the four-point scalar integral can be decomposed on sectors labelled 
by three indices and a three dimensional integral over the first octant of $\mathds{R}^3$ is 
associated to each sector. Then, two cases are distinguished depending on the sectors. 
In the first case, presented in~\ref{subsect52}, 
the determinant of the one-pinched kinematical matrix $\cals$ vanishes but not the internal 
mass associated to this sector: this case is met when a soft divergence appears. In the 
second case, presented in subsec. \ref{casDelta20Dt0}, both the determinant of the 
one-pinched $\cals$ matrix and the internal mass associated to this sector vanish: this case 
is met when a collinear or a soft and collinear divergence shows up. In sec.~\ref{examplefourpoint}, 
the infrared divergent part of the scalar four-point integral is shown to be proportional 
to a three-point scalar integral as it should be. Some explicit examples are given and 
compared to the results found in the literature. We then conclude.
Various appendices gather a number of utilities removed from the main text to facilitate its reading. 
Accordingly, in appendix~\ref{appendJ}, we complete appendix~\ref{P1-appendJ} of \cite{paper1} and 
appendix~\ref{P2-appendJ} of \cite{paper2} by giving a missing case where the power of 
the integration variable is not an integer as required by
dimensional regularisation. Appendix~\ref{calculJx1x2} shows how to compute an integral 
appearing in the three-point case in closed form. 
Appendix \ref{herba} collects a bunch of integrals required to compute the three- and 
four-point functions having soft and/or collinear divergences in the case of general complex masses.
Then, appendix~\ref{direcway3pIR} 
goes through the examples given in subsec. \ref{exp_exemp_ir} and explains in detail how 
the results obtained in the latter subsection can be found again from those derived in 
the ``indirect way'' case. Appendix~\ref{appF} provides the way to compute the last integration 
in closed form in terms of dilogarithms for the case of infrared divergent integrals. 
It complements the appendix~\ref{P1-appF} of \cite{paper1} and appendix~\ref{P2-appF} of \cite{paper2}. 
Lastly, appendix~\ref{ir-lambda} proves a tricky point used in sec.~\ref{sectfourpointir}: for 
the real mass case, the sign of the vanishing imaginary part of the denominator in the last integral can be safely changed.

\section{Three-point function with infrared divergences}\label{3point_ir}

\begin{figure}[h]
\centering
\parbox[c][43mm][t]{80mm}{\begin{fmfgraph*}(60,80)
  \fmfleftn{i}{1} \fmfrightn{o}{1} \fmftopn{t}{1}
  \fmf{fermion,label=$p_1$}{t1,v1}
  \fmf{fermion,label=$p_2$}{i1,v2}
  \fmf{fermion,label=$p_3$}{o1,v3}
  \fmf{fermion,tension=0.5,label=$q_1$}{v1,v2}
  \fmf{fermion,tension=0.5,label=$q_2$}{v2,v3}
  \fmf{fermion,tension=0.5,label=$q_3$,label.side=right}{v3,v1}
\end{fmfgraph*}}
\caption{\footnotesize 
The triangle picturing the one-loop three-point function.}
\label{fig1} 
\end{figure}
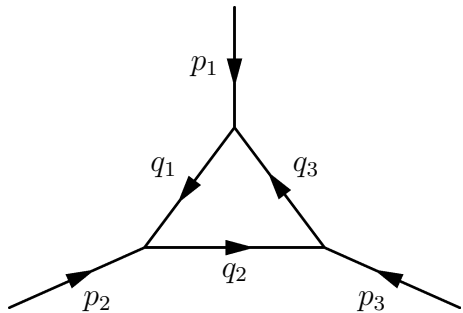

\noindent
When some internal masses vanish, 
divergences of collinear or soft origin appear and the approach shall be
revisited. We regularise these divergences 
using dimensional re\-gularisation, shifting the dimension of the space time by a 
small positive amount from 4 to $n = 4 - 2 \, \varepsilon$ with
$\varepsilon < 0$. After performing the loop momentum integral, 
instead of eq. (\ref{P1-eqSTARTINGPOINT3}) of ref. \cite{paper1} we get\footnote{As in refs. \cite{paper1,paper2}, we assume that the elements of
the kinematic matrix $\cals$ have been made dimensionless by an appropriate
rescaling.}:
\begin{equation}
I_3^n 
= 
- \Gamma(1+\varepsilon) \, 
\int \prod_{i=1}^3 \, dz_i \, 
\delta(1-\sum_{i=1}^3 z_i)\, 
\left(
 - \, \frac{1}{2} \, z^T \cdot \cals \cdot z - i \, \lambda
\right) ^{-1-\varepsilon}
\label{eqdefi3n}
\end{equation}
To appropriately shift the power of the denominator in eq. (\ref{eqdefi3n}) 
so as to apply the Stokes identity (\ref{P1-eqDEFREL1}) of ref. \cite{paper1} as we did in the massive
case, we use the  following modified integral representation instead of
identity (\ref{P1-eqFOND1-simple}) of the previous reference (cf. appendix~\ref{P1-ap2} of \cite{paper1}):
\[
\frac{1}{D^{1+\varepsilon}} 
= 
\frac{\nu}{B(2-1/\nu,1/\nu)} \; 
\int^{+\infty}_{0} \, \frac{d \xi}{(D+\xi^{\nu})^2}
\]
with $\nu = 1/(1-\varepsilon)$. Instead of eq. (\ref{P1-eqI341}) of ref. \cite{paper1} we now get:
\begin{align}
I_3^n 
&= - \, 2^{1+\varepsilon} \, \frac{\Gamma(1+\varepsilon)}{1-\varepsilon} \, 
\frac{1}{B(1+\varepsilon,1-\varepsilon)} \, 
\int^{+\infty}_0 d \xi \, \int_{\Sigma_{bc}}  
\frac{dx_b \, dx_c}{(D^{(a)}(x_b,x_c) + \xi^{\nu} - i \, \lambda)^2}
\label{eqdefi3n1}
\end{align}
We otherwise proceed as in subsec. \ref{P1-sect3pstep1} of ref. \cite{paper1}. 
The counterpart of eq. (\ref{P1-eqI345}) of the same reference now reads:
\begin{align}
I_3^n 
&= 2^{\varepsilon} \, \frac{\Gamma(1+\varepsilon)}{1-\varepsilon} \, 
\frac{1}{B(1+\varepsilon,1-\varepsilon)} \, 
\notag\\
&\quad {} \times
\sum_{i \in S_3} \, \frac{\bbar_i}{\det(G)} \, 
\int^{+\infty}_0 \frac{d \xi}{\Delta_2 - \xi^{\nu}+ i \, \lambda} \, 
\int^1_0 \, \frac{dx}{D^{\{i\}(j)}(x)+ \xi^{\nu} - i \, \lambda}
\label{eqdefi3n2}
\end{align}
where $j \in S_3 \setminus \{i\}$ ($S_3 = \{1,2,3\}$). More precisely, we assume that $j$ is chosen to be $1 + (i \; \mbox{modulo} \; 3)$.
Similarly to what we did for the three-point function in the massive case, 
one can also consider both a ``direct way" and an ``indirect way" in the IR 
case. We first focus on the ``direct way" which provides a more 
straightforward and synthetic
discussion of the various cases at hand. We then illustrate how these
cases are involved in a few examples. The ``indirect way" instead
leads to a cumbersome split-up discussion. Notwithstanding the latter has 
its own interest. 
The calculation of the four-point one-loop 
integral relying on the approach described in this article proceeds 
along the ``indirect way'' as we found no extension of the ``direct way'' approach in
this case. 
In refs. \cite{Binoth:2005ff,Binoth:1999sp} it was 
shown on general grounds using the decomposition\footnote{This decomposition has been discovered before and used for different purposes, see \cite{vanNeerven:1983vr,Kotikov:1991pm,Bern:1993kr,Tarasov:1996br}.}
\begin{equation}\label{decomp-golem}
\dets \, I_4^n(\cals) 
= 
\sum_{i=1}^{4} \bbar_{i} \, I_3^n(\cals^{\{i\}}) - \detg \, (1 - 2 \, \varepsilon) \, I_4^{n+2}(\cals)
\end{equation}
that the infrared structure of any IR divergent four-point one-loop 
integral is carried by IR divergent three-point one-loop functions resulting 
from appropriate iterated pinchings. 
Therefore the comparison of the IR structures 
in both sides of eq. (\ref{decomp-golem}) proceeds most conveniently via a term
by term comparison using the three-point one-loop functions decomposed according
to the ``indirect way'' as well. In anticipation, we hereby give the key 
ingredients to perform this comparison, as well as the general recombination
of these ``indirect way'' ingredients into the more compact expression obtained 
from the ``direct way'', thereby checking their equivalence.
The extensive collection of expressions computed in closed form which enable 
to perform detailed case-by-case comparisons is gathered in appendix 
\ref{direcway3pIR} to lighten the presentation.

\subsection{Direct way}\label{dirway}

\noindent
Soft and/or collinear divergences are caused by some vanishing masses which 
make $\dets$ vanish so that $\Delta_2 = 0$, whereas the other internal masses 
may or may not vanish as well - and may even be complex.
We will keep the $- \, i \, \lambda$ prescription having in mind that 
it is ineffective in the case of complex masses.

\vspace{0.3cm}

\noindent
Starting from eq. (\ref{eqdefi3n2}) and performing the $\xi$ integration using
eq. (\ref{eqmodifk}), we end up with:
\begin{align}
I_3^n 
&= \frac{2^{\varepsilon}}{\varepsilon} \, \Gamma(1+\varepsilon) \, 
\sum_{i \in S_3} \, \frac{\bbar_i}{\det(G)} \, 
\int^1_0 dx \, 
\left( D^{\{i\}(j)}(x) - i \, \lambda \right)^{-1-\varepsilon} 
\label{eqdirei3n1}
\end{align}
In the general case $D^{\{i\}(j)}(x)$ depends on two internal 
masses squared $m_j^2$ and $m_k^2$ such that $m_j^2 = D^{\{i\}(j)}(0)/2 = \tD_{ik}/2$ and $m_k^2 = D^{\{i\}(j)}(1)/2 = \tD_{ij}/2$,
cf.\ sec.\ \ref{P1-sectthreepoint} of \cite{paper1}.
We introduced the label $k$ which is the only element of the complement of $\{i,j\}$ in $S_3$.
With our assumption on $j$, this implies that $k \equiv 1 + ((i+1) \; \mbox {modulo} \;3)$.
Let us focus on the function $W$ given by:
\begin{equation}
  W\left(\detgj{i},\tD_{ij}, \tD_{ik}\right) = \frac{2^{\varepsilon}}{\varepsilon} \, \Gamma(1+\varepsilon) \, \int^1_0 dx \, 
\left( D^{\{i\}(j)}(x) - i \, \lambda \right)^{-1-\varepsilon}
\label{eqdefwi0}
\end{equation}
We remind (cf.\ eqs.~(\ref{P1-Db}), (\ref{P1-Dc}) and (\ref{P1-Da}) of ref.\ \cite{paper1}):
\begin{equation}
D^{\{i\}(j)}(x) 
= 
G^{\{i\}(j)} \, x^2 - 2 \, V^{\{i\}(j)} \, x - C^{\{i\}(j)}
\label{eqremd1}
\end{equation}
with
\begin{align}
  G^{\{i\}(j)} &= - \cals_{kk} + 2 \, \cals_{kj} - \cals_{jj}  = \detgj{i} \notag \\
  V^{\{i\}(j)} &= \cals_{kj} - \cals_{jj} = \frac{1}{2} \left[ \detgj{i} - \tD_{ij} + \tD_{ik} \right] \label{eqremd2}\\
  C^{\{i\}(j)} &= \cals_{jj}  = - \tD_{ik} \notag
\end{align}
The Gram matrix $G^{\{i\}(j)}$ is built from the one-pinched $\cals$ matrix $\cals^{\{i\}}$, it is a real 
matrix which depends only on a squared external momentum in the three-point case. Notice that, in this 
case, $G^{\{i\}(j)}$ is a $1 \times 1$ matrix and $V^{\{i\}(j)}$ a one-dimensional vector, this explains 
the notations\footnote{Let us remind that $\cals_{jk} = \cals^{\{i\}}_{jk}$ for $j,k \ne i$.} used in 
eqs.~(\ref{eqremd1}) and (\ref{eqremd2}). 
The knowledge of $\detgj{i}$, $\tD_{ij}$ and $\tD_{ik}$ fully determines the polynomial $D^{\{i\}(j)}(x)$.
These two internal masses may or may not vanish, hence three
cases to be considered. 

\vspace{0.3cm}

\noindent
{\bf a) Neither $m_j^2$ nor $m_k^2$ vanishes}\\
We perform a Taylor expansion\footnote{Here and below, only the
terms in the $\varepsilon$-expansion providing the divergent and finite terms
in the limit $\varepsilon \to 0$ are kept.} of $W\left(\detgj{i},\tD_{ij}, \tD_{ik}\right)$ in $\varepsilon$:
\begin{align}
\hspace{2em}&\hspace{-2em}W\left(\detgj{i},\tD_{ij}, \tD_{ik}\right) \notag \\
&=  \frac{2^{\varepsilon}}{\varepsilon} \, \Gamma(1+\varepsilon) \, \left[ \int^1_0 \frac{dx}{D^{\{i\}(j)}(x) - i \, \lambda} 
- 
 \varepsilon  
 \int^1_0 dx  
 \frac{\ln \left( D^{\{i\}(j)}(x) - i \, \lambda \right)}
 {D^{\{i\}(j)}(x) - i \, \lambda} \right]
\label{eqdirei3n2}
\end{align}
Let us note $x_1$ and $x_2$ the two roots of $D^{\{i\}(j)}(x) - i \lambda$, 
given by, cf. eqs. (\ref{eqremd1}), (\ref{eqremd2}):
\begin{align}
x_{\underset{2}{1}} 
&= 
\frac{
  \detgj{i} - \tD_{ij} + \tD_{ik}
 \pm 
 \sqrt{ {\cal K}\left( \detgj{i},\tD_{ij},\tD_{ik} \right) + i \, \lambda \, S_G}
}{2 \, \detgj{i}}
\label{eqroot12}
\end{align}
where ${\cal K}$ is the K\"all\'en function:
\begin{equation}
  {\cal K}(x,y,z) = x^2 + y^2 + z^2 - 2 \, x \, y - 2 \, x \, z - 2 \, y \, z
  \label{eqkallenfunc}
\end{equation}
and $S_G = \sign(\detgj{i})$. Then, we introduce $J(x_1,x_2)$ and 
$K(x_1,x_2)$ defined by:
\begin{align}
K(x_1,x_2) 
&=   \int^1_0 dx \, \frac{1}{(x - x_1) \, (x - x_2)} 
\label{eqcompk1} \\
J(x_1,x_2) 
&= \int^1_0 dx \, 
\frac{\ln\left( (x-x_1) \, (x-x_2) \right)}{(x - x_1) \, (x - x_2)}
\label{eqcompj1}
\end{align}
As it will become clear in the forthcoming paragraph on the origin of infrared
singularities, only
the case with $m_j^2$ and $m_k^2$ both real matters in practice, which makes 
the explicit calculation of $J(x_1,x_2)$ somewhat simpler\footnote{With real 
masses, $x_1$ and $x_2$ in
eq. (\ref{eqroot12}) have imaginary parts of opposite signs. This namely
simplifies splittings and recombinations of logarithms of ratios in the
explicit calculation of the function $J(x_1,x_2)$ computed in appendix 
\ref{calculJx1x2}.}. The latter is provided in appendix \ref{calculJx1x2}.
The $x$ integration in the function $K(x_1,x_2)$ straightforwardly gives:
\begin{align}
 K(x_1,x_2) 
&= \frac{1}{x_1-x_2} \, 
\left[ 
 \ln \left( \frac{x_1-1}{x_1} \right) 
 - 
 \ln \left( \frac{x_2-1}{x_2} \right) 
\right]
\label{eqcompk11}
\end{align}
Thus $W\left(\detgj{i},\tD_{ij}, \tD_{ik}\right)$ reads:
\begin{align}
\hspace{2em}&\hspace{-2em}W\left(\detgj{i},\tD_{ij}, \tD_{ik}\right) \notag \\
&= 
\frac{1}{\varepsilon} \, \frac{\Gamma(1+\varepsilon)}{\detgj{i}} 
\left\{
 \left[
   1 - \varepsilon \, \ln \left(\frac{\detgj{i}}{2} - i \, \lambda \right)
 \right] \, K(x_1,x_2) 
 - \varepsilon \, J(x_1,x_2) 
\right\}
\label{eqcompintlog1}
\end{align}

\vspace{0.3cm}

\noindent
{\bf b) One and only one of $m_j^2$ and $m_k^2$  vanishes}\\
Let us assume that the vanishing internal mass is $m_j^2$. $D^{\{i\}(j)}(x)$ 
becomes:
\begin{equation}
D^{\{i\}(j)}(x) 
= x \, \left( G^{\{i\}(j)} \, x - 2 \, V^{\{i\}(j)} \right)
\label{eqnewquad1}
\end{equation}
From eqs.~(\ref{eqdefwi0}) and (\ref{eqremd2}), $W\left(\detgj{i},\tD_{ij}, 0\right)$ is thus of the form
\begin{align}
W\left(\detgj{i},\tD_{ij}, 0\right)
&= \frac{2^{\varepsilon}}{\varepsilon} \, \Gamma(1+\varepsilon) \, \int^1_0 dx \, x^{-1-\varepsilon} \, (a \, x + z)^{-1-\varepsilon}
\label{eqdefyint1}
\end{align}
where $a = \detgj{i}$ is real and 
$z = - \detgj{i} + \tD_{ij} - i \, \lambda$ is complex.
As $z$ and $a \, x + z$ have imaginary parts of the same sign 
$(a \, x + z)^{-1-\varepsilon}$ can be split as follows:
\[
(a \, x + z)^{-1-\varepsilon} 
= 
z^{-1-\varepsilon} \, \left( 1 + \frac{a}{z} \, x \right)^{-1-\varepsilon}
\]
The r.h.s. of eq. (\ref{eqdefyint1}) involves the Gauss hypergeometric function 
$_{2}F_{1}$:
\begin{equation}
W\left(\detgj{i},\tD_{ij}, 0\right) = 
- \, \frac{2^{\varepsilon}}{\varepsilon^2} \, \Gamma(1+\varepsilon)  \, z^{-1-\varepsilon} \,
_{2}F_{1} 
\left( 1+\varepsilon,-\varepsilon;1-\varepsilon;- \, \frac{a}{z} \right)
  \label{eqdefyint2}
\end{equation}
We use the identity \cite{abramowitz} 
\begin{align*}
  _{2}F_{1}(a,b;c;w) &= \frac{\Gamma(c) \, \Gamma(c-a-b)}{\Gamma(c-a) \, \Gamma(c-b)} \; _{2}F_{1}(a,b;a+b-c+1;1-w) \\
  &\quad {} + (1-w)^{c-a-b} \, \frac{\Gamma(c) \, \Gamma(a+b-c)}{\Gamma(a) \, \Gamma(b)} \; _{2}F_{1}(c-a,c-b;c-a-b+1;1-w)
\end{align*}
and the Pfaff identity
\[
_{2}F_{1}(a,b;c;w) 
= 
(1-w)^{-b} \, _{2}F_{1} \left( c-a,b;c;\frac{w}{w-1} \right)
\]
to rewrite:
\begin{align}
&_{2}F_{1} 
\left( 1+\varepsilon,-\varepsilon;1-\varepsilon;- \, \frac{a}{z} \right)
\notag\\
&= 
2 \, \frac{\Gamma^2(1-\varepsilon)}{\Gamma(1- 2 \, \varepsilon)} \, 
\left( - \frac{a}{z} \right)^{\varepsilon} 
- \left( \frac{a+z}{z} \right)^{-\varepsilon} \, 
\left( - \frac{a}{z} \right)^{2 \, \varepsilon} \, 
_{2}F_{1} 
\left( - 2 \, \varepsilon, - \varepsilon; 1-\varepsilon; \frac{a+z}{a} \right)
\label{eqsplit2F1}
\end{align}
Performing a Taylor expansion in $\varepsilon$ we get:
\[
_{2}F_{1} 
\left( - 2 \, \varepsilon, - \, \varepsilon ;1 - \varepsilon; \tau \right)
= 
1 + 2 \, \varepsilon^2 \, \dilog(\tau) 
\]
and splitting $\ln( (a+z)/z) = \ln(a+z) - \ln(z)$, we rewrite $W\left(\detgj{i},\tD_{ij}, 0\right)$as:
\begin{align}
W\left(\detgj{i},\tD_{ij}, 0\right) 
&= \frac{2^{\varepsilon}}{\varepsilon^2} \, \Gamma(1+\varepsilon) \, \frac{1}{z} \,
\left\{ 
 \left( a+z \right)^{-\varepsilon} \, 
 \left( - \frac{a}{z} \right)^{2 \, \varepsilon} \, 
 \left[ 1 + 2 \, \varepsilon^2 \, \dilog \left( \frac{a+z}{a} \right) \right] 
\right. 
\notag \\
&\qquad \qquad \qquad \qquad {} 
- 
\left. 
 2 \; \frac{\Gamma^2(1-\varepsilon)}{\Gamma(1- 2 \, \varepsilon)} \, 
 \left( z \right)^{-\varepsilon} \, 
 \left( - \frac{a}{z} \right)^{\varepsilon} 
\right\}
\label{eqdefyint3}
\end{align}
Making explicit $z = - \detgj{i} + \tD_{ij}  - i \, \lambda$,  
$a+z = \tD_{ij}  - i \, \lambda$ we get:
\begin{align}
\hspace{2em}&\hspace{-2em}W\left(\detgj{i},\tD_{ij}, 0 \right) \notag \\
  &= 
  - \frac{2^{\varepsilon}}{\varepsilon^2} \, \Gamma(1+\varepsilon)  \; \frac{1}{\detgj{i} - \tD_{ij}} \notag \\
&\quad {} \times
   \left\{ 
     \left( \frac{\detgj{i}}{\detgj{i} - \tD_{ij} + i \, \lambda } \right)^{2 \, \varepsilon}  \,  \, \left(\tD_{ij} - i \, \lambda \right)^{-\varepsilon} \, \left[ 1 + 2 \, \varepsilon^2 \, \dilog \left( \frac{\tD_{ij} - i \, \lambda}{\detgj{i}} \right) \right] 
  \right. \notag \\
    &\qquad \qquad {} - \left. 
    2 \,  \frac{\Gamma^2(1-\varepsilon)}{\Gamma(1 - 2 \, \varepsilon)} \, \left( \frac{\detgj{i}}{\detgj{i} - \tD_{ij} + i \, \lambda } \right)^{\varepsilon}  \, \left( \tD_{ij} - \detgj{i} - i \, \lambda \right)^{- \varepsilon}
  \vphantom{\dilog \left( \frac{- \cals_{kk} - i \, \lambda}{G^{\{i\}(j)}_{kk}} \right)} \right\}
  \label{eqcompwcasb4}
\end{align}
This formula is manifestly well-behaved as $\tD_{ij} \rightarrow 0$ ($m_k^2 \rightarrow 0$) yet it is not
handy to expand around $\varepsilon=0$. A more practical alternative
may be obtained as follows. Firstly, we use the identities relating $\dilog(1-w)$,
 $\dilog(w)$ and $\dilog(1/w)$ to change the argument of the $\dilog$
function, and the following relations:
\begin{align}
\ln \left( \frac{a}{z} \right) &= \ln \left( - \frac{a}{z} \right) - i \, \pi \, S(a \, z) \\
  \ln \left( \frac{a+z}{a} \right) &= \ln \left( -\frac{a+z}{a} \right) + i \, \pi \, S(a \, z) \\
  \ln \left( -\frac{a+z}{a} \right) &= \ln \left( \frac{a+z}{z} \right) - \ln \left( - \frac{a}{z} \right)
  \label{eqrelalamormoilenoeud}
\end{align}
with
\begin{equation}
  S(z) = \sign\left( \Im(z) \right)
  \label{eqdeffuncS0}
\end{equation}
so that the $\dilog$ function can be rewritten as:
\begin{align}
  \dilog \left( \frac{a+z}{a} \right) &= \dilog \left( - \frac{a}{z} \right) - \frac{\pi^2}{6} - \frac{1}{2} \, \ln^2 \left( - \frac{a}{z} \right) + \ln \left( \frac{a+z}{z} \right) \, \ln \left( - \frac{a}{z} \right)
  \label{eqchangdilog1}
\end{align}
Secondly, we Taylor expand around $\varepsilon=0$ the $(-a/z)$ terms in eq.
(\ref{eqdefyint3}). We thus get:
\begin{align}
W\left(\detgj{i},\tD_{ij}, 0\right)
&= -\,\frac{2^{\varepsilon}}{\varepsilon} \, \frac{\Gamma(1+\varepsilon)}{z} \, 
\left[ 
  \frac{2}{\varepsilon} \, (z)^{-\varepsilon} 
  - 
  \frac{1}{\varepsilon} \, (a+z)^{-\varepsilon}  
  - 
  2 \, \varepsilon \, \dilog \left( - \frac{a}{z} \right) 
\right]
\label{eqdefyint6}
\end{align}
i.e. making explicit $z$ and $a$ in terms of $\detgj{i}$ and $\tD_{ij}$:
\begin{align}
W\left(\detgj{i},\tD_{ij}, 0\right)
&= \frac{1}{\varepsilon} \, \frac{\Gamma(1+\varepsilon)}{\detgj{i} - \tD_{ij}} \notag \\
&\quad {} \times
\left\{ 
 \frac{2}{\varepsilon} \, 
 \left[ \frac{1}{2} \, \left( \tD_{ij} - \detgj{i} \right) - i \, \lambda \right]^{-\varepsilon} - 
 \frac{1}{\varepsilon} \, \left( \frac{\tD_{ij}}{2}  - i \, \lambda \right)^{-\varepsilon}
\right. 
\notag \\
&\qquad \quad {} - 
\left. 
 2 \, \varepsilon \, 
\dilog 
\left( 
\frac{\detgj{i}}{\detgj{i} - \tD_{ij} + i \, \lambda } 
\right) 
\right\}
\label{eqdirei3n3}
\end{align}
which is both well behaved when $\tD_{ij} \rightarrow 0$ ($m_k^2 \rightarrow 0$) and more compact.

\vspace{0.3cm}

\noindent
{\bf c) Both $m_j^2$ and $m_k^2$ vanish}

\noindent
The function $D^{\{i\}(j)}(x)$ becomes:
\begin{equation}
D^{\{i\}(j)}(x) = - \, G^{\{i\}(j)} \, x \, (1-x)
\label{eqnewquad2}
\end{equation}
and we immediately get:
\begin{align}
W\left(\detgj{i},0,0\right)
&= -\frac{1}{\varepsilon^2} \, \Gamma(1+\varepsilon)  \, 
\frac{\Gamma^2(1-\varepsilon)}
{\Gamma(1 - 2 \, \varepsilon)} \, 
\left( - \, \frac{\detgj{i}}{2} - i \, \lambda \right)^{-1-\varepsilon}
\label{eqdirei3n4}
\end{align}
In the limit $\tD_{ij} \rightarrow 0$ ($m_k^2 \rightarrow 0$), 
eq.~(\ref{eqdirei3n3}) or eq.~(\ref{eqcompwcasb4}) smoothly becomes eq. (\ref{eqdirei3n4}) as expected.

\subsection{Practical implementation of the preceding cases and 
explicit examples}\label{exp_exemp_ir}

The various cases reviewed above may or may not be involved 
in a specific computation 
because some coefficients weighing the $W\left(\detgj{i},\tD_{ij},\tD_{ik}\right)$ may vanish. 
In particular, as seen on eq. (\ref{eqdirei3n1}), 
when $\Delta_2=0$, the three-point function in 
dimension $4 - 2 \, \varepsilon$ is 
the sum of three two-point functions in
dimension $2 - 2 \, \varepsilon$
These 
two-point functions correspond to the three distinct pinchings
of the 
internal propagators of the three-point function. At first sight,
one should worry that some of these two-point functions in low dimensions 
may badly diverge due to a threshold singularity which is however not present 
in the three-point function! 
For example, one of the pinchings of a three-point function having IR/collinear
singularities would lead to a two-point function with the external legs on the
mass shell of one of the propagators whereas the other propagator is massless. 
This would lead to a polynomial $D^{\{i\}(j)}(x) \propto x^2$ or $(1-x)^2$. 
Fortunately the corresponding $\bbar$ coefficients weighting 
such pathological terms identically vanish, and the discussion which follows, 
illustrated with examples, elucidates why it happens so.
Let us note $p_i^2 = s_i$ with $i=1,2,3$.

\vspace{0.3cm}

\noindent
{\bf 1.} A soft divergence occurs 
when the kinematic matrix $\cals$ has a vanishing line (and corresponding 
column). This happens whenever a massless propagator connects
two vertices in which enter external momenta on the 
mass shells of the two other propagators.
As the external momenta are real this case can occur
only when the non vanishing internal masses are real.
Let us assume, cf. fig. 1, that the internal mass squared $m_1^2$ vanishes 
whereas the external four-momenta $p_1$ and $p_2$ satisfy the mass shell 
conditions $s_1=m_3^2$, $s_2=m_2^2$.
The $\cals$ matrix has the following texture:
\begin{equation}
  \cals^{soft} =
  \left(
  \begin{array}{ccc}
    0 & 0     & 0 \\
    0 & -2 \, m_2^2 & s_3 - m_2^2 - m_3^2 \\
    0 & s_3 - m_2^2 - m_3^2 & - 2 \, m_3^2
  \end{array}
  \right)
  \label{eqcalssoft}
\end{equation}
If one singles out row and column $1$ in $\cals^{soft}$, 
the two-component vector $V^{(1)}$ is readily seen to vanish and so do the 
coefficients  $\bbar_2$ and $\bbar_3$ which are proportional to the two 
components of $(G^{(1)})^{-1} \cdot V^{(1)}$: 
thus only $\bbar_1$ differs from zero, cf. eq. (\ref{P1-eqdefdelta2}) of ref. \cite{paper1}.

\vspace{0.3cm}

\noindent
To illustrate this point, let us consider the case where $m_1=0$, $m_2=m_3=m$,
$s_1=s_2=m^2$ and $s_3$ arbitrary. In this case, since $\bbar_1$ is the
only non vanishing coefficient, the polynomial $D^{\{1\}(2)}(x)$ involved in 
$I_3^n$
corresponds to the one appearing in the two-point function obtained by pinching 
the internal line with four-momentum $q_1$ (cf.\ fig.~\ref{fig1}). This polynomial involves two 
masses (equal here) and this example corresponds to case {\bf a} of the 
preceding section. In this simple case, the two roots of the polynomial $D^{\{1\}(2)}(x)$ is given by:
\begin{equation}
  x_{1,2} = \frac{1}{2} \pm \frac{1}{2} \, \sqrt{ 1 - \frac{4 \, ( m^2 - i \,\lambda)}{s_3}}
  \label{eqroot12ex}
\end{equation}
with the property:
\[
  1 - x_1 = x_2
\]
Injecting this property into eqs.\ (\ref{eqcompk11}) and (\ref{eqcompj5}), we get for the functions $J(x_1,x_2)$ and $K(x_1,x_2)$ in eq.~(\ref{eqcompintlog1}):
\begin{align}
  J(x_1,x_2) &= \frac{2}{x_1-x_2} \, \left\{ \dilog \left( \frac{x_2}{x_1} \right) - \dilog \left( \frac{x_1}{x_2} \right) \right. \notag \\
  &\quad + \left. \ln \left( -\frac{x_2}{x_1} \right) \, \ln \left( - (x_1-x_2)^2 \right)  \right\} \label{eqcompj6} \\
  K(x_1,x_2) &=  \frac{2}{x_1-x_2} \, \ln \left( -\frac{x_2}{x_1} \right) \label{eqcompk2}
\end{align}
We check numerically that we recover the result of \cite{Ellis:2007qk}.

\vspace{0.3cm}

\noindent
{\bf 2.} A collinear divergence occurs 
when two internal masses vanish whereas the external four-momentum which 
enters into the vertex connecting the two adjacent massless propagators is 
lightlike (massless collinear splitting at this vertex).
Note that the non vanishing internal mass can be real or complex.
Let us assume that the labels of the two massless propagators are 
$1$ and $2$, with $s_2=0$: the $\cals$ matrix has the following texture:
\begin{equation}
  \cals^{coll} =
  \left(
  \begin{array}{ccc}
    0 & 0 & s_1 - m_3^2 \\
    0 & 0 & s_3 - m_3^2 \\
    s_1 - m_3^2 & s_3 - m_3^2 & - 2 \, m_3^2
  \end{array}
  \right)
  \label{eqcalscoll}
\end{equation}
If one singles out row and column $3$ in $\cals^{coll}$, the Gram matrix 
$G^{(3)}$ and the vector $V^{(3)}$ read:
\begin{align}
  G^{(3)} &= \left(
  \begin{array}{cc}
    2 \, s_1  & s_1+s_3\\
    s_1+s_3 & 2 \, s_3
  \end{array}
  \right)
  \label{eqdefg3coll} \\
  V^{(3)} &= \left(
  \begin{array}{c}
    s_1 + m_3^2 \\
    s_3 + m_3^2
  \end{array}
  \right)
  \label{eqdefv3coll}
\end{align}
A simple calculation yields:
\begin{align}
\sum_{i \in S_3 \setminus \{3\}} \, 
\left[ \left( G^{(3)} \right) ^{-1} \cdot V^{(3)} \right]_i 
&= 1
\label{eqsumgm1v}
\end{align}
so that $\bbar_3$ given by eq. (\ref{P1-eqdefdelta2}) of ref. \cite{paper1}
vanishes, whereas 
$\bbar_{1}$ and $\bbar_{2}$ generically differ from zero.

\vspace{0.3cm}

\noindent
Let us compute $I_3^n$ for this specific case. As $\bbar_3 = 0$, the 
relevant polynomials $D^{\{i\}(j)}(x)$ are those of the two-point 
functions obtained in the two pinching configurations 
(cf.\ figure (\ref{fig1})) where 
either the internal line with four-momentum $q_1$ 
or the one with the four-momentum $q_2$ shall be pinched. 
As these two lines are associated to vanishing 
masses and the third propagator is associated to a non vanishing mass, 
the two polynomials both have one vanishing mass, this corresponds 
to case {\bf b} of the previous section. 
Starting with eq. (\ref{eqdirei3n1}) $I_3^n$ reads:
\begin{align}
I_3^n 
&= \frac{2^{\varepsilon}}{\varepsilon} \, \Gamma(1+\varepsilon) \, 
\left[ 
 \frac{\bbar_1}{\det(G)} \, 
 \int^1_0 dx \, 
 \left( D^{\{1\}(2)}(x) - i \, \lambda \right)^{-1-\varepsilon} 
\right. 
\notag \\
&
\;\;\;\;\;\;\;\;\;\;\;\;\;\;\;\;\;\;\;\;\;\;
\left. 
\, + \frac{\bbar_2}{\det(G)} \, 
\int^1_0 dx \, 
\left( D^{\{2\}(3)}(x) - i \, \lambda \right)^{-1-\varepsilon} \right]
\label{eqdirei3n5}
\end{align}
It is better to change $x \leftrightarrow 1-x$ in the second integral in order to have 
a polynomial of the type (\ref{eqnewquad1}) and write $I_3^n$ as:
\begin{align}
I_3^n 
&= \frac{2^{\varepsilon}}{\varepsilon} \, \Gamma(1+\varepsilon) \, 
\left[
 \, 
 \frac{\bbar_1}{\det(G)} \, 
 \int_0^1 dx \, 
 \left( D^{\{1\}(2)}(x) - i \, \lambda \right)^{-1-\varepsilon} 
\right. 
\notag \\
&
\;\;\;\;\;\;\;\;\;\;\;\;\;\;\;\;\;\;\;\;\;\;
\left. 
  + \, 
 \frac{\bbar_2}{\det(G)} \, 
 \int_0^1 dx \, 
 \left( D^{\{2\}(1)}(x) - i \, \lambda \right)^{-1-\varepsilon} 
\right] \notag \\
&= \frac{\bbar_1}{\det(G)} \, W\left( \detgj{1},\tD_{12},0 \right) + \frac{\bbar_2}{\det(G)} \, W\left( \detgj{2},\tD_{21},0 \right)
\label{eqdirei3n6}
\end{align}
Noting that $\bbar_1 = (s_3-m_3^2) \, (s_1-s_3)$ and 
$\bbar_2 = (s_1-m_3^2) \, (s_3-s_1)$, determining $\detgj{1}$, $\detgj{2}$ and $\tD_{12}$ from the $\cals$ matrix 
elements (cf.\ eq.~(\ref{eqremd2})) and applying directly the result of 
eq. (\ref{eqdirei3n3}), we get:
\begin{align}
I_3^n 
&= \frac{\Gamma(1+\varepsilon)}{s_1-s_3} \, 
\left\{ - \, \frac{1}{\varepsilon^2} 
 \left[ 
  \left( {} - s_3+m_3^2 - i \, \lambda \right)^{-\varepsilon} 
\;\; - \;\;
  \left( {} -s_1+m_3^2 -i \, \lambda \right)^{-\varepsilon} 
 \right] 
\right. 
\notag \\
&\;\;\;\;\;\;\;\;\;\;\;\;\;\;\;\;\;\;\;\;\;\;\;\;\;\;\;\;
+ 
\left.  
 \dilog \left( \frac{s_3}{s_3-m_3^2+i \, \lambda} \right) 
 \; - \;\; 
 \dilog \left( \frac{s_1}{s_1-m_3^2+i \, \lambda} \right) \;
\right\}
\label{eqdirei3n8}
\end{align}
Using the Landen identity (\ref{eqlanden}) we recover the formula (4.8) of ref. \cite{Ellis:2007qk} 
after some algebra.

\vspace{0.3cm}

\noindent
{\bf 3.} Both a soft and a collinear divergence may occur at the same time,
thereby proceeding from both cases {\bf 1.} and {\bf 2.} above.

\vspace{0.3cm}

\noindent
Let us take the example of case {\bf 2} and specify $s_1 = m_3^2$. 
The texture of the $\cals$ matrix becomes:
\begin{equation}
  \cals^{cs} =
  \left(
  \begin{array}{ccc}
    0 & 0 & 0 \\
    0 & 0 & s_3 - m_3^2 \\
    0 & s_3 - m_3^2 & - 2 \, m_3^2
  \end{array}
  \right)
  \label{eqcalscs}
\end{equation}
Only $\bbar_1$ does not vanish, so $I_3^n$ reads simply:
\begin{align}
I_3^n 
&= \frac{2^{\varepsilon}}{\varepsilon} \, \Gamma(1+\varepsilon) \, 
\frac{\bbar_1}{\det(G)} \, 
\int^1_0 dx \, 
\left( D^{\{1\}(2)}(x) - i \, \lambda \right)^{-1-\varepsilon} \notag \\
&= \frac{\bbar_1}{\det(G)} \, W\left( \detgj{1},\tD_{12},0 \right)
\label{eqdirei3n9}
\end{align}
Using  eq. (\ref{eqdirei3n3}) and expressing $\detgj{1}$ and $\tD_{12}$ in terms of
the $\cals$ matrix elements (cf.\ eq.~(\ref{eqremd2})), we get:
\begin{align}
I_3^n &= 
\frac{\Gamma(1+\varepsilon)}{m_3^2-s_3} \, 
\left\{ 
 - \, \frac{1}{\varepsilon^2} 
 \left( {} - s_3+m_3^2 - i \, \lambda \right)^{-\varepsilon}  
 + \frac{1}{2 \, \varepsilon^2} \, 
 \left( m_3^2 - i \, \lambda \right)^{-\varepsilon}
\right. 
\notag \\
&\quad \qquad \qquad \qquad \qquad \qquad \quad {} 
\left.
 + \dilog \left( \frac{s_3}{s_3-m_3^2+i \, \lambda} \right) 
\right\}
  \label{eqdirei3n10}
\end{align}
After some algebra, we recover\footnote{In contrast to the present eq. 
(\ref{eqdirei3n10}), eq. (4.11) of ref. \cite{Ellis:2007qk} contains 
a factor $\Gamma^2(1-\varepsilon)/\Gamma(1-2\varepsilon)$ and an extra term
$+\pi^2/12$ inside the brackets. Yet they cancel against each other when 
performing the $\varepsilon$-expansion at the appropriate order.} 
formula (4.11) of ref. \cite{Ellis:2007qk}.

\vspace{0.3cm}

\subsection{Indirect way}\label{indirway}

The present subsection provides the calculation according to the 
``indirect way". The results presented are valid for both real 
and complex masses; unless explicitly specified the $i \, \lambda$ prescription 
is kept having in mind that it is ineffective in the case of complex masses.
Let us start from eq. (\ref{eqdefi3n2}).
The $x$ integration is traded for a $\rho$ integration in a way very similar to
the four-dimensional case (see subsubsec.~\ref{P1-subsubsectindway} of ref.\ \cite{paper1}) and we get:
\begin{align}
I_3^n 
&= \sum_{i \in S_3} \, \frac{\bbar_i}{\det(G)} \, 
\sum_{j \in S_3 \setminus \{i\}} \, \frac{\bbj{j}{i}}{\detgj{i}} \, 
L_3^n \left( 0, \Delta_{1}^{\{i\}}, \tD_{ij} \right)
\label{eqdef3n3}
\end{align}
with:
\begin{align}
L_3^n \left( 0, \Delta_{1}^{\{i\}}, \tD_{ij} \right)
&= \kappa_{_{IR}} \, 
\int^{+\infty}_0 \frac{d \xi}{\xi^{\nu} - i \, \lambda}  
\notag\\
&\quad {} \quad {} 
\times
\int^{+\infty}_0 
\frac{d \rho}{
 (\xi^{\nu} + \rho^2 - \Delta_1^{\{i\}} - i \, \lambda) 
 (\xi^{\nu} + \rho^2+ \tD_{ij} - i \, \lambda)^{1/2}
}
\label{eqlijsoft1}
\end{align}
and:
\[
  \kappa_{_{IR}} =2^{\varepsilon} \, \frac{\Gamma(1+\varepsilon)}{(1-\varepsilon)} \, 
  \frac{1}{B(1+\varepsilon,1-\varepsilon)}, \quad \nu = \frac{1}{1-\varepsilon}
\]

\noindent
To handle the cases with soft and/or collinear divergences, the two relevant configurations are 
1) $\tD_{ij} \neq 0$ and 2) $\tD_{ij} = 0$.
Note that when both $\Delta_2$ and $\Delta_1^{\{i\}}$ vanish 
$L_3^n ( 0, 0, \tD_{ij} )$ is 
weighting a vanishing $\bbar$, therefore it shall not be considered (cf.\ subsec.~\ref{exp_exemp_ir}).

\vspace{0.3cm}

\noindent
The $\rho$ integration can be done using appendix \ref{P2-appendJ} as in the four 
dimensional case and as the outcome of this integration, we shall distinguish two cases depending on the sign of 
$\Im(\Delta_1^{\{i\}})$.

\vspace{0.3cm}

\noindent
{\bf 1) $\tD_{ij} \neq 0$}

\vspace{0.3cm}

\noindent
{\bf 1.a) $\Im(\Delta_1^{\{i\}}) > 0$}\\
This case covers in particular real masses. After the $\rho$ integration, 
$L_3^n (0, \Delta_{1}^{\{i\}}, \tD_{ij} )$ becomes:
\begin{align}
&L_3^n (0, \Delta_{1}^{\{i\}}, \tD_{ij} )
\notag\\
&=  \kappa_{_{IR}} \,\int^{+\infty}_0 d \xi \, 
\int^{1}_0 d z \, 
\frac{1}
{
 (\xi^{\nu} - i \, \lambda) 
 (\xi^{\nu} - (1-z^2) \, \Delta_1^{\{i\}} + z^2 \, \tD_{ij} - i \, \lambda)
}
\label{eqlijsoft3}
\end{align}
The $\xi$ integration is performed first, using
eq.~(\ref{eqmodifk}) of the appendix \ref{appendJ} 
and $L_3^n (0, \Delta_{1}^{\{i\}}, \tD_{ij} )$ becomes:
\begin{align}
L_3^n (0, \Delta_{1}^{\{i\}}, \tD_{ij} ) 
&=   - \frac{2^{\varepsilon}}{\varepsilon} \, \Gamma(1+\varepsilon) \, 
\int^1_0 dz 
\left(
 z^2 \, (\tD_{ij}+\Delta_1^{\{i\}}) - \Delta_1^{\{i\}} - i \, \lambda 
\right)^{-1-\varepsilon}
\label{eqlijsoft40}
\end{align}
Expanding\footnote{Let us remind that only the terms in the 
$\varepsilon$-expansion which provide the divergent and finite terms in the 
limit $\varepsilon \to 0$ are kept.} 
the r.h.s. of eq. (\ref{eqlijsoft40}) around $\varepsilon=0$ then gives:
\begin{align}
L_{3}^{n} \left( 0,\Delta_{1}^{\{i\}},\tD_{ij} \right) 
&=  2^{\varepsilon} \, \Gamma(1+\varepsilon) \, 
\left[ 
 - \, \frac{1}{\varepsilon} \, 
 \int^1_0 
 \frac{dz}
 {z^2 \, (\tD_{ij}+\Delta_1^{\{i\}}) - \Delta_1^{\{i\}} - i \, \lambda} 
\right.
\notag\\
& \quad {} \quad {} \quad {} \quad {} \quad {} \quad {} \quad {} 
\left.
 + \int^1_0 dz \, 
 \frac{\ln(z^2 \,(\tD_{ij}+\Delta_1^{\{i\}})-\Delta_1^{\{i\}} - i \,\lambda)}
 {z^2 \, (\tD_{ij}+\Delta_1^{\{i\}}) - \Delta_1^{\{i\}} - i \, \lambda} 
\right]
\label{eqlijsoft41}
\end{align}

\vspace{0.3cm}

\noindent
{\bf 1.b) $\Im(\Delta_1^{\{i\}}) < 0$}\\
This case occurs only with complex masses,  
in a way such that the $i \, \lambda$ prescriptions are overshadowed and
ineffective: the latter are therefore dropped in this part. 
After $\rho$ integration performed as in the four-dimensional case,
we get (see eq.~(\ref{eqdeffuncj7}) of appendix~\ref{appendJ}):
\begin{align}
&L_3^n (0, \Delta_{1}^{\{i\}}, \tD_{ij} ) 
\notag \\
&= 
-  \kappa_{_{IR}} \,\int^{+\infty}_0  
\frac{d \xi}{\xi^{\nu} - i \, \lambda} \, 
\left[ 
 i  \int^{+\infty}_0 
 \frac{dz}{- \xi^{\nu} + \tD_{ij} \, z^2 + (1+z^2) \, \Delta_1^{\{i\}}} 
\right.
\notag \\
& \quad {}\quad {}\quad {}\quad {}\quad {}\quad {} \quad {}\quad {}
\quad {}\quad {}\quad {}
\left.
+ \int^{+\infty}_1 \frac{dz}
 {\xi^{\nu} + \tD_{ij} \, z^2 - (1-z^2) \, \Delta_1^{\{i\}}} 
\right]
\label{eqlijsoft5}
\end{align}
Using identity (\ref{eqmodifk}), 
$L_3^n (0, \Delta_{1}^{\{i\}}, \tD_{ij} )$ reads:
\begin{align}
L_3^n (0, \Delta_{1}^{\{i\}}, \tD_{ij} )
&=  \frac{2^{\varepsilon}}{\varepsilon} \, \Gamma(1+\varepsilon) \, 
\left[ 
 - \, i \int^{+\infty}_0 dz 
 \left(
  -z^2 \, (\tD_{ij}+\Delta_1^{\{i\}}) - \Delta_1^{\{i\}} 
 \right)^{-1-\varepsilon} 
\right. 
\notag \\
&\quad {}\quad {}\quad {}\quad {} \quad {} \quad {} \quad {}
+
\left. 
 \int^{+\infty}_1 dz \;
 \left(\, 
  z^2 \, (\tD_{ij}+\Delta_1^{\{i\}}) - \Delta_1^{\{i\}} 
 \right)^{-1-\varepsilon} 
\;
\right]
\label{eqlijsoft60}
\end{align}
Expanding the terms in the square bracket in eq.
(\ref{eqlijsoft60}) around $\varepsilon=0$ then gives:
\begin{align}
&L_{3}^{n} \left( 0,\Delta_{1}^{\{i\}},\tD_{ij} \right) 
\notag \\
&=  2^{\varepsilon} \, \Gamma(1+\varepsilon) \, 
\left\{ 
 \;
 \frac{1}{\varepsilon} \,
 \left[ 
  \; i \, \int^{+\infty}_0  
  \frac{dz}{z^2 \, (\tD_{ij}+\Delta_1^{\{i\}}) + \Delta_1^{\{i\}}} 
   + 
  \int^{+\infty}_1 
  \frac{dz}{z^2 \, (\tD_{ij}+\Delta_1^{\{i\}}) - \Delta_1^{\{i\}}} 
 \;
 \right] 
\right. 
\notag \\ 
&\quad \quad \quad \quad \quad \quad \quad {}
 -i \int^{+\infty}_0 dz \, 
 \frac{\ln(-z^2 \, (\tD_{ij}+\Delta_1^{\{i\}}) - \Delta_1^{\{i\}})}
 {z^2 \, (\tD_{ij}+\Delta_1^{\{i\}}) + \Delta_1^{\{i\}}}
\notag\\
&\quad \quad \quad \quad \quad \quad \quad {}
\left. 
 - \;\; \int^{+\infty}_1 dz \,
 \frac{\ln(z^2 \, (\tD_{ij}+\Delta_1^{\{i\}}) - \Delta_1^{\{i\}})}
 {z^2 \, (\tD_{ij}+\Delta_1^{\{i\}}) - \Delta_1^{\{i\}}} 
\right\}
\label{eqlijsoft61}
\end{align}
For eqs. (\ref{eqlijsoft41}) and (\ref{eqlijsoft61}), the $z$ integration 
can be then performed using appendix \ref{appF}.\\

\vspace{0.3cm}

\noindent
Since the cuts of 
$( z^2 \, (\tD_{ij}+\Delta_1^{\{i\}}) -\Delta_1^{\{i\}})^{-1-\varepsilon}$ and
$\ln ( z^2 \, (\tD_{ij}+\Delta_1^{\{i\}}) - \Delta_1^{\{i\}} )$ are the same, 
the discussion carried to extend the ``indirect way'' to the general complex mass 
case holds likewise here (cf.\ sec.~\ref{P2-sectthreepoint} of ref.\ \cite{paper2}). 
Eq. (\ref{eqlijsoft60}) can be rewritten as:
\begin{align}
&L_3^n (0, \Delta_{1}^{\{i\}}, \tD_{ij} )
\notag\\
&= - \frac{2^{\varepsilon}}{\varepsilon} \, \Gamma(1+\varepsilon) \, 
\left\{  \int_{0}^{i \, \infty} + \int_{+\infty}^{1} \right\} dz  
\left(
 \, z^2 \, (\tD_{ij}+\Delta_1^{\{i\}}) - \Delta_1^{\{i\}} 
\right)^{-1-\varepsilon}
  \label{eqlijsoft62}
\end{align}
In eq. (\ref{eqlijsoft62}) let us add a vanishing contribution along
the ``contour at $\infty$" in the north-east quadrant 
$\{\Re(z) > 0$, $\Im(z) > 0\}$ so as to concatenate the two 
contributions. The connected contour thus obtained can in turn be deformed into 
a finite contour $\widehat{(0,1)}_{i,j}$ stretched from 0 to 1 as pictured on 
fig. \ref{contour} thereby unifying eqs. (\ref{eqlijsoft40}) and 
(\ref{eqlijsoft62}): 

\begin{figure}[h]
\begin{center}
\includegraphics[scale=0.8]{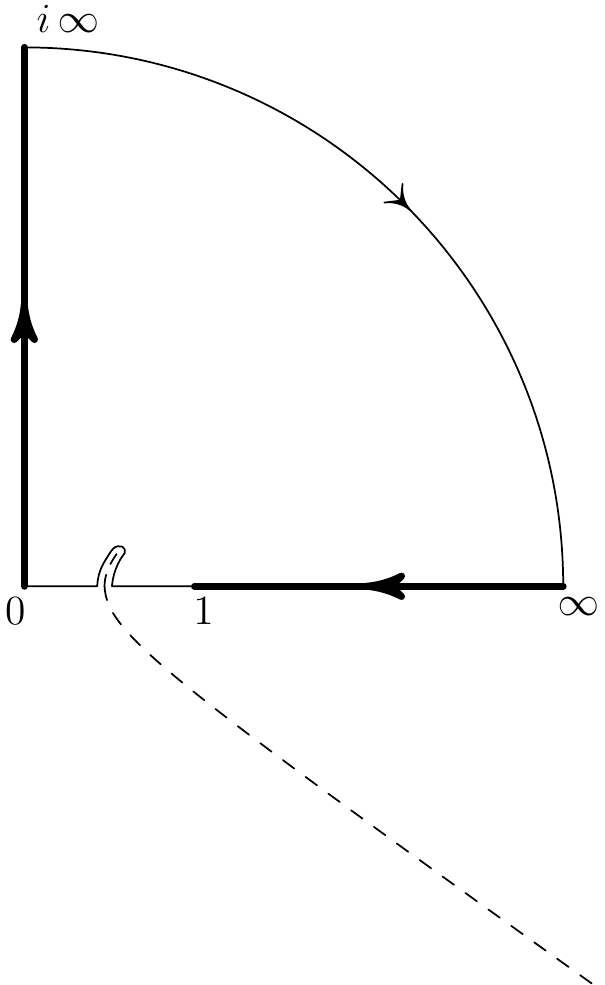}
\end{center}
\caption{\footnotesize Location of the relevant discontinuity cut 
${\cal C}_{i,j}$ with respect to the two half straight lines $[0, + i \infty[$
and $[1,+\infty[$ and deformation of the contour $\widehat{(0,1)}$ partly 
wrapping the extremity of the cut.}\label{contour}
\end{figure}

\begin{align}
L_3^n (0, \Delta_{1}^{\{i\}}, \tD_{ij} )
&= - \frac{2^{\varepsilon}}{\varepsilon} \, \Gamma(1+\varepsilon) \, 
\int_{\widehat{(0,1)}_{i,j}} dz  
\left(
 \, z^2 \, (\tD_{ij}+\Delta_1^{\{i\}}) - \Delta_1^{\{i\}} 
\right)^{-1-\varepsilon}
  \label{eqlijsoft63}
\end{align}

\vspace{0.3cm}

\noindent
{\bf 2) $\tD_{ij} = 0$}\\
Here again we shall distinguish two cases depending on the sign of 
$\Im(\Delta_1^{\{i\}})$.

\vspace{0.3cm}

\noindent
{\bf 2.a) $\Im(\Delta_1^{\{i\}}) > 0$}\\
This case covers in particular real masses.
The calculation initially amounts to setting $\tD_{ij} = 0$ in eq. 
(\ref{eqlijsoft40}), which becomes:
\begin{align}
L_{3}^{n} \left( 0,\Delta_{1}^{\{i\}},0 \right)  
&=  - \, \frac{2^{\varepsilon}}{\varepsilon} \, \Gamma(1+\varepsilon) \, 
\left( - \Delta_1^{\{i\}} - i \, \lambda \right)^{-1-\varepsilon} \, 
\int^1_0 dz \left(1-z^2 \right)^{-1-\varepsilon}
\label{eqlijsoft7}
\end{align}
The integration over $z$ is performed using eq.~(\ref{thirdinty}) and
$L_{3}^{n}(0,\Delta_{1}^{\{i\}},0) $ becomes:
\begin{align}
L_{3}^{n} \left( 0,\Delta_{1}^{\{i\}},0 \right)  
&= \frac{1}{\varepsilon^2} \, \Gamma(1+\varepsilon) \, 
\frac{\Gamma(1-\varepsilon)^2}{\Gamma(1-2 \, \varepsilon)} \,
\left( - 2 \, \Delta_1^{\{i\}} - i \, \lambda \right)^{-1-\varepsilon}
\label{eqlijsoft8}
\end{align}

\noindent
{\bf 2.b) $\Im(\Delta_1^{\{i\}}) < 0$} \\
Here again, this case occurs only with complex masses,  
with the $i \, \lambda$ prescriptions overshadowed and therefore dropped. 
The calculation initially amounts anew to setting  $\tD_{ij} = 0$ in eq.
(\ref{eqlijsoft60}) which becomes:
\begin{align}
L_{3}^{n} \left( 0,\Delta_{1}^{\{i\}},0 \right)
&=  \frac{2^{\varepsilon} \, \Gamma(1+\varepsilon)}{\varepsilon} \, 
\left[ 
 - i \, \left( - \Delta_1^{\{i\}} \right)^{-1-\varepsilon} \, 
 \int^{+\infty}_0 dz \left(1+z^2 \right)^{-1-\varepsilon} 
\right.
\notag\\
& \quad {}\quad {}\quad {}\quad {}\quad {}\quad {}\quad {}\quad {}
\left.
 - \left( \Delta_1^{\{i\}} \right)^{-1-\varepsilon} \, 
 \int^{+\infty}_1 dz \left(z^2-1 \right)^{-1-\varepsilon} 
\right]
\label{eqlijsoft9}
\end{align}
The $z$ integrals are computed in appendix \ref{herba} (eqs.~(\ref{firstinty}) and (\ref{secondinty})).
Hence for $L_{3}^{n}(0,\Delta_{1}^{\{i\}},0)$:
\begin{align}
L_{3}^{n} \left( 0,\Delta_{1}^{\{i\}},0 \right)
&=  
- \, \frac{2^{-\varepsilon}}{2 \, \varepsilon^2} \, \Gamma(1+\varepsilon) \, 
\frac{\Gamma^{2}(1-  \varepsilon)}{\Gamma(1 - 2 \,\varepsilon)} \, 
\frac{1}{\cos(\pi \, \varepsilon)}
\notag \\
&
\;\;\;\;\;\;\;\; {} \times
\left[ 
 i \, \sin(\pi \, \varepsilon) \, 
\left( - \Delta_1^{\{i\}} \right)^{-1-\varepsilon} \,   
 + 
\left( \Delta_1^{\{i\}} \right)^{-1-\varepsilon} \,  
\right]
\label{eqlijsoft10}
\end{align}
Since $\Im(\Delta_1^{\{i\}}) < 0$, $( \Delta_1^{\{i\}} )^{-1-\varepsilon}$ may
be rewritten as $- \, \text{e}^{\, i \, \pi \, \varepsilon}\, 
( - \, \Delta_1^{\{i\}} )^{-1-\varepsilon}$, so that 
$L_{3}^{n}(0,\Delta_{1}^{\{i\}},\tD_{ij})$ simplifies into:
\begin{align}
L_{3}^{n} \left( 0,\Delta_{1}^{\{i\}},0 \right)
&=  
\frac{1}{\varepsilon^2} \, \Gamma(1+\varepsilon) \, 
\frac{\Gamma^{2}(1-  \varepsilon)}{\Gamma(1 - 2 \,\varepsilon)} \, 
 \left( - 2 \, \Delta_1^{\{i\}} \right)^{-1-\varepsilon} \,   
\label{eqlijsoft10bis}
\end{align}
which coincides with eq. (\ref{eqlijsoft8}). 

\vspace{0.3cm}

\noindent
Let us show now the equivalence between the ``indirect way'' and the ``direct way'' starting 
from the integral representation of $L_3^n(0,\Delta_1^{\{i\}},\tD_{ij})$ and disregarding 
the fact that some $\tD_{ij}$ may or may not vanish.
Thus, $I_3^n$ now reads:
\begin{align}
I_3^n 
&= -  \frac{2^{\varepsilon}}{\varepsilon} \, \Gamma(1+\varepsilon) \, 
\sum_{i \in S_3} \, \frac{\bbar_i}{\det(G)} \, 
\sum_{j \in S_3 \setminus \{i\}} \, \frac{\bbj{j}{i}}{\detgj{i}}
\notag \\
&\quad \quad \quad \quad
\times 
\int_{\widehat{(0,1)}_{i,j}} dz  \,
\left(
 \, z^2 \, (\tD_{ij}+\Delta_1^{\{i\}}) - \Delta_1^{\{i\}} 
\right)^{-1-\varepsilon}
 \label{eqlijsoft64}
\end{align}
Sticking to the general complex mass case, we perform the following change of
variable: $s = \bbj{j}{i} \, z$ in such a way that the two integrands
corresponding to the sum over $j$ (at $i$ fixed) in eq. (\ref{eqlijsoft64}) are
the same (use eqs.~(\ref{P1-eqtruc1}) and (\ref{P1-eqtruc3}) of \cite{paper1}). 
Specifying the two elements of $S_3 \setminus \{i\}$ to be 
$k \equiv 1 + ((i+1)$ modulo $3)$ and $l \equiv 1 + (i$ modulo $3)$, the two 
integrals are concatenated into a single one integrated along the contour 
${\cal I}^{(i)}_{k,l} \equiv  -\bbj{k}{i}\widehat{(0,1)}_{i,k} \cup
\bbj{l}{i}\widehat{(0,1)}_{i,l}$ in  the complex $s$-plane:
\begin{align}
I_3^n 
&= -  \frac{2^{\varepsilon}}{\varepsilon} \, \Gamma(1+\varepsilon) \, 
\sum_{i \in S_3} \, \frac{\bbar_i}{\det(G) \detgj{i}} \, 
\notag \\
&\quad \quad \quad
\times 
\int_{{\cal I}^{(i)}_{k,l}} ds  
\left( 
\frac{s^2 + \detsj{i}}{\detgj{i}} 
\right)^{-1-\varepsilon}
\label{eqlijsoft65}
\end{align}

\begin{figure}[h]
\begin{center}
\includegraphics[scale=0.4]{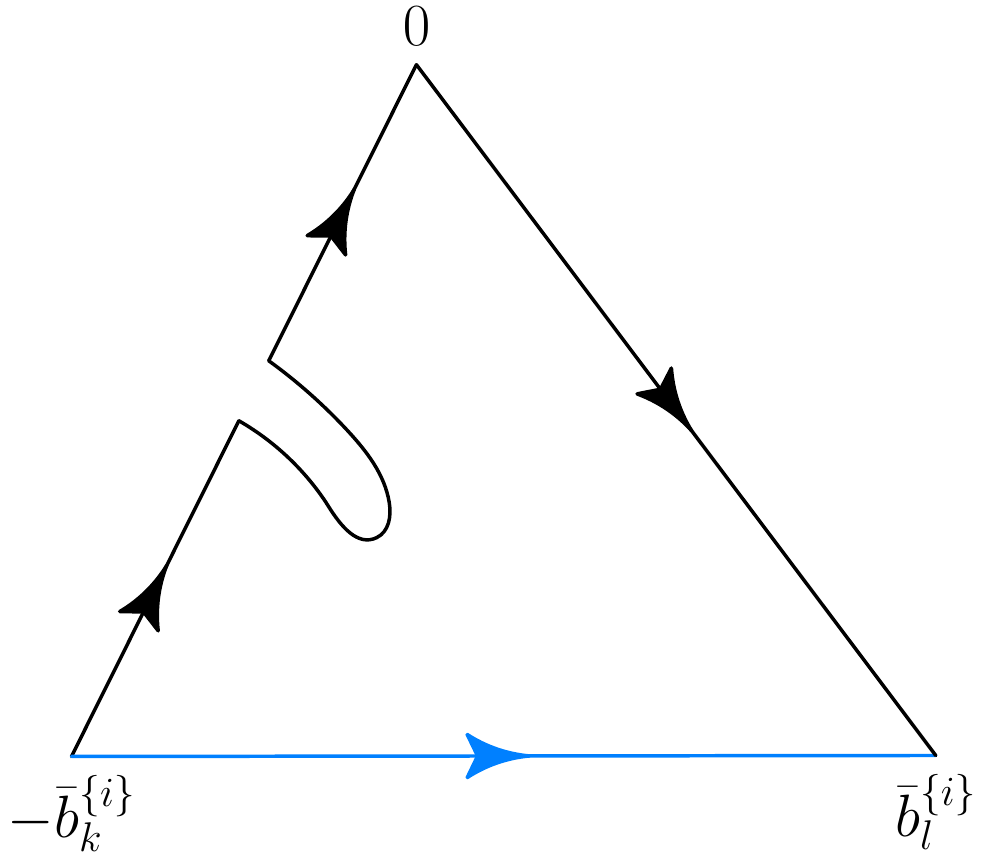}
\end{center}
\caption{\footnotesize Example of a contour deformation involving a 
triangle with one distorted side, for which no cut crosses the straight base 
$[ -\bbj{k}{i},\bbj{l}{i}]$.}\label{triangle1}
\end{figure}

\vspace{0.3cm}

\noindent
As in the case $\Delta_2 \ne 0$ treated in \cite{paper2}, the contour ${\cal I}^{(i)}_{k,l}$ can be 
deformed into the straight line $\left[ -\bbj{k}{i},\bbj{l}{i} \right]$ as depicted on figure \ref{triangle1} :
\begin{align}
I_3^n &= -  \frac{2^{\varepsilon}}{\varepsilon} \, \Gamma(1+\varepsilon) \, 
\sum_{i \in S_3} \, \frac{\bbar_i}{\det(G) \detgj{i}} \, 
\notag \\
&\quad 
\times 
\int_{- \bbj{k}{i}}^{\bbj{l}{i}} ds  
\left( 
\frac{s^2 + \detsj{i}}{\detgj{i}} 
\right)^{-1-\varepsilon}
\label{eqlijsoft66}
\end{align}
Performing the change of variable: 
$s = -\bbj{k}{i} - \detgj{i} \,u$ and following the procedure given by eqs.~(\ref{P1-cvar1}) to (\ref{P1-eqdefgi}) of ref.\ \cite{paper1} leads to:
\begin{align}
I_3^n &=  \frac{2^{\varepsilon}}{\varepsilon} \, \Gamma(1+\varepsilon) \, 
\sum_{i \in S_3} \, \frac{\bbar_i}{\det(G)} \,
\int_{0}^{1} du  \left( D^{\{i\} \, (l)}(u) \right)^{-1-\varepsilon}
\nonumber
\end{align}
which is namely eq. (\ref{eqdirei3n1}).

\vspace{0.3cm}

\noindent
It is instructive to recover the results of the ``direct way'' from the ones of 
the ``indirect way'' using for the latter the closed form formulae. These formulae 
can be obtained for the case $\tD_{ij} \ne 0$ by using results of appendix \ref{appF} 
and for $\tD_{ij} = 0$ using eq.~(\ref{eqlijsoft8}). This exercise is performed 
with great details in appendix~\ref{direcway3pIR}.

\section{Four-point function with Infrared divergences}\label{sectfourpointir}

\begin{figure}[h]
\centering
\parbox[c][43mm][t]{40mm}{\begin{fmfgraph*}(60,40)
  \fmfleftn{i}{2} \fmfrightn{o}{2}
  \fmf{fermion,label=$p_1$}{i2,v1}
  \fmf{fermion,label=$p_2$}{i1,v2}
  \fmf{fermion,label=$p_3$}{o1,v3}
  \fmf{fermion,label=$p_4$}{o2,v4}
  \fmf{fermion,tension=0.5,label=$q_1$}{v1,v2}
  \fmf{fermion,tension=0.5,label=$q_2$}{v2,v3}
  \fmf{fermion,tension=0.5,label=$q_3$}{v3,v4}
  \fmf{fermion,tension=0.5,label=$q_4$}{v4,v1}
\end{fmfgraph*}}
\caption{The box picturing the one-loop four-point function.}
\label{fig2} 
\end{figure}
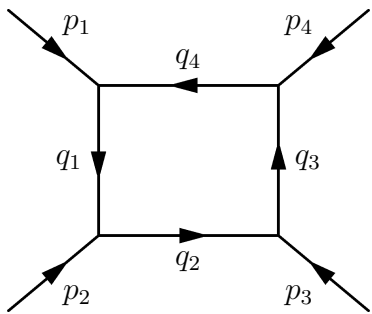

\noindent
In the case where infrared divergences appear, the latter can be regularised by 
dimensional regularisation shifting the space-time dimension 
$n = 4 - 2 \varepsilon$ slightly above 4 ($\varepsilon < 0$). The Feynman 
parametrisation of $I_4^n$ reads:
\begin{eqnarray} 
I_4^n 
& = & 
\Gamma\left(2 + \varepsilon \right) 
\int_0^1 \, \prod_{i=1}^4 \, \delta(1- \sum_{i=1}^4 z_i)  
\left(  -\frac{1}{2} \, Z^{\;T} \cdot
\cals \cdot Z - i \, \lambda  \right)^{-2 - \varepsilon} 
\label{eqstartingpointir}
\end{eqnarray} where $Z$ is a column 4-vector whose components are the $z_{i}$. 
The power $- 2 - \varepsilon$ in 
eq. (\ref{eqstartingpointir}) is not an integer. Notwithstanding, the 
tricks and techniques
elaborated in section \ref{P1-sectfourpoint} of ref. \cite{paper1} can be used with a slight adaptation.
Let us sketch the different steps for this special case.

\subsection{Computation of $I_4^{n}$}\label{compI4n}

We make use of identity (\ref{P1-eqFOND1biss}) of ref. \cite{paper1} to shift the power of the
denominator in the integrand, choosing $\mu = 5/2$ and 
$\nu = 2/(1-2 \varepsilon)$ so that
$I_4^{n}$ is recast as:
\begin{eqnarray}
  I_4^{n} 
& = & 
\frac{2^{3+\varepsilon}}{B(2 + \varepsilon,1/2-\varepsilon)} \, 
\frac{\Gamma(2 + \varepsilon)}{(1-2 \, \varepsilon)} \, 
\nonumber \\
&& \mbox{} \times
\int_0^{+\infty} d \xi \,
\int_{\Sigma_{bcd}}
\frac{dx_b \, d x_c \, d x_d}
{(D^{(a)}(x_b,x_c,x_d)  + \xi^{\nu} - i \, \lambda)^{5/2}}
\label{eqI4b1}
\end{eqnarray}
Step 1 is very similar to the case $n=4$ and 
we get:
\begin{eqnarray}
I_4^{n} 
& = & 
\frac{2^{3+\varepsilon}}{3 \, B(2 +\varepsilon,1/2 - \varepsilon)} \, 
\frac{\Gamma(2 +\varepsilon)}{(1-2 \, \varepsilon)} \, 
\nonumber \\
&& \mbox{} \times
\sum_{i=1}^{4} \, \frac{\bbar_i}{\detg} \,
\int_0^{+\infty} 
d \xi \, \frac{1}{\Delta_3-\xi^{\nu} + i \, \lambda} 
\nonumber\\ 
& & 
\quad{} \quad{} \quad{} \quad{}
\times \int_{\Sigma_{kl}}
\frac{dx_k \, d x_l}
{(D^{\{i\} \, (i^{\prime})}(x_k,x_l) 
 + \xi^{\nu} - i \, \lambda)^{3/2}}
  \label{eqI4b2}
\end{eqnarray}
with the same notational conventions as in eqs. (\ref{P1-eqI442}) and 
(\ref{P1-eqI448}) of sec. \ref{P1-sectfourpoint} of ref. \cite{paper1}.
Likewise, steps 2 and 3 are identical to section \ref{P1-sectfourpoint} of ref. \cite{paper1} but 
the power of the variable $\xi$: $\nu$ instead of 2, and will not be repeated 
here.
Regarding step 4, the integration shall be performed over the variables 
$\xi$, $\rho$ then $\sigma$ in the corresponding $L_4^n(\Delta_3,\Delta_2^{\{i\}},\Delta_1^{\{i,j\}},\tD_{ijk})$ now given by: 
\begin{align}
L_4^n(\Delta_3,\Delta_2^{\{i\}},\Delta_1^{\{i,j\}},\tD_{ijk})
&= 
\kappa \,\int^{+\infty}_0 d \xi  \, \int^{+\infty}_0 d \rho \, 
\int^{+\infty}_0 d \sigma 
\frac{1}{\xi^{\nu} - \Delta_3 - i \, \lambda} \, 
\nonumber \\
&\quad 
\mbox{} \times \frac{1}{\xi^{\nu} + \rho^2 - \Delta_2^{\{i\}} - i \, \lambda} \, 
\frac{1}{\xi^{\nu} + \rho^2 + \sigma^2 - \Delta_1^{\{i,j\}} - i \, \lambda} \notag \\ 
&\quad {} \times \frac{1}{(\tD_{ijk} + \xi^{\nu} + \rho^2 + \sigma^2 - i \, \lambda)^{1/2}}
\label{eqdefnl1}
\end{align}
with
\[
\kappa = 
\frac{2^{4+\varepsilon}}
{3 \, B(2+\varepsilon,1/2-\varepsilon) \, B(3/2,1/2) \, B(1,1/2)}
\frac{\Gamma(2+\varepsilon)}{(1-2 \, \varepsilon)} 
\]
reminiscent of eq. (\ref{P1-eqDEFK2}) of sec. 
\ref{P1-sectfourpoint} of ref. \cite{paper1}.
Infrared divergences in the four-point function are not dominant Landau-type
singularities but subleading ones. Such an infrared divergence corresponds to 
the vanishing determinant $\detsj{i}$ of some reduced $\cals$ matrix (or 
some $\Delta_2^{\{i\}}$), associated 
with some three-point functions which are obtained from the four-point function 
considered by one pinching \cite{Binoth:2005ff}. 
The various cases of vanishing kinematic matrices associated
with three-point functions plagued with infrared soft or collinear 
singularities have been evoked in sec. \ref{3point_ir}.
Let us note that the method developed in section \ref{P1-sectfourpoint} of ref. \cite{paper1} is still
valid because we never divide by $\Delta_2^{\{i\}}$ {\em per se} but by
$\Delta_2^{\{i\}}- \xi^{\nu}- \rho^2$.
The divergences will show up when performing the integrations over the
parameters of eq. (\ref{eqdefnl1}). Note also that if we face a case when 
IR divergences arise whereas there are some non vanishing internal masses, 
only some of the contributions (let us call them sector)``$i,j,k$", not all, are plagued with IR 
divergences; the other ones, to which corresponds $\Delta_2^{\{i\}} \ne 0$,
will be treated as in the $n=4$ case.
Besides, in any IR-divergent sector ``$i$'' for which 
$\Delta_2^{\{i\}}=0$, we do not have to sum over all three sub-sectors 
$j \in S_{4} \setminus \{i\}$ because, as was seen in the 
IR-divergent three-point function case, some of the $\bbj{i}{j}$ vanish.

\vspace{0.3cm}

\noindent
Let us consider a sector $i$ which diverges in the IR region. 
We have to distinguish two cases : 1) when $\Delta_2^{\{i\}} = 0$ whereas 
$\tD_{ijk} \ne 0$ in which case there are only soft divergences,
2) when both $\Delta_2^{\{i\}} = 0$ and $\tD_{ijk} = 0$ in which case there 
are collinear or both soft and collinear divergences. 
In this section, we will treat at once
the real and complex mass case. Some of the
cases correspond to real or complex masses, others to
complex masses only. For those corresponding real or complex 
masses, we keep the imaginary part $- i \, \lambda$ explicitly having in mind 
that with complex masses this $- i \, \lambda$ is ineffective.
Let us stress in passing that we also keep an infinitesimal prescription 
$- i \, \lambda$ in the pole term where we put $\Delta_{2}^{\{i\}} = 0$.

\subsection{$\Delta_2^{\{i\}} = 0$ and $\tD_{ijk} \ne 0$}\label{subsect52}

\begin{align}
L_4^n(\Delta_3,0,\Delta_1^{\{i,j\}},\tD_{ijk}) &=
  \kappa \,\int^{+\infty}_0 d \xi  \, \int^{+\infty}_0 d \rho \, 
\int^{+\infty}_0 d \sigma 
\frac{1}
{(\xi^{\nu} - \Delta_3 - i \, \lambda) \, (\xi^{\nu} + \rho^2  - i \, \lambda)} 
\nonumber \\
&\quad {} \quad {}
\times  
\frac{1}
{(\xi^{\nu} + \rho^2 + \sigma^2 - \Delta_1^{\{i,j\}} - i \, \lambda) \, 
(\tD_{ijk} + \xi^{\nu} + \rho^2 + \sigma^2 - i \, \lambda)^{1/2}}
\label{eqdefnlir2}
\end{align}
In the complex mass case, $\Im(\Delta_3)$ and 
$\Im(\Delta_1^{\{i,j\}})$ have arbitrary signs and $\Im(\tD_{ijk})$ is 
negative, so we have to distinguish between different cases.

\vspace{0.3cm}

\noindent
Let us define the function $M_2(\xi^{\nu})$ as:
\begin{align}
  L_4^n(\Delta_3,0,\Delta_1^{\{i,j\}},\tD_{ijk}) &= \kappa \,\int^{+\infty}_0 d \xi  \, \frac{1}{\xi^{\nu} - \Delta_3 - i \, \lambda} \, M_2(\xi^{\nu})  
  \label{eqdefM2}
\end{align}
The integrations over $\sigma$ and $\rho$ are identical with those appearing in the massive cases,
we thus borrow the results derived in subsec.\ \ref{P1-fourpointstep4} of ref.\ \cite{paper1} and in subsec.~\ref{P2-sectfourpointcomp} of ref.~\cite{paper2} with 
$\Delta_2^{\{i\}} = 0$, explicitly
  \footnote{
    The integration over $\sigma$ is of the ``second kind'' (cf.\ eqs.~(\ref{eqdeffuncj2}) and (\ref{eqdeffuncj7})) 
    while the integration over $\rho$ is of the ``first kind'' (cf.\ eq.~(\ref{eqdefk1ext})). 
    For both integrations, the power $\nu$ appearing in the eqs.~(\ref{eqdefk1ext}), (\ref{eqdeffuncj2}) and (\ref{eqdeffuncj7}) is taken equal to $2$.}: 
\begin{itemize}
\item[-] for $\Im(\Delta_1^{\{i,j\}}) \geq 0$,
\begin{align}
M_2(\xi^{\nu}) 
&= 
\frac{1}{2} \, B(1/2,1/2) \, 
\int^1_0 \frac{d u}{u^2 \, (\tD_{ijk}+\Delta_1^{\{i,j\}}) -\Delta_1^{\{i,j\}}} 
\notag \\
&\quad {} \times 
\left[ 
 \frac{1}{\left(\xi^{\nu}-i \, \lambda \right)^{1/2}}
 - 
 \frac{1}
 {\left(\xi^{\nu}+ u^2 \, (\tD_{ijk}+\Delta_1^{\{i,j\}})
  - 
  \Delta_1^{\{i,j\}} - i \, \lambda \right)^{1/2}} 
\right]
\label{eqisigir3}
\end{align}

\item[-] for $\Im(\Delta_1^{\{i,j\}}) < 0$,
\begin{align}
M_2(\xi^{\nu}) 
&= 
- \frac{1}{2} \, B(1/2,1/2) \, 
\left\{ 
  i \, \int^{+\infty}_0 
 \frac{d u}{u^2 \, (\tD_{ijk}+\Delta_1^{\{i,j\}}) +\Delta_1^{\{i,j\}}} 
\right. 
\notag \\
&\quad {} \times 
 \left[ 
  \frac{1}{\left(\xi^{\nu}-i \, \lambda \right)^{1/2}} - 
  \frac{1}{\left(\xi^{\nu}- u^2 \, (\tD_{ijk}+\Delta_1^{\{i,j\}})
  -\Delta_1^{\{i,j\}} \right)^{1/2}} 
 \right] 
\notag \\
&\quad {} 
+ \int^{+\infty}_1 
\frac{d u}{u^2 \, (\tD_{ijk}+\Delta_1^{\{i,j\}}) -\Delta_1^{\{i,j\}}} 
\notag \\
&\quad {} \times 
\left. 
 \left[ 
  \frac{1}{\left(\xi^{\nu}-i \, \lambda \right)^{1/2}} 
  - 
  \frac{1}
  {\left(\xi^{\nu}+ u^2 \, (\tD_{ijk}+\Delta_1^{\{i,j\}})-\Delta_1^{\{i,j\}} 
  \right)^{1/2}} 
 \right] 
\right\}
\label{eqisigir4}
    \end{align}
\end{itemize}
The integration over $\xi$ to obtain $L_4^n$ is of the ``second kind'' (cf.\ eqs.~(\ref{eqdeffuncj2}) and (\ref{eqdeffuncj7}) with $\nu = 2/(1 - 2 \, \varepsilon)$).
Let us go through the different cases with respect to the sign
of the imaginary part of $\Delta_3$ and $\Delta_1^{\{i,j\}}$. Sticking with the
notations of section \ref{P1-sectfourpoint} of ref. \cite{paper1}, we introduce:
\begin{align*}
 P_{ijk} &=  \tD_{ijk} + \Delta_1^{\{i,j\}} \\
 R_{ij}  &=  - \, \Delta_1^{\{i,j\}}   \\
 T       &=  - \, \Delta_3 
\end{align*}
The strategy for computing the different integrals is the same for all cases
and very similar to  section \ref{P2-sectfourpoint} of ref. \cite{paper2}. 
We display the successive steps for the first case tackled only and 
we give the final result for the three others.

\subsubsection{$\Im(\Delta_3) > 0$, $\Im(\Delta_1^{\{i,j\}}) > 0$\label{subsubcase1}}

In this case, the three complex numbers $\Delta_3$, $\Delta_1^{\{i,j\}}$ and $\tD_{ijk}$ have an imaginary part of the same sign,
We start with eq.~(\ref{eqisigir3}) for $M_2(\xi^{\nu})$ and use eq.~(\ref{eqdeffuncj2}) for the $\xi$ integration to get:
\begin{align}
L_4^n(\Delta_3,0,\Delta_1^{\{i,j\}},\tD_{ijk})
&
= F(\varepsilon) \, \int^1_0 \frac{d u}{u^2 \, P_{ijk} + R_{ij}} \notag \\
&\qquad \qquad \quad {} \times \left[ 
 \int^1_0 \frac{d z}
 {\left( (1-z^2) \, T - i \, \lambda \right)^{1+\varepsilon}} 
\right. 
\label{eqlijkir10} \\
&\qquad \qquad \qquad \quad {}
- 
\left. 
 \int^1_0 \frac{d z}
 {\left( 
   z^2 \, (u^2 \, P_{ijk} + R_{ij})+ (1-z^2) \, T  - i \, \lambda 
  \right)^{1+\varepsilon}
 } 
\right]
\notag
\end{align}
where 
\begin{align}
F(\varepsilon) 
&= 2^{1+\varepsilon} \, \Gamma(1+\varepsilon)
\label{eqdeffepsilon}
\end{align}
To facilitate the reading, we will define some steps in a similar way as in the subsec. \ref{P1-fourpointstep4} of ref.\ \cite{paper1}. Let us proceed along them. \\

\noindent
{\bf 1)} We change $u=\sqrt{y/x}$ and $z = \sqrt{x}$ and exchange the $y$ and the $x$ integration so that:
\begin{align}
L_4^n(\Delta_3,0,\Delta_1^{\{i,j\}},\tD_{ijk})
&= 
- \frac{F(\varepsilon)}{4} \, 
\int^1_0 \frac{d y}{\sqrt{y}} \, \int^1_y d x \, 
\frac{1}{y \, P_{ijk} + x \, R_{ij}} \notag \\
&\qquad \qquad {} \times \left[  
 \frac{1}{\left( y \, P_{ijk} + x \, (R_{ij} - T) + T 
 - i \, \lambda \right)^{1+\varepsilon}} 
\right. 
\notag \\
&\qquad  \qquad  \qquad  {}
- 
\left.  
\frac{1}{\left( (1-x) \, T - i \, \lambda \right)^{1+\varepsilon}}  
\right]
\label{eqlijkir12}
\end{align}

\noindent
{\bf 2)} We set $y = u^2$ and perform a partial fraction decomposition on the variable $x$ to write 
$L_4^n(\Delta_3,0,\Delta_1^{\{i,j\}},\tD_{ijk})$ in the following form:
\begin{align}
\hspace{2em}&\hspace{-2em}L_4^n(\Delta_3,0,\Delta_1^{\{i,j\}},\tD_{ijk}) \notag \\
&= 
- \frac{F(\varepsilon)}{2} \, 
\int^1_0 \frac{du}{u^2 \, P_{ijk} T + R_{ij} (T - i \, \lambda)}
\int^1_{u^2} d x \, 
\notag \\
&\qquad \qquad {}
\times 
\Biggl\{ 
  (T - R_{ij}) \, 
   \left[ 
    u^2 \, P_{ijk} + x \, (R_{ij} - T) + T - i \, \lambda 
   \right]^{-1-\varepsilon} \notag \\
   &\qquad \qquad \qquad {}
   - 
   T \, 
  \left[ (1-x) \, T - i \, \lambda \right]^{-1-\varepsilon}  
  + \frac{R_{ij}}{u^2 \, P_{ijk} + x \, R_{ij}}
\notag \\ 
&\qquad {} \qquad \qquad {}
 \times 
 \Biggl[  
  \left[ 
   u^2 \, P_{ijk} + x \, (R_{ij} - T) + T - i \, \lambda 
  \right]^{-\varepsilon}  
  - 
  \left[ (1-x) \, T - i \, \lambda \right]^{-\varepsilon} \, 
 \Biggr] 
\Biggr\} 
\label{eqlijkir131}
\end{align}
In eq. (\ref{eqlijkir131}), the last line provides a contribution of order
$\varepsilon$ only: for the computation of the one-loop four-point function it
can thus be dropped\footnote{Similar truncations of $\varepsilon$ expansions
will be performed everywhere throughout this section. In the perspective of
computing generalised one-loop building blocks to be used in computations beyond
one-loop, one might be led to keep further terms evanescent with $\varepsilon$
to the appropriate order, whenever such terms would hit $1/\varepsilon$ poles
generated by the extra integrations, cf. introduction of \cite{paper1}.}.\\ 

\noindent
{\bf 3)} Thus, ignoring these terms, the integration in $x$ is readily done providing:
\begin{align}
L_4^n(\Delta_3,0,\Delta_1^{\{i,j\}},\tD_{ijk})
&= 
\frac{2^{\varepsilon}}{T} \, \frac{\Gamma(1+\varepsilon)}{\varepsilon}
\int^1_0   
\frac{d u}{u^2 \, P_{ijk} + R_{ij} - i \, \lambda \, \sigma_0} \,
\label{eqlijkir14} \\
& 
\quad {}\quad {}\quad {}\quad {}\quad {}\quad {}
\times 
\Biggl\{
 \left[ 
   u^2 \, (P_{ijk} + R_{ij} - T) + T - i \, \lambda 
 \right]^{-\varepsilon}  
\notag\\
&
\quad {}\quad {}\quad {}\quad {}\quad {}\quad {}\quad {}\quad {}
 - 
 \left[ 
   u^2 \, P_{ijk} + R_{ij} - i \, \lambda 
 \right]^{-\varepsilon}  
 -  (T - i \, \lambda)^{-\varepsilon} \, (1-u^2)^{-\varepsilon} 
\Biggr\}
\notag 
\end{align}
where we have introduced $\sigma_0 = \sign(R_{ij}/T)$ in the real mass case.
In the complex mass case all ``$-i \lambda$" and``$- i \sigma_0 \, \lambda$" 
contour prescriptions are 
ineffective and irrelevant, and all three terms in eq. (\ref{eqlijkir14}) 
can be straightforwardly expanded in powers of $\varepsilon$. In contrast the 
real mass case requires a more cautious treatment which we elaborate below.
We note that, when $u^2 \to - R_{ij}/P_{ijk}$, 
\begin{align}
u^2 \, (P_{ijk} + R_{ij} - T) + T & \to (T - R_{ij}) \frac{\tD_{ijk}}{P_{ijk}}
\neq 0
\notag\\
1-u^2 & \to \frac{\tD_{ijk}}{P_{ijk}} \neq 0
\notag
\end{align} 
i.e. the pole at $u^2 = - R_{ij}/P_{ijk}$ is distinct from the branch points
of the first and third functions in the numerator. Therefore we can readily perform an
expansion around $\varepsilon=0$ for the first and third term as:
\begin{align}
\frac{1}{\varepsilon} \,
\left[ 
 u^2 \, (P_{ijk} + R_{ij} - T) + T - i \, \lambda 
\right]^{-\varepsilon}  
& =
\frac{1}{\varepsilon}  -
\ln
\left[ 
 u^2 \, (P_{ijk} + R_{ij} - T) + T - i \, \lambda 
\right] + {\cal O}(\varepsilon)
\notag\\
\frac{1}{\varepsilon} \,
(T - i \, \lambda)^{-\varepsilon} \, (1-u^2)^{-\varepsilon} 
& = 
- \frac{1}{\varepsilon}  - 
\ln(T - i \, \lambda) - \ln(1-u^2) + {\cal O}(\varepsilon)
\notag
\end{align}
The $1/\varepsilon$ poles cancel between these two contributions leaving only 
logarithms. On the other hand the contribution coming from the second term requires 
some care since pole and branch point coincide. If this singularity lies outside
the integration region the contour prescription for the pole is irrelevant and
can be dropped. If the singularity lies inside $[0,1]$ and $\sigma_0=+$ the contour
prescription for the pole and cut are the same, there is no pinching. The
integral over $u$ can be performed after an expansion in $\varepsilon$ using
appendix \ref{appF} (cf.\ eq.~(\ref{eqdefh05})). If the singularity lies inside $[0,1]$ and 
$\sigma_0 = -$ however, a pinching occurs at the singular point in the limit 
$\lambda \to 0^{+}$. A too early expansion in powers
of $\varepsilon$ before performing the integration over $u$ would lead to a
divergence order by order in $\varepsilon$. On the other hand, we note that
for $u^2 = - R_{ij}/P_{ijk}$ the numerator of the second
term - i.e. the pole residue - is $(-  \, i \, \lambda)^{-\varepsilon} \to 0$ 
with $\lambda \to 0^{+}$ for any fixed $\varepsilon <0$. Therefore, up to terms
vanishing $\propto \lambda^{- \varepsilon}$ we can make the replacement
\[
\int_{0}^{1} du \,
\frac{\left[  u^2 \, P_{ijk} + R_{ij} - i \, \lambda \right]^{-\varepsilon}}
{u^2 \, P_{ijk} + R_{ij} + i \, \lambda}   
\to 
\int_{0}^{1} du \,
\frac{\left[  u^2 \, P_{ijk} + R_{ij} - i \, \lambda \right]^{-\varepsilon}}
{u^2 \, P_{ijk} + R_{ij} - i \, \lambda}   
\]
Details are provided in appendix \ref{ir-lambda}.
Finally, we perform a partial expansion around $\varepsilon=0$ and we get:
\begin{align}
\hspace{2em}&\hspace{-2em}L_4^n(\Delta_3,0,\Delta_1^{\{i,j\}},\tD_{ijk}) = \frac{2^{\varepsilon}}{T} \, \Gamma(1+\varepsilon) \notag \\
& {} \times 
\left\{ 
 \int^1_0 \, \frac{d u }{u^2 \, P_{ijk} + R_{ij} - i \, \lambda \, \sigma_0} 
\left[
 \ln 
 \left(
  (T - i \lambda) \, (1-u^2)
 \right) -    
 \ln 
 \left(
  u^2 \, (P_{ijk} + R_{ij} - T) + T - i \, \lambda 
 \right)   
 \right]
\right. 
\notag\\
&\qquad{}
\left.
 - 
 \frac{1}{\varepsilon} \, 
 \int^1_0 \, \frac{d u }{u^2 \, P_{ijk} + R_{ij} - i \, \lambda}  
 + 
 \int^1_0 d u \, 
 \frac{\ln(u^2 \, P_{ijk} + R_{ij} - i \, \lambda)}
{u^2 \, P_{ijk} + R_{ij} - i \, \lambda} 
\right\}
  \label{eqlijkir15}
\end{align}
Let us notice that the imaginary part of the argument of each logarithm in eq.~(\ref{eqlijkir15}) keeps a constant sign when $u$ spans $[0,1]$.
This is obviously true for the real mass case and in the case of complex masses, this is easily verified keeping in mind the assumptions: $\Im(T) < 0$ and $\Im(R_{ij}) < 0$. 
For the sake of coherence with respect to ref. \cite{paper2}, the pole residue contributions will 
be added and subtracted for the two logarithms making up the first term inside the curly 
brackets of eq.~(\ref{eqlijkir15}). This latter equation recast in the form:
\begin{align}
\hspace{2em}&\hspace{-2em}L_4^n(\Delta_3,0,\Delta_1^{\{i,j\}},\tD_{ijk}) = \frac{2^{\varepsilon}}{T} \, \Gamma(1+\varepsilon) \notag \\
& {} \times
\left\{ 
 - 
 \frac{1}{\varepsilon} \, 
 \int^1_0 \, \frac{d u }{u^2 \, P_{ijk} + R_{ij} - i \, \lambda}  
 + 
 \int^1_0 d u \, 
 \frac{\ln(u^2 \, P_{ijk} + R_{ij} - i \, \lambda)}
{u^2 \, P_{ijk} + R_{ij} - i \, \lambda} 
 \right.
 \notag \\
&\qquad {}
 + \int^1_0 \, \frac{d u }{u^2 \, P_{ijk} + R_{ij} - i \, \lambda \, \sigma_0} \notag\\
&\qquad \quad{}
\times
 \left[ \ln \left( (T - i \lambda) \, (1-u^2) \right) - \ln \left( \frac{T \, (P_{ijk} + R_{ij})}{P_{ijk}} - i \, \lambda\right) \right.
\notag\\
&\qquad \quad{}
 - 
 \left. \ln \left( u^2 \, (P_{ijk} + R_{ij} - T) + T - i \, \lambda  \right) + \ln \left( \frac{(P_{ijk}+R_{ij}) \, (T - R_{ij})}{P_{ijk}} - i \, \lambda \right) \right]
\notag\\
&\qquad {}
\left.
- 
 \int^1_0 \, \frac{d u }{u^2 \, P_{ijk} + R_{ij} - i \, \lambda \, \sigma_0} \, 
 \ln \left( \frac{T - R_{ij} - i \, \lambda \, \sigma_1}{T - i \, \lambda \, \sigma_1} \right)
\right\}
  \label{eqlijkir15r}
\end{align}
with $\sigma_1 = \sign( (P_{ijk} + R_{ij})/P_{ijk} )$ for the real mass case and:
\begin{align}
\hspace{2em}&\hspace{-2em}L_4^n(\Delta_3,0,\Delta_1^{\{i,j\}},\tD_{ijk}) = \frac{2^{\varepsilon}}{T} \, \Gamma(1+\varepsilon) \notag \\
& {} \times
\left\{ 
 \int^1_0 \, \frac{d u }{u^2 \, P_{ijk} + R_{ij} } \, 
 \left[ \ln \left( T  \, (1-u^2) \right) - \ln \left( \frac{T \, (P_{ijk} + R_{ij})}{P_{ijk}} \right) \right]
 \right.
 \notag \\
&\qquad {} \;
 - 
 \int^1_0 \, \frac{d u }{u^2 \, P_{ijk} + R_{ij} } \, 
 \left[ \ln \left( u^2 \, (P_{ijk} + R_{ij} - T) + T   \right) - \ln \left( \frac{(P_{ijk}+R_{ij}) \, (T - R_{ij})}{P_{ijk}}  \right) \right]
\notag\\
&\qquad {} \;
 - 
 \frac{1}{\varepsilon} \, 
 \int^1_0 \, \frac{d u }{u^2 \, P_{ijk} + R_{ij} }  
 + 
 \int^1_0 d u \, 
 \frac{\ln(u^2 \, P_{ijk} + R_{ij} )}
{u^2 \, P_{ijk} + R_{ij} } 
\notag\\
&\qquad {} \;
\left.
- 
 \int^1_0 \, \frac{d u }{u^2 \, P_{ijk} + R_{ij} } \, 
 \left[ \ln \left( \frac{T - R_{ij}}{T} \right) + \eta \left( \frac{T \, (P_{ijk} + R_{ij})}{P_{ijk}}, \frac{T - R_{ij}}{T} \right) \right]
\right\}
  \label{eqlijkir15c}
\end{align}
for the complex mass case. The relevant integrals are given in appendices \ref{appF} of this paper, \ref{P1-appF} of ref. \cite{paper1} and \ref{P2-appF} of ref. \cite{paper2}.
\vspace{0.3cm}

\noindent
Eq. (\ref{eqlijkir15c}) matches eq. (\ref{P2-eqcas1eclate}) of ref. \cite{paper2}
obtained in the corresponding general complex mass case 1.(a), considering the 
latter in the limit $\Re(\Delta_{2}^{\{i\}}) = \Re(Q_{i}+T) \to 0$ while keeping
an infinitesimal positive imaginary part $\Im(\Delta_{2}^{\{i\}}) = \lambda$. 
One then formally gets:
\begin{align}
\text{r.h.s.} \, (\ref{P2-eqcas1eclate}) \, \text{of \cite{paper2}} 
&\to \frac{1}{T} \,
\left\{ 
 \int^1_0 \, \frac{d u }{u^2 \, P_{ijk} + R_{ij} } \, 
 \left[ \ln \left( T  \, (1-u^2) \right) - \ln \left( \frac{T \, (P_{ijk} + R_{ij})}{P_{ijk}} \right) \right.
 \right.
 \notag \\
&\qquad {} \;
 - 
 \left. \ln \left( u^2 \, (P_{ijk} + R_{ij} - T) + T   \right) + \ln \left( \frac{(P_{ijk}+R_{ij}) \, (T - R_{ij})}{P_{ijk}}  \right) \right]
\notag\\
&\qquad {} \;
 - 
 \ln\left( Q_i + T \right) \, 
 \int^1_0 \, \frac{d u }{u^2 \, P_{ijk} + R_{ij} }  
 + 
 \int^1_0 d u \, 
 \frac{\ln(u^2 \, P_{ijk} + R_{ij} )}
{u^2 \, P_{ijk} + R_{ij} } 
\notag\\
&\qquad {} \;
\left.
- 
 \int^1_0 \, \frac{d u }{u^2 \, P_{ijk} + R_{ij} } \, 
 \left[ \ln \left( \frac{T - R_{ij}}{T} \right) + \eta \left( \frac{T \, (P_{ijk} + R_{ij})}{P_{ijk}}, \frac{T - R_{ij}}{T} \right) \right]
\right\}
 \notag
\end{align}
where the divergent term $\ln ( Q_{i}+T )$ corresponds to the ``dressed pole" 
$(2 \, e^{- \gamma_{E}})^{\varepsilon}/\varepsilon$.

\subsubsection{$\Im(\Delta_3) > 0$, $\Im(\Delta_1^{\{i,j\}}) < 0$}\label{subsubseccas2}

Compared to the previous case, one uses eq.~(\ref{eqisigir4}) for $M_2(\xi^{\nu})$ and the $\xi$
integration is carried out with the help of eqs.~(\ref{eqdeffuncj2}) and (\ref{eqdeffuncj7}) depending on the different terms.
Then, performing the first step described in subsubsec.\ \ref{subsubcase1}, we obtain:
\begin{align}
\hspace{2em}&\hspace{-2em}L_4^n(\Delta_3,0,\Delta_1^{\{i,j\}},\tD_{ijk}) \notag \\
&= - \frac{F(\varepsilon)}{4} \left\{ i \, \int_0^{\infty} \frac{dy}{\sqrt{y}} \, \int_0^1 \, \frac{dx}{y \, P_{ijk} - x \, R_{ij}} \, \left( T (1-x) \right)^{-1-\varepsilon} \right. \notag \\
&\qquad \qquad \quad {} + e^{i \, \pi \, \varepsilon} \, \int_0^{\infty} \frac{dy}{\sqrt{y}} \, \int_0^{\infty} \, \frac{dx}{y \, P_{ijk} - x \, R_{ij}} \, \left( - y \, P_{ijk} + x \, R_{ij} - T \, (1+x) \right)^{-1-\varepsilon} \notag \\
&\qquad \qquad \quad {} + i \, \int_0^{\infty} \frac{dy}{\sqrt{y}} \, \int_1^{\infty} \, \frac{dx}{y \, P_{ijk} - x \, R_{ij}} \, \left(  - y \, P_{ijk} + x \, R_{ij} + T \, (1-x) \right)^{-1-\varepsilon} \notag \\
&\qquad \qquad \quad {} + \int_1^{\infty} \frac{dy}{\sqrt{y}} \, \int_0^1 \, \frac{dx}{y \, P_{ijk} + x \, R_{ij}} \, \left[ \left( T \, (1-x) \right)^{-1-\varepsilon} \right. \notag \\
&\qquad \qquad \qquad \qquad \qquad \qquad \qquad \qquad \qquad \quad {} - \left. \left( y \, P_{ijk} + x \, R_{ij} + T \, (1-x) \right)^{-1-\varepsilon} \right] \notag \\
&\qquad \qquad \quad {} + \int_0^{1} \frac{dy}{\sqrt{y}} \, \int_0^y \, \frac{dx}{y \, P_{ijk} + x \, R_{ij}} \, \left[ \left( T \, (1-x) \right)^{-1-\varepsilon} \right. \notag \\ 
&\qquad \qquad \qquad \qquad \qquad \qquad \qquad \qquad \qquad {} - \left. \left. \left( y \, P_{ijk} + x \, R_{ij} + T \, (1-x) \right)^{-1-\varepsilon} \right] \vphantom{\frac{dx}{y \, P_{ijk} - x \, R_{ij}}} \right\}
  \label{eqlijkir160}
\end{align}
We set $y = u^2$, perform a partial fraction decomposition on the variable $x$ and expand around $\varepsilon=0$. The $x$ integration is readily done and $L_4^n(\Delta_3,0,\Delta_1^{\{i,j\}},\tD_{ijk})$ is written as:
\begin{align}
\hspace{2em}&\hspace{-2em}L_4^n(\Delta_3,0,\Delta_1^{\{i,j\}},\tD_{ijk}) \notag \\
&= - \frac{2^{\varepsilon}}{T} \, \Gamma(1+\varepsilon) 
\left\{ 
 - \frac{T^{-\varepsilon}}{\varepsilon} \, 
 \left[ 
  i \, \int^{+\infty}_0 \, \frac{d u }{u^2 \, P_{ijk} - R_{ij}} 
  + \int^{+\infty}_1 \, 
  \frac{d u }{u^2 \, P_{ijk} + R_{ij}} 
 \right] 
\right. 
\notag \\
&\qquad \qquad \qquad \qquad {} 
 + i \, \int^{+\infty}_0 \, 
 \frac{d u}{u^2 \, P_{ijk} - R_{ij}} \, 
 \left[ 
  \ln \left(\frac{u^2 \, P_{ijk} - R_{ij}}{u^2 \, P_{ijk}} \right) 
  - 
  \ln \left( \frac{R_{ij}-T}{R_{ij}} \right) 
 \right] 
\notag \\ 
&\qquad \qquad \qquad \qquad {} 
- \int^{+\infty}_0 \, \frac{d u }{u^2 \, P_{ijk} + R_{ij}} \, 
 \left[ 
  \ln \left(\frac{- u^2 \, P_{ijk}}{ R_{ij}} \right) 
  - 
  \ln \left( \frac{P_{ijk} \, u^2 + T}{T-R_{ij}} \right) 
 \right] 
\notag \\ 
&\qquad \qquad \qquad \qquad {} 
+ \int^{+\infty}_1 \, \frac{d u }{u^2 \, P_{ijk} + R_{ij}} \, 
 \ln \left(\frac{u^2 \, P_{ijk} + R_{ij}}{u^2 \, P_{ijk} + T} \right) 
\notag \\ 
&\qquad \qquad \qquad \qquad {} 
+ 
 \int^1_0 \, \frac{d u }{u^2 \, P_{ijk} + R_{ij}} \, 
 \left[ 
  \ln \left(\frac{u^2 \, (P_{ijk} + R_{ij} - T) + T}{u^2 \, P_{ijk} + T} \right) \right. \notag \\ 
  &\qquad \qquad \qquad \qquad \qquad \qquad \qquad \qquad \qquad {} - \left. \left. \ln(1-u^2) 
  \vphantom{\ln \left(\frac{u^2 \, (P_{ijk} + R_{ij} - T) + T}{u^2 \, P_{ijk} + T} \right)} 
 \right]  
\right\}
\label{eqlijkir16}
\end{align}
In this case, we have $\Im(R_{ij}) > 0$, $\Im(P_{ijk}) < 0$ and 
$\Im(R_{ij}-T) > 0$. Furthermore, 
\begin{itemize}
\item
$u^2 \, P_{ijk} - R_{ij} = (1+u^2) \, \Delta_1^{\{i,j\}} + u^2 \, \tD_{ijk}$ \\
thus $\Im(u^2 \, P_{ijk} - R_{ij}) < 0$ when $u \in [0,\infty[$, 
\item
$u^2 \, P_{ijk} + R_{ij} 
= u^2 \, \tD_{ijk} + (u^2 - 1) \, \Delta_1^{\{i,j\}}$ \\
thus $\Im(u^2 \, P_{ijk} + R_{ij}) < 0$ when $u \in [1,\infty[$, 
\item
$u^2 \, P_{ijk} + T 
= u^2 \, (\tD_{ijk} +  \Delta_1^{\{i,j\}}) - \Delta_3$ \\
thus $\Im(u^2 \, P_{ijk} + T) < 0$ when $u \in [0,\infty[$, 
\item
$u^2 \, (P_{ijk}+R_{ij}-T) + T = u^2 \, \tD_{ijk} - (1-u^2) \, \Delta_3$ \\
thus $\Im(u^2 \, (P_{ijk}+R_{ij}-T) + T) < 0$ when $u \in [0,1]$. 
\end{itemize}
Rearrangements and simplifications similar to those done in the massive case 
2.(a) of ref.~\cite{paper2} can be performed, which lead to the following alternative expression:
\begin{align}
\hspace{2em}&\hspace{-2em}L_4^n(\Delta_3,0,\Delta_1^{\{i,j\}},\tD_{ijk}) = \frac{2^{\varepsilon}}{T} \, \Gamma(1+\varepsilon) \notag \\
&  {} \times \Bigg\{ 
 - 
 \frac{1}{\varepsilon} \, 
 \int_{\usebox{\Gammap}} \, \frac{d u}{u^2 \, P_{ijk} + R_{ij}}  
 \quad + \quad
 \int_{\usebox{\Gammap}} d u \, 
 \frac{\ln(u^2 \, P_{ijk} + R_{ij})}{u^2 \, P_{ijk} + R_{ij}} 
\notag\\
&
\notag\\
&\qquad {} + 
 \int^1_0 \, \frac{d u }{u^2 \, P_{ijk} + R_{ij}} \, 
 \Bigg[ 
   \ln \left( T \, (1-u^2) \right) - \ln \left( \frac{T \, (P_{ijk} + R_{ij})}{P_{ijk}} \right)
\notag\\
&\qquad {}
  -
  \ln \left( u^2 \, (P_{ijk} + R_{ij} - T) + T \right) + \ln \left( \frac{(P_{ijk} + R_{ij}) \, (T - R_{ij})}{P_{ijk}} \right)
\notag\\
&\qquad {}
  - \eta \left( \frac{T \, (P_{ijk} + R_{ij})}{P_{ijk}}, \frac{T - R_{ij}}{T} \right)\Biggr]   
 -
 \int_{\usebox{\Gammap}} \, \frac{d u }{u^2 \, P_{ijk} + R_{ij}} \, 
 \ln \left( \frac{T - R_{ij}}{T} \right) 
\Biggl\}
\label{eqlijkir16reframed}
\end{align}

\vspace{0.3cm}

\noindent
The contour \raisebox{0.8ex}{$\usebox{\Gammap}$} can be deformed into a contour 
$\widehat{(0,1)}^{+}$ stretched from 0 to 1 and which eventually wraps 
from above the cut of $\ln(u^2 \, P_{ijk} + R_{ij})$ emerging from 
the branch point $u_{0} = \sqrt{- R_{ij}/P_{ijk}}$, whenever the latter lies 
in the ``north-east" quadrant $\{\Re(u) > 0, \Im(u) > 0\}$. 

\vspace{0.3cm}

\noindent
Eq. (\ref{eqlijkir16reframed}) can be compared with eq. (\ref{P2-eqcas2aa}) of \cite{paper2} 
obtained in the corresponding general complex mass case 2.(a), considering the 
latter in the limit $\Re(\Delta_{2}^{\{i\}}) = \Re(Q_{i}+T) \to 0$ while keeping
an infinitesimal positive imaginary part $\Im(\Delta_{2}^{\{i\}}) = \lambda$. 
Whereas it appeared
convenient to formulate eq. (\ref{P2-eqcas2aa}) of ref. \cite{paper2} in terms of manifestly vanishing
pole residues, this is no longer the case for eq. (\ref{eqlijkir16reframed}) 
since the pole has become also the branch point of 
$\ln(u^2 \, P_{ijk} + (R_{ij} +Q_{i} +T))$
in the limit $(Q_{i}+T)  \to 0$. For the purpose of the comparison the $\eta$ 
functions introduced in eq. (\ref{P2-eqcas2aa}) of ref. \cite{paper2} containing $Q_{i} + T$ shall thus be made explicit in 
terms of constant logarithms, part of which then cancel against the constant
logarithms which were subtracted so as to build the explicitly vanishing pole
residues. One then formally gets:
\begin{align}
\hspace{2em}&\hspace{-2em}\text{r.h.s.\ (\ref{P2-eqcas2aa}) of \cite{paper2}} \to \frac{1}{T}  \notag \\
&  {} \times \Bigg\{ 
 - 
 \ln \left( Q_i + T \right) \, 
 \int_{\widehat{(0,1)}^{+}} \, \frac{d u}{u^2 \, P_{ijk} + R_{ij}}  
 \quad + \quad
 \int_{\widehat{(0,1)}^{+}} d u \, 
 \frac{\ln(u^2 \, P_{ijk} + R_{ij})}{u^2 \, P_{ijk} + R_{ij}} 
\notag\\
&\qquad {} + 
 \int^1_0 \, \frac{d u }{u^2 \, P_{ijk} + R_{ij}} \, 
 \Bigg[ 
   \ln \left( T \, (1-u^2) \right) - \ln \left( \frac{T \, (P_{ijk} + R_{ij})}{P_{ijk}} \right)
\notag\\
&\qquad {}
  -
  \ln \left( u^2 \, (P_{ijk} + R_{ij} - T) + T \right) + \ln \left( \frac{(P_{ijk} + R_{ij}) \, (T - R_{ij})}{P_{ijk}} \right)
\notag\\
&\qquad {}
  - \eta \left( \frac{T \, (P_{ijk} + R_{ij})}{P_{ijk}}, \frac{T - R_{ij}}{T} \right)\Biggr]   
 -
 \int_{\widehat{(0,1)}^{+}} \, \frac{d u }{u^2 \, P_{ijk} + R_{ij}} \, 
 \ln \left( \frac{T - R_{ij}}{T} \right) 
\Biggl\}
\notag
\end{align}
where the divergent term $\ln ( Q_{i}+T )$ corresponds to the ``dressed pole" 
$(2 \, e^{- \gamma_{E}})^{\varepsilon}/\varepsilon$.

\subsubsection{$\Im(\Delta_3) < 0$, $\Im(\Delta_1^{\{i,j\}}) > 0$}

In this case, we start with eq.~(\ref{eqisigir3}) for $M_2(\xi^{\nu})$ and use eq.~(\ref{eqdeffuncj7}) for all the $\xi$ integrations.
We give the result for $L_4^n(\Delta_3,0,\Delta_1^{\{i,j\}},\tD_{ijk})$ after the intermediate step 1 of subsubsec. \ref{subsubcase1}.
\begin{align}
\hspace{2em}&\hspace{-2em}L_4^n(\Delta_3,0,\Delta_1^{\{i,j\}},\tD_{ijk}) \notag \\
&= \frac{F(\varepsilon)}{4} \left\{ i \, e^{- i \, \pi \, \varepsilon} \, \int_0^{\infty} \frac{dy}{\sqrt{y}} \, \int_y^{\infty} \, \frac{dx}{y \, P_{ijk} + x \, R_{ij}} \, \left[ \left( y \, P_{ijk} + x \, R_{ij} - T \, (1+x) \right)^{-1-\varepsilon} \right. \right. \notag \\
&\qquad \qquad \qquad \qquad \qquad \qquad \qquad \qquad \qquad \qquad \quad {} - \left. \left( - T \, (1+x) \right)^{-1-\varepsilon} \right] \notag \\
&\qquad \qquad \quad {} + \int_1^{\infty} \frac{dy}{\sqrt{y}} \, \int_y^{\infty} \, \frac{dx}{y \, P_{ijk} + x \, R_{ij}} \, \left[ \left( y \, P_{ijk} + x \, R_{ij} + T \, (1-x) \right)^{-1-\varepsilon} \right. \notag \\
&\qquad \qquad \qquad \qquad \qquad \qquad \qquad \qquad \qquad \quad {} - \left. \left( T \, (1-x) \right)^{-1-\varepsilon} \right] \notag \\
&\qquad \qquad \quad {} + \int_0^{1} \frac{dy}{\sqrt{y}} \, \int_1^{\infty} \, \frac{dx}{y \, P_{ijk} + x \, R_{ij}} \, \left[ \left( y \, P_{ijk} + x \, R_{ij} + T \, (1-x) \right)^{-1-\varepsilon} \right. \notag \\
&\qquad \qquad \qquad \qquad \qquad \qquad \qquad \qquad \qquad \quad {} - \left. \left. \left( T \, (1-x) \right)^{-1-\varepsilon} \right] \vphantom{\frac{dx}{y \, P_{ijk} + x \, R_{ij}}} \right\} 
  \label{eqeqlijkir170}
\end{align}
Then, performing the steps 2 and 3 of this subsubsec.\ yields:
\begin{align}
\hspace{2em}&\hspace{-2em}L_4^n(\Delta_3,0,\Delta_1^{\{i,j\}},\tD_{ijk}) \notag \\
&= \frac{2^{\varepsilon}}{T} \, \Gamma(1+\varepsilon) 
\left\{ 
 - \frac{(-T)^{-\varepsilon}}{\varepsilon} \,  
 \int^{1}_0 \, \frac{d u }{u^2 \, P_{ijk} + R_{ij}}  
\right. 
\notag \\
&\qquad \qquad \qquad \quad {} 
 + i \, \int^{+\infty}_0 \, \frac{d u }{u^2 \, P_{ijk} - R_{ij}} \, 
 \left[ 
  \ln \left( \frac{u^2 \, (P_{ijk} + R_{ij}-T) - T}{R_{ij}-T} \right) 
  - 
  \ln \left( u^2+1 \right) \right] 
\notag \\ 
&\qquad \qquad \qquad \quad {} 
+ \int^{+\infty}_1 \, \frac{d u }{u^2 \, P_{ijk} + R_{ij}} \, 
 \left[ 
  \ln \left(\frac{u^2 \, (P_{ijk} + R_{ij}-T) + T}{R_{ij}-T} \right) 
  - 
  \ln \left( u^2 -1 \right) 
 \right] 
\notag \\ 
&\qquad \qquad \qquad \quad {} 
+ 
\left. 
\int^1_0 \, \frac{d u }{u^2 \, P_{ijk} + R_{ij}} \, 
\ln \left( \frac{u^2 \, P_{ijk} + R_{ij}}{R_{ij} - T} \right)  
\right\}
 \label{eqlijkir17}
\end{align}
In this case, we have $\Im(R_{ij}-T) < 0$.
Furthermore, 
\begin{itemize}
\item
$u^2 \, P_{ijk} + R_{ij} 
= u^2 \, \tD_{ijk} - (1 - u^2) \, \Delta_1^{\{i,j\}}$ \\
thus $\Im(u^2 \, P_{ijk} + R_{ij}) < 0$ when $u \in [0,1]$, 
\item
$u^2 \, (P_{ijk}+R_{ij}-T) - T = u^2 \, \tD_{ijk} + (1+u^2) \, \Delta_3$ \\
thus $\Im(u^2 \, (P_{ijk}+R_{ij}-T) - T) < 0$ when $u \in [0,\infty[$. 
\item
$u^2 \, (P_{ijk}+R_{ij}-T) + T = u^2 \, \tD_{ijk} + (u^2-1) \, \Delta_3$ \\
thus $\Im(u^2 \, (P_{ijk}+R_{ij}-T) + T) < 0$ when $u \in [1,\infty[$. 
\end{itemize}
Rearrangements and simplifications similar to those done in the massive case 
1.(c) of \cite{paper2} can be performed, which lead to the following alternative expression:
\begin{align}
\hspace{2em}&\hspace{-2em}L_4^n(\Delta_3,0,\Delta_1^{\{i,j\}},\tD_{ijk}) = \frac{2^{\varepsilon}}{T} \, \Gamma(1+\varepsilon) \notag \\
& {} \times \Biggl\{ 
 - 
 \frac{1}{\varepsilon} \, 
 \int_{0}^{1} \, \frac{d u}{u^2 \, P_{ijk} + R_{ij}}  
 \quad {} + \quad {}
 \int_{0}^{1} d u \, 
 \frac{\ln(u^2 \, P_{ijk} + R_{ij})}{u^2 \, P_{ijk} + R_{ij}} 
\notag\\
&\qquad  {} +
 \int_{\usebox{\Gammap}} \, \frac{d u }{u^2 \, P_{ijk} + R_{ij}} \, 
 \Biggl[ 
  \ln \left( T \, (1-u^2) \right) - \ln \left( \frac{T \, (P_{ijk} + R_{ij})}{P_{ijk}} \right)
  \notag \\
&\qquad \qquad  {} 
  -
  \ln \left( u^2 \, (P_{ijk} + R_{ij} - T) + T \right) + \ln \left( \frac{(P_{ijk} + R_{ij}) \, (T - R_{ij})}{P_{ijk}} \right)
\notag\\
&\qquad \qquad{}
  - \eta \left( \frac{T \, (P_{ijk} + R_{ij})}{P_{ijk}}, \frac{T - R_{ij}}{T} \right)
 \Biggr]    
 - \int_{0}^1 \, \frac{d u }{u^2 \, P_{ijk} + R_{ij}} \, 
 \ln \left( \frac{T - R_{ij}}{T} \right)
\Biggr\}
\label{eqlijkir17reframed}
\end{align}
The deformation of the contour \raisebox{0.8ex}{$\usebox{\Gammap}$} into a contour
$\widehat{(0,1)}^{+}$ stretched from 0 to 1 may eventually wrap 
from above the cut of $\ln(u^2 \, (P_{ijk} + R_{ij} -T) +T)$ emerging from 
the branch point $u_{0} = \sqrt{- T/(P_{ijk} + R_{ij} -T)}$, whenever the 
latter lies in the ``north-east" quadrant.

\vspace{0.3cm}

\noindent
In a way similar to the previous case, eq. (\ref{eqlijkir17reframed}) 
matches eq. (\ref{P2-eqcas4eclatesoustrait}) of \cite{paper2} obtained in the corresponding 
general complex mass case 1.(c), considering the latter in the limit 
$\Re(\Delta_{2}^{\{i\}}) = \Re(Q_{i}+T) \to 0$ while keeping
an infinitesimal positive imaginary part $\Im(\Delta_{2}^{\{i\}}) = \lambda$.

\subsubsection{$\Im(\Delta_3) < 0$, $\Im(\Delta_1^{\{i,j\}}) < 0$}

One uses eq.~(\ref{eqisigir4}) for $M_2(\xi^{\nu})$ and the $\xi$
integration is carried out with the help of eqs.~(\ref{eqdeffuncj2}) and (\ref{eqdeffuncj7}) depending on the terms. Then, the step 1 
(cf.\ subsubsec.\ \ref{subsubcase1}) leads to:
\begin{align}
\hspace{2em}&\hspace{-2em}L_4^n(\Delta_3,0,\Delta_1^{\{i,j\}},\tD_{ijk}) \notag \\
&= \frac{F(\varepsilon)}{4} \, \left\{ i \, \left(-T\right)^{-1-\varepsilon} \, \int_0^{\infty} \frac{dy}{\sqrt{y}} \, \int_1^{\infty} \, \frac{dx}{y \, P_{ijk} - x \, R_{ij}} \, (x-1)^{-1-\varepsilon} \right. \notag \\
&\qquad \qquad \quad {} - e^{-i \, \pi \, \varepsilon} \, \left( - T \right)^{-1-\varepsilon} \, \int_0^{\infty} \frac{dy}{\sqrt{y}} \, \int_0^{\infty} \, \frac{dx}{y \, P_{ijk} - x \, R_{ij}} \, (1+x)^{-1-\varepsilon} \notag \\
&\qquad \qquad \quad {} + i \, \int_0^{\infty} \frac{dy}{\sqrt{y}} \, \int_0^{1} \, \frac{dx}{y \, P_{ijk} - x \, R_{ij}} \, \left( - y \, P_{ijk} + x \, R_{ij} + T \, (1-x) \right)^{-1-\varepsilon} \notag \\
&\qquad \qquad \quad {} - i \, e^{- i \, \pi \, \varepsilon} \, \int_0^{\infty} \frac{dy}{\sqrt{y}} \, \int_0^{y} \, \frac{dx}{y \, P_{ijk} + x \, R_{ij}} \, \left[ \left( y \, P_{ijk} + x \, R_{ij} - T \, (1+x) \right)^{-1-\varepsilon} \right. \notag \\
&\qquad \qquad \qquad \qquad \qquad \qquad \qquad \qquad \qquad \qquad \qquad {} - \left. \left( -T \, (1+x) \right)^{-1-\varepsilon} \right] \notag \\
&\qquad \qquad \quad {} - \int_1^{\infty} \frac{dy}{\sqrt{y}} \, \int_1^{y} \, \frac{dx}{y \, P_{ijk} + x \, R_{ij}} \, \left[ \left( y \, P_{ijk} + x \, R_{ij} + T \, (1-x) \right)^{-1-\varepsilon} \right. \notag \\
&\qquad \qquad \qquad \qquad \qquad \qquad \qquad \qquad \qquad \quad {} - \left. \left. \left( T \, (1-x) \right)^{-1-\varepsilon} \right] \vphantom{\frac{dx}{y \, P_{ijk} - x \, R_{ij}}} \right\}
  \label{eqlijkir180}
\end{align}
At the end of the step 5, we get;
\begin{align}
\hspace{2em}&\hspace{-2em}L_4^n(\Delta_3,0,\Delta_1^{\{i,j\}},\tD_{ijk}) \notag \\
&= 
- \frac{2^{\varepsilon}}{T} \, \Gamma(1+\varepsilon) 
\left\{
 - \frac{(-T)^{-\varepsilon}}{\varepsilon} \, 
 \left[ 
  i \, \int^{+\infty}_0 \, \frac{d u }{u^2 \, P_{ijk} - R_{ij}} 
  + \int^{+\infty}_1 \, \frac{d u }{u^2 \, P_{ijk} + R_{ij}} 
 \right] 
\right. 
\notag \\
&\qquad \qquad \qquad \quad {} 
+ i \, \int^{+\infty}_0 \, \frac{d u }{u^2 \, P_{ijk} - R_{ij}} \, 
\left[ 
 \ln \left(\frac{R_{ij} - u^2 \, P_{ijk}}{R_{ij}} \right) 
 - 
 \ln \left( \frac{P_{ijk} \, u^2 - T}{u^2 \, P_{ijk}} \right) 
\right] 
\notag \\ 
&\qquad \qquad \qquad \quad {} 
- i \, \int^{+\infty}_0 \, \frac{d u }{u^2 \, P_{ijk} - R_{ij}} \, 
\left[ 
 \ln \left( \frac{u^2 \, (P_{ijk} + R_{ij} - T) - T}{u^2 \, P_{ijk} - T} \right) 
 - 
 \ln \left( u^2+1 \right) 
\right] 
\notag \\ 
&\qquad \qquad \qquad \quad {} 
- 
\int^{+\infty}_1 \, \frac{d u }{u^2 \, P_{ijk} + R_{ij}} \, 
 \left[ 
  \ln 
  \left(
   \frac{u^2 \, (P_{ijk} + R_{ij} - T) + T}{u^2 \, P_{ijk} + R_{ij}} 
  \right) 
  - \ln(u^2-1) 
 \right]  
\notag \\ 
&\qquad \qquad \qquad \quad {} 
- 
\left. 
\int^{+\infty}_0 \, \frac{d u }{u^2 \, P_{ijk} + R_{ij}} \, 
\ln \left( \frac{- u^2 \, P_{ijk}}{ R_{ij}} \right)  
\right\}
\label{eqlijkir18}
\end{align}
In this case, we have $\Im(R_{ij}) > 0$, $\Im(P_{ijk}) < 0$.
Furthermore, 
\begin{itemize}
\item
$u^2 \, P_{ijk} - R_{ij} = u^2 \, \tD_{ijk} + (u^2+1) \, \Delta_1^{\{i,j\}}$ \\
thus $\Im(u^2 \, P_{ijk} - R_{ij}) < 0$ when $u \in [0,\infty[$, 
\item
$u^2 \, P_{ijk} + R_{ij} 
= u^2 \, \tD_{ijk} + (u^2- 1) \, \Delta_1^{\{i,j\}}$ \\
thus $\Im(u^2 \, P_{ijk} + R_{ij}) < 0$ when $u \in [1,\infty[$, 
\item
$u^2 \, P_{ijk} - T 
= u^2 \, (\tD_{ijk} +  \Delta_1^{\{i,j\}}) + \Delta_3$ \\
thus $\Im(u^2 \, P_{ijk} - T) < 0$ when $u \in [0,\infty[$, 
\item
$u^2 \, (P_{ijk}+R_{ij}-T) - T = u^2 \, \tD_{ijk} + (u^2+1) \, \Delta_3$ \\
thus $\Im(u^2 \, (P_{ijk}+R_{ij}-T) - T) < 0$ when $u \in [0,\infty[$. 
\item
$u^2 \, (P_{ijk}+R_{ij}-T) + T = u^2 \, \tD_{ijk} + (u^2-1) \, \Delta_3$ \\
thus $\Im(u^2 \, (P_{ijk}+R_{ij}-T) + T) < 0$ when $u \in [1,\infty[$. 
\end{itemize}
Rearrangements and simplifications similar to those done in the massive case 
2.(c) of \cite{paper2} can be performed, which lead to the following alternative expression:
\begin{align}
L_4^n(\Delta_3,0,\Delta_1^{\{i,j\}},\tD_{ijk})
&= 
\frac{2^{\varepsilon}}{T} \, \Gamma(1+\varepsilon) \, \int_{\usebox{\Gammap}} \, \frac{d u}{u^2 \, P_{ijk} + R_{ij}}
\notag\\
&\quad {} \times
\Biggl\{  
 - 
 \frac{1}{\varepsilon} \, 
  + 
 \ln(u^2 \, P_{ijk} + R_{ij})
+
  \ln \left( T \, (1-u^2) \right) - \ln \left( \frac{T \, (P_{ijk} + R_{ij})}{P_{ijk}} \right)
\notag\\
& \quad {} \quad 
-
 \ln \left( u^2 \, (P_{ijk} + R_{ij} - T) + T \right) + \ln \left( \frac{(P_{ijk} + R_{ij}) \, (T - R_{ij})}{P_{ijk}} \right)
 \notag \\
& \quad {} \quad 
- \eta \left( \frac{T \, (P_{ijk} + R_{ij})}{P_{ijk}}, \frac{T - R_{ij}}{T} \right) - \ln \left( \frac{T - R_{ij}}{T} \right)
\Biggr\}
\label{eqlijkir18reframed}
\end{align}
As in the previous cases, the contours  \raisebox{0.8ex}{$\usebox{\Gammap}$} can be deformed into contours 
$\widehat{(0,1)}^{+}_{1,2}$ stretched from 0 to 1. They eventually wrap 
from above the cuts of $\ln(u^2 \, P_{ijk} + R_{ij})$ emerging from 
the branch point $\sqrt{- R_{ij}/P_{ijk}}$ and of
$\ln(u^2 \, (P_{ijk} + R_{ij} -T) +T)$ emerging from the branch point \linebreak 
$\sqrt{- T/(P_{ijk} + R_{ij} -T)}$, respectively, whenever either of these 
branch points or both lie in the ``north-east" quadrant. 

\vspace{0.3cm}

\noindent
In a way similar to the previous case, eq. (\ref{eqlijkir18reframed}) 
matches eq. (\ref{P2-eqcas2cc}) of \cite{paper2} obtained in the corresponding 
general complex mass case 2.(c), considering the latter in the limit 
$\Re(\Delta_{2}^{\{i\}}) = \Re(Q_{i}+T) \to 0$ while keeping
an infinitesimal positive imaginary part $\Im(\Delta_{2}^{\{i\}}) = \lambda$.

\subsection{$\Delta_2^{\{i\}} = 0$ and $\tD_{ijk} = 0$}\label{casDelta20Dt0}

In this case, we have $P_{ijk} = - R_{ij}$ and $Q_i = -T$, so that
eqs. (\ref{eqisigir3}) and (\ref{eqisigir4}) become:
\begin{itemize}
\item[-] for $\Im(\Delta_1^{\{i,j\}} > 0)$,
\begin{align}
M_2(\xi^{\nu}) 
&= 
\frac{1}{2} \, B(1/2,1/2) \, 
\int^1_0 \frac{d z}{\Delta_1^{\{i,j\}} \, (z^2 - 1)}
\notag \\
&\quad {} \times 
\left[ 
 \frac{1}{\left(\xi^{\nu}-i \, \lambda \right)^{1/2}} 
 - 
 \frac{1}{\left(\xi^{\nu}+ \Delta_1^{\{i,j\}} \, (z^2-1) 
         - i \, \lambda \right)^{1/2}} 
\right]
\label{eqisigir3p}
\end{align}
\item[-] for $\Im(\Delta_1^{\{i,j\}} < 0)$,
\begin{align}
M_2(\xi^{\nu}) 
&= 
- \frac{1}{2} \, B(1/2,1/2) \, 
\left\{ 
 i \, \int^{+\infty}_0 \frac{d z}{\Delta_1^{\{i,j\}} \, (z^2+1)} 
\right. 
\notag \\
&\quad {} 
\times 
\left[ 
 \frac{1}{\left(\xi^{\nu}-i \, \lambda \right)^{1/2}} 
 - 
 \frac{1}{\left(\xi^{\nu}- \Delta_1^{\{i,j\}} \, (1+z^2) \right)^{1/2}} 
\right] 
\notag \\     
&\quad {} 
+ \int^{+\infty}_1 \frac{d z}{\Delta_1^{\{i,j\}} \, (z^2-1)} 
\notag \\
&\quad {} 
\times 
\left. 
 \left[ 
  \frac{1}{\left(\xi^{\nu}-i \, \lambda \right)^{1/2}} 
  - 
  \frac{1}{\left(\xi^{\nu}+ \Delta_1^{\{i,j\}} \, (z^2-1) \right)^{1/2}} 
 \right] 
\right\}
\label{eqisigir4p}
\end{align}
\end{itemize}
Here again, the $\xi$ integration will be of the type (\ref{eqdeffuncj2}) or (\ref{eqdeffuncj7}) 
(cf.\ appendix (\ref{appendJ})).
Let us go through the different cases according the signs of 
$\Im(\Delta_3)$ and $\Im(\Delta_1^{\{i,j\}})$.

\subsubsection{$\Im(\Delta_3) > 0$, $\Im(\Delta_1^{\{i,j\}}) > 0$}\label{531}

We proceed along the two first steps described in subsubsec. \ref{subsubcase1}. We borrow the result obtained in eq.~(\ref{eqlijkir12}), set $P_{ijk} = -R_{ij}$
and make the following change of variables $x = y + (1-y) \, v$.
The quantity $L_4^n(\Delta_3,0,\Delta_1^{\{i,j\}},0)$ reads now:
\begin{align}
  L_4^n(\Delta_3,0,\Delta_1^{\{i,j\}},0) &= - \frac{F(\varepsilon)}{4 \, R_{ij}} \, 
\int^1_0 \frac{d v}{v} \, 
\left[ 
 \frac{1}{[ v \, R_{ij} +(1-v) \, T \,   - i \, \lambda]^{1+\varepsilon}} 
 - 
 \frac{1}{[(1-v) \, T \, - i \, \lambda]^{1+\varepsilon}} 
\right]
\notag\\
&  \quad {}\quad {}\quad {} \quad {}
\times 
\int^1_0 d y \, y^{-1/2} \, (1-y)^{-1-\varepsilon}
\label{eqisigir8p}
\end{align}
The integrals over $y$ and $v$ are unnested and are computed easily using eqs.~(\ref{thirdinty0}), (\ref{thirdinty}) and (\ref{eqsecondontv3}).
Notice that $\Im(v \, R_{ij} +(1-v) \, T \,   - i \, \lambda)$ never changes its sign when $v$ spans $[0,1]$, this is obviously true in the real mass case and due to the fact that $\Im(T)$ and $\Im(R_{ij})$ have a same sign imaginary part in the complex mass case.
Inserting the explicit results for the $y$ and $v$ integrals,
we get:
\begin{eqnarray}
L_4^n(\Delta_3,0,\Delta_1^{\{i,j\}},0)
& = & 
\frac{1}{2} \, \Gamma(1+\varepsilon) \, 
\frac{\Gamma^2(1-\varepsilon)}{\Gamma(1- 2 \, \varepsilon)} \, 
\frac{1}{R_{ij} \, T} \, 
\nonumber \\
& & {} \times 
\left[ 
 \frac{1}{\varepsilon^2} \, (2 \, R_{ij} - i \, \lambda)^{-\varepsilon} 
 + 
 \dilog \left( \frac{T-R_{ij}}{T - i \, \lambda} \right) 
 - \frac{\pi^2}{6} 
\right]
\label{eqdefnl6}
\end{eqnarray}
Equation (\ref{eqdefnl6}) displays explicitly the singularity of $L_4^n(\Delta_3,0,\Delta_1^{\{i,j\}},0)$ 
when $\varepsilon \to 0$. In eq. (\ref{eqdefnl6}) we use the property $\dilog(z) + \dilog(1-z) = \pi^2/6 - \ln(z) \, \ln(1-z)$
to obtain the following alternative form  
suitable for further comparisons:
\begin{align}
L_4^n(\Delta_3,0,\Delta_1^{\{i,j\}},0)
&= 
\frac{1}{2} \, \Gamma(1+\varepsilon) \, 
\frac{\Gamma^2(1-\varepsilon)}{\Gamma(1- 2 \, \varepsilon)} \, 
\frac{1}{R_{ij} \, T} \, 
\nonumber \\
&\quad {} \times
\left\{ 
 \frac{1}{\varepsilon^2} \, (2 \, R_{ij} - i \lambda)^{-\varepsilon} 
  - \dilog \left( \frac{R_{ij}- i \lambda}{T- i \lambda} \right) 
\right.
\notag\\
& \quad {} \quad {}
\left. 
 -
 \left[
  \ln \left(R_{ij}- i \lambda \right) - \ln \left( T- i \lambda \right)
 \right] \, \ln \left( \frac{T - R_{ij}}{T- i \lambda} \right)
\right\}
\label{eqdefnl6bis}
\end{align}

\subsubsection{$\Im(\Delta_3) > 0$, $\Im(\Delta_1^{\{i,j\}}) < 0$}

The starting point is eq.~(\ref{eqlijkir160}) with $P_{ijk} = - R_{ij}$.
In the first line of eq. (\ref{eqlijkir160}) we rescale $y = x \, u^2$ so
that the double integral factorises into a product of two unnested integrals
over $u$ and over $x$. 
The integrals of the second and third lines of eq. (\ref{eqlijkir160}) yield no
divergences when $\varepsilon \rightarrow 0$, we thus take $\varepsilon=0$ in
them. We then make the following change of variables: 
$x = y - (y-1) \, v$ in the penultimate line, and $x = y -(1-y) \, v$ in the 
last line. We obtain:
\begin{align}
\hspace{2em}&\hspace{-2em}L_4^n(\Delta_3,0,\Delta_1^{\{i,j\}},0) \notag \\
&= \frac{F(\varepsilon)}{4 \, R_{ij}} \, 
\left\{
 i \, T^{-1-\varepsilon} \, \int_0^{+\infty} \frac{2 \, du}{1+u^2} \, 
\int_{0}^1 \, \frac{dx}{\sqrt{x}} \, (1-x)^{-1-\varepsilon} 
\right. 
\notag \\
&\quad {} 
 + 
 \quad {} 
 \int_0^{+\infty} \frac{dy}{\sqrt{y}} \, \int_0^{+\infty} 
 \frac{dx}{(y+x) \; ( (x+y) \, R_{ij} - T \, (1+x) )} 
\notag \\
&\quad {} 
 + i \, \int_0^{+\infty} \frac{dy}{\sqrt{y}} \, 
 \int_1^{+\infty} \frac{dx}{(y+x) \; ( (x+y) \, R_{ij} + T \, (1-x) )} 
\label{eqdefnl92} \\
&\quad {} 
 + \int_1^{+\infty} \frac{dy}{\sqrt{y}} \, (y-1)^{-1-\varepsilon} \, 
 \int_1^{\frac{y}{y-1}} \, \frac{dv}{v} \, 
 \left[ 
  T^{-1-\varepsilon} \, (v-1)^{-1-\varepsilon} 
  -  
  \left( - v \, R_{ij} - T \, (1-v) \right)^{-1-\varepsilon} 
 \right] 
\notag \\
&\quad {} 
\left. 
 + \int_0^{1} \frac{dy}{\sqrt{y}} \, (1-y)^{-1-\varepsilon} \, 
 \int_0^{\, \frac{y}{1-y}} \frac{dv}{v} \, 
 \left[ 
   T^{-1-\varepsilon} \, (1+v)^{-1-\varepsilon} 
  - 
  \left( - v \, R_{ij} + T \, (1+v) \right)^{-1-\varepsilon} 
 \right]  
\right\} \notag
\end{align}
Let us compute the different terms. Using eq.~(\ref{thirdinty0}), the first integral is readily given by:
\begin{align}
\int_0^{+\infty} \frac{2 \, du}{1+u^2} \, 
\int_{0}^1 \, \frac{dx}{\sqrt{x}} \, (1-x)^{-1-\varepsilon} 
&=
\pi \, B \left( \frac{1}{2}, - \, \varepsilon \right)
\label{eqfirstintyv}
\end{align}
In the second and third lines of eq. (\ref{eqdefnl92}) the $x$ integration is
easily performed after a partial fraction decomposition on the $x$ variable.
We get:
\begin{align}
\hspace{2em}&\hspace{-2em}
\int_0^{+\infty} \frac{dy}{\sqrt{y}} \, \int_0^{+\infty} 
\frac{dx}{(y+x) \; ( (x+y) \, R_{ij} - T \, (1+x) )} 
\notag \\
& \quad {}\quad {}\quad {}\quad {}\quad {}\quad {}
= \frac{1}{T} \, \int_0^{+\infty} \frac{dy}{\sqrt{y}} \, \frac{1}{y-1} 
\left[ \ln \left( \frac{y \, R_{ij} -T}{R_{ij} - T} \right) - \ln(y) \right]
\label{eqsecondintyv} \\
\hspace{2em}&\hspace{-2em}
\int_0^{+\infty} \frac{dy}{\sqrt{y}} \, \int_1^{+\infty} 
\frac{dx}{(y+x) \; ( (x+y) \, R_{ij} + T \, (1-x) )} 
\notag \\
& \quad {}\quad {}\quad {}\quad {}\quad {}\quad {}
= \frac{1}{T} \, \int_0^{+\infty} \frac{dy}{\sqrt{y}} \, \frac{1}{y+1} \, 
\ln \left( \frac{R_{ij}}{R_{ij} - T} \right)
  \label{eqthirdintyv}
\end{align}
We write the last two integrals of eq. (\ref{eqdefnl92}) as:
\begin{align}
\hspace{2em}&\hspace{-2em}
\int_1^{+\infty} \frac{dy}{\sqrt{y}} \, (y-1)^{-1-\varepsilon} \, 
\int_1^{\frac{y}{y-1}} \, \frac{dv}{v} \, 
\left[ 
 T^{-1-\varepsilon} \, (v-1)^{-1-\varepsilon} 
 -  
 \left( - v \, R_{ij} - T \, (1-v) \right)^{-1-\varepsilon} 
\right] 
\notag \\
&= \int_1^{+\infty} \frac{dy}{\sqrt{y}} \, (y-1)^{-1-\varepsilon} \, 
\left[ E_1(y)-E_1(1^{+})\right] \, 
\; + \; E_1(1^{+}) \, 
B \left( \frac{1}{2} + \varepsilon, - \, \varepsilon \right)
\label{eqfourthintyv} \\
\hspace{2em}&\hspace{-2em}
\int_0^{1} \frac{dy}{\sqrt{y}} \, (1-y)^{-1-\varepsilon} \, 
\int_0^{\, \frac{y}{1-y}} \frac{dv}{v} \, 
\left[ 
 T^{-1-\varepsilon} \, (1+v)^{-1-\varepsilon} 
 - 
 \left( - v \, R_{ij} + T \, (1+v) \right)^{-1-\varepsilon} 
\right] 
\notag \\
&= \int_0^{1} \frac{dy}{\sqrt{y}} \, (1-y)^{-1-\varepsilon} \, 
\left[ E_2(y)-E_2(1^{-}) \right] 
\; + \; E_2(1^{-}) \, B\left( \frac{1}{2}, - \, \varepsilon \right)
\label{eqfifthintyv}
\end{align}
with:
\begin{align}
E_1(y) 
&= \int_1^{\frac{y}{y-1}} \, \frac{dv}{v} \, 
\left[ 
 T^{-1-\varepsilon} \, (v-1)^{-1-\varepsilon} 
 -  
 \left( - v \, R_{ij} - T \, (1-v) \right)^{-1-\varepsilon} 
\right] 
\label{eqdeffunce1} \\
E_2(y) 
&= \int_0^{\, \frac{y}{1-y}} \frac{dv}{v} \, 
\left[ 
 T^{-1-\varepsilon} \, (1+v)^{-1-\varepsilon} 
 - 
 \left( - v \, R_{ij} + T \, (1+v) \right)^{-1-\varepsilon} 
\right]
\label{eqdeffunce2}
\end{align}
When $y \to 1^{+}$, the upper bound of the integral
defining the function $E_1(y)$ goes to $+ \infty$ and likewise for $E_2(y)$ 
when $y \to 1^{-}$. The quantities 
$E_1(y)-E_1(1^{+})$ and  $E_2(y)-E_2(1^{-})$ are given by integrals between $y/(y-1)$
and $+\infty$ and between $y/(1-y)$ and $+\infty$ respectively. 
As, in $E_1(y)-E_1(1^{+})$, $y$ is greater than $1$ and so is the
lower bound $y/(y-1)$, the singular support $v=1$ of the distribution
$(v-1)^{-1-\varepsilon}$ lies outside the range of integration thus the first 
term of the r.h.s. of eq. (\ref{eqdeffunce1}) can be taken at $\varepsilon=0$. 
In $E_2(y)-E_2(1^{-})$, the integrand is non singular either and the first term of
the r.h.s. of eq. (\ref{eqdeffunce2}) can also be taken at $\varepsilon=0$. 
These two terms give:
\begin{align}
\int_1^{+\infty} \frac{dy}{\sqrt{y}} \, (y-1)^{-1-\varepsilon} \, 
\left[ E_1(y)-E_1(1^{+}) \right] 
&= - \, \frac{1}{T} \, \int_1^{+\infty} \frac{dy}{\sqrt{y}} \, \frac{1}{y-1} \, 
\ln \left( \frac{y \, R_{ij} - T}{R_{ij} - T} \right) 
\label{eqfourthintyv1t} \\
\int_0^{1} \frac{dy}{\sqrt{y}} \, (1-y)^{-1-\varepsilon} \, 
\left[ E_2(y)-E_2(1^{-}) \right]
&= - \, \frac{1}{T} \, \int_0^1 \frac{dy}{\sqrt{y}} \, \frac{1}{y-1} \, 
\ln \left( \frac{y \, R_{ij} - T}{R_{ij} - T} \right)
 \label{eqfifthintyv1t}
\end{align}
The combined r.h.s. of eqs. (\ref{eqfourthintyv1t}) and (\ref{eqfifthintyv1t})
cancel against the first term in eq. (\ref{eqsecondintyv}). 
The expressions of $E_1(1^{+})$ and $E_2(1^{-})$ are computed using eqs. 
(\ref{eqresulK3}) and ( \ref{eqresulK4}):
\begin{align}
E_1(1^{+}) 
&= - \, K_4(T-R_{ij},-T) 
\notag \\
&= \frac{1}{T} \, 
\left\{ 
 - \, \frac{1}{\varepsilon} \, \left( -R_{ij} \right)^{-\varepsilon} 
 + \ln (T) - \ln (T-R_{ij}) 
\right. 
\notag \\
&\quad {} \quad {} \quad {}
\left. 
 + \, \varepsilon \, 
 \left[ 
  \dilog \left( \frac{R_{ij}}{T} \right) 
  + \ln \left( -R_{ij} \right)\, \ln \left( \frac{T-R_{ij}}{T} \right) 
 \right] 
\right\}
\label{eqdefe3de1}\\
E_2(1^{-}) 
&= - \, K_3(T-R_{ij},T) 
\notag \\
&= \frac{1}{T} \, \ln \left( \frac{T-R_{ij}}{T} \right) \, 
\left[ 1 - \varepsilon \, \ln(T) \right] 
\label{eqdefe2de1}
\end{align}
Putting everything together, we get for $L_4^n(\Delta_3,0,\Delta_1^{\{i,j\}},0)$:
\begin{align}
\hspace{2em}&\hspace{-2em}L_4^n(\Delta_3,0,\Delta_1^{\{i,j\}},0) \notag \\
&= \frac{F(\varepsilon)}{4 \, R_{ij} \, T} \, 
\left\{ 
 i \, \pi \, B \left( \frac{1}{2}, - \, \varepsilon \right) \,
 \left( 1 - \varepsilon \ln(T) \right) 
\right.
\notag\\
& \quad {} \quad {} \quad {} \quad {} \quad {}
- \int_0^{+\infty} \frac{dy}{\sqrt{y}} \, \frac{1}{y-1} \, \ln(y) 
 + i \, \int_0^{+\infty} \frac{dy}{\sqrt{y}} \, \frac{1}{y+1} \, 
 \ln \left( \frac{R_{ij}}{R_{ij}-T} \right) 
\notag \\
&\quad {} \quad {} \quad {} \quad {} \quad {}
 + B \left( \frac{1}{2}, - \, \varepsilon \right) \, 
 \left[ 
  \ln \left( \frac{T - R_{ij}}{T} \right) \, 
  \left( 1 - \varepsilon \, \ln(T) \right) 
 \right] 
\notag \\
&\quad {} \quad {} \quad {} \quad {} \quad {}
 + B \left( \frac{1}{2} + \varepsilon, - \, \varepsilon \right)
 \left[ 
  - \, \frac{1}{\varepsilon} \left( -R_{ij} \right)^{-\varepsilon} 
  + \ln (T) - \ln (T-R_{ij}) 
 \right. 
\notag \\
&\quad \quad \quad \quad \quad \quad \quad
\quad \quad \quad \quad \qquad {}
+ \left. 
 \left.
   \varepsilon 
  \left[ 
   \dilog\left( \frac{R_{ij}}{T} \right) 
   + \ln \left( -R_{ij} \right)\, \ln \left( \frac{T-R_{ij}}{T} \right) 
  \right] \right] 
\right\}
\label{eqdefnl93}
\end{align}
Extracting the Euler Beta functions from eqs. (\ref{secondinty0}), (\ref{secondinty}) and (\ref{thirdinty0}), (\ref{thirdinty}) and using the fact that 
$\ln(- \, R_{ij}) = \ln(R_{ij}) - i \, \pi$,
eq. (\ref{eqdefnl93}) can be cast in the following form:
\begin{align}
\hspace{2em}&\hspace{-2em}L_4^n(\Delta_3,0,\Delta_1^{\{i,j\}},0) \notag \\
&= \frac{1}{2} \, \Gamma(1+\varepsilon) \, 
\frac{\Gamma^2(1-\varepsilon)}{\Gamma(1-2 \, \varepsilon)} \, 
\frac{1}{R_{ij} \, T} \, 
\notag\\
& \quad {} \times
\left\{ 
 \frac{1}{\varepsilon^2} \, \left( 2 \, R_{ij} \right)^{-\varepsilon} 
 - \dilog\left( \frac{R_{ij}}{T} \right) 
 - \left[ \ln \left(R_{ij}\right) - \ln \left( T \right) \right] \, 
 \ln \left( \frac{T - R_{ij}}{T} \right) 
\right\}
\label{eqdefnl95}
\end{align}
Eqs.(\ref{eqdefnl95}) and (\ref{eqdefnl6bis}) have the same analytic 
expression.

\subsubsection{$\Im(\Delta_3) < 0$, $\Im(\Delta_1^{\{i,j\}}) > 0$}

We start with eq.~(\ref{eqeqlijkir170}) with $P_{ijk} = - R_{ij}$.
We then set $x = y + (1+y) \, v$ in the first integral of eq.
(\ref{eqeqlijkir170}), $x = y + (y-1) \, v$ in the second integral and 
$x = y + (1-y) \, v$ in the third one and we get:
\begin{align}
L_4^n(\Delta_3,0,\Delta_1^{\{i,j\}},0)
&= \frac{F(\varepsilon)}{4 \, R_{ij}} \; 
\left\{ 
  i \, \text{e}^{-i \, \pi \, \varepsilon} \, \int^{+\infty}_0  \frac{d y}{\sqrt{y}} \, 
 (1+y)^{-1-\varepsilon} \, 
\right.
\notag\\
& \quad {}\quad {}\quad {}\quad {}\quad {}\quad {}\quad {}\quad {}
 \times 
 \int^{+\infty}_0 \frac{d v}{v} \, 
 \left[ 
  \frac{1}{[ v \, R_{ij} -  (1+v) \, T]^{1+\varepsilon}} 
  - 
  \frac{1}{[- (1+v) \, T]^{1+\varepsilon}} 
 \right] 
\notag \\
&\quad {} \quad {}\quad {}\quad {}\quad {}\quad {}
 + \quad {}
 \int^{+\infty}_1  \frac{d y}{\sqrt{y}} \, (y-1)^{-1-\varepsilon} \, 
\notag\\
& \quad {}\quad {}\quad {}\quad {}\quad {}\quad {}\quad {}\quad {}
 \times 
 \int^{+\infty}_0 \frac{d v}{v} \, 
 \left[ 
  \frac{1}{[ v \, R_{ij} - (1+v) \, T]^{1+\varepsilon}} 
  - 
  \frac{1}{[- (1+v) \, T]^{1+\varepsilon}} 
 \right] 
\notag \\
& \quad {} \quad {}\quad {}\quad {}\quad {}\quad {}
 + \quad {} 
 \int^{1}_0  \frac{d y}{\sqrt{y}} \, (1-y)^{-1-\varepsilon} \, 
\notag\\
& \quad {}\quad {}\quad {}\quad {}\quad {}\quad {}\quad {}\quad {}
\left. 
 \times 
 \int^{+\infty}_1 \frac{d v}{v} \, 
 \left[ 
  \frac{1}{[ v \, R_{ij} + (1-v) \, T ]^{1+\varepsilon}} 
  - 
  \frac{1}{[(1-v) \, T]^{1+\varepsilon}} 
 \right] 
\right\}
\label{eqisigir8s}
\end{align}
Here again the integrals over $y$ and $v$ are unnested. 
Since the signs of 
$\Im(R_{ij})$ and $\Im(T)$ are mutually opposite, let us note that
the imaginary parts of each of the terms raised to the power 
$1+\varepsilon$ in denominators in the $v$ integrals remain constant
over the corresponding ranges of integration over $v$. 
The first two integrals on $v$ of eq.~(\ref{eqisigir8s}) are given by eq.~(\ref{eqresulK3})
with $A = R_{ij} - T$ and $B = - T$ while the last one is given by eq.~(\ref{eqresulK4}) with
$A^{\prime} = R_{ij} - T$ and $B^{\prime} = T$. As for the $y$ integration, they can be read from
eqs.~(\ref{firstinty0}) to (\ref{thirdinty}).
All ingredients combine into:
\begin{align}
\hspace{2em}&\hspace{-2em}L_4^n(\Delta_3,0,\Delta_1^{\{i,j\}},0) \notag \\
&= 
\frac{1}{2} \, \Gamma(1+\varepsilon) \, 
\frac{\Gamma^2(1-\varepsilon)}{\Gamma(1- 2 \, \varepsilon)} \, 
\frac{1}{R_{ij} \, T} \, 
\nonumber \\
&\quad {} \times
\left\{ 
 \frac{1}{\varepsilon^2} \, (2 \, R_{ij})^{-\varepsilon} 
 - \dilog \left( \frac{R_{ij}}{T} \right) - 
 \left[ 
  \ln \left( R_{ij} \right) - \ln \left( - \, T \right) \, - i \, \pi   
 \right] \, 
 \ln \left( \frac{T-R_{ij}}{T} \right) 
\right\}
\label{eqdefnl8}
\end{align}
In the present case $\Im(- \, R_{ij})$ and $\Im(T)> 0$ so that
$\ln \left( - \,T \right) = \ln \left( T \right) - i \, \pi$
thus eq. (\ref{eqdefnl8}) can be rewritten:
\begin{align}
L_4^n(\Delta_3,0,\Delta_1^{\{i,j\}},0)
&= 
\frac{1}{2} \, \Gamma(1+\varepsilon) \, 
\frac{\Gamma^2(1-\varepsilon)}{\Gamma(1- 2 \, \varepsilon)} \, 
\frac{1}{R_{ij} \, T} \, 
\label{eqdefnl8bis} \\
&\quad {} \times
\left\{ 
 \frac{1}{\varepsilon^2} \, (2 \, R_{ij})^{-\varepsilon} 
  - \dilog \left( \frac{R_{ij}}{T} \right) -
  \left[ \ln \left( R_{ij} \right) - \ln \left( T \right) \right] \, 
 \ln \left( \frac{T - R_{ij}}{T} \right)
\right\}
\nonumber
\end{align}
i.e. again the same analytic form as the previous two cases, cf.
eqs. (\ref{eqdefnl6bis}) and (\ref{eqdefnl95}).

\subsubsection{$\Im(\Delta_3) < 0$, $\Im(\Delta_1^{\{i,j\}}) < 0$}

The starting point is now eq.~(\ref{eqlijkir180}) with $P_{ijk} = - R_{ij}$.
In the first two integrals we rescale $y = x \, u^2$ to factorise the double integral into a product of two unnested integrals over $u$ and over $x$.
Using results given in eqs.~(\ref{firstinty0}) and (\ref{secondinty0}), these unnested integrals are readily performed yielding:
\begin{align}
\hspace{2em}&\hspace{-2em}L_4^n(\Delta_3,0,\Delta_1^{\{i,j\}},0) \notag \\
&= \frac{F(\varepsilon)}{4 \, R_{ij}} \, 
\left\{ 
 i \, \pi \, \left( -T \right)^{-1-\varepsilon}
 \left[   
   - i \, \text{e}^{- i \, \pi \, \varepsilon} \, 
  B \left( \frac{1}{2}, \frac{1}{2}+ \varepsilon \right)
  - \,
  B \left( \frac{1}{2}+ \varepsilon, - \, \varepsilon \right) 
\right] 
\right. 
\notag \\
&\quad {} \quad {}
- i \, \int_0^{+\infty} \frac{dy}{\sqrt{y}} \, \int_0^1 \frac{dx}{y+x} 
\left( R_{ij} \, (x+y) + T \, (1-x) \right)^{-1-\varepsilon} 
\notag \\
&\quad {}\quad {} 
+ i \, \text{e}^{- i \, \pi \, \varepsilon} \, 
 \int_0^{+\infty} \frac{dy}{\sqrt{y}} \int_0^y \frac{dx}{y-x} 
\notag \\
&\qquad \qquad \qquad {} \times
 \left[ 
  \left( - R_{ij} \, (y-x) - T \, (1+x) \right)^{-1-\varepsilon} 
 - \left( -T \right)^{-1-\varepsilon} \, (1+x)^{-1-\varepsilon} \right] 
\notag \\
&\quad {}\quad {} 
 + \int_1^{+\infty} \frac{dy}{\sqrt{y}} \int_1^y \frac{dx}{y-x} 
\notag \\
&\qquad \qquad \qquad {} \times
\left.  
 \left[ 
  \left( - R_{ij} \, (y-x) + T \, (1-x) \right)^{-1-\varepsilon} 
  - 
  \left( -T \right)^{-1-\varepsilon} \, (x-1)^{-1-\varepsilon} 
 \right] 
 \vphantom{\frac{1}{2}} \right\}
\label{eqdefnl101}
\end{align}
In the three remaining integrals in eq. (\ref{eqdefnl101}), the first two ones 
remain finite when $\varepsilon \to 0$ and are thus computed in this limit,
which yields:
\begin{align}
&\int_0^{+\infty} \frac{dy}{\sqrt{y}} \, \int_0^1 \frac{dx}{y+x} 
\left( R_{ij} \, (x+y) + T \, (1-x) \right)^{-1}
\notag\\
& \quad {} \quad {} \quad {} \quad {}
 = 
\frac{1}{T} \, \int_0^{+\infty} \frac{dy}{\sqrt{y}} \, \frac{1}{1+y} \, 
\left[ \ln\left( \frac{y \, R_{ij} + T}{R_{ij}} \right) - \ln(y) \right] 
&
\label{i1}\\
& \int_0^{+\infty} \frac{dy}{\sqrt{y}} \int_0^y \frac{dx}{y-x} 
\left[ 
 \left( 
  - R_{ij} \, (y-x) - T \, (1+x) \right)^{-1-\varepsilon} 
  - 
  \left( -T \right)^{-1} \, (1+x)^{-1} 
\right] 
\notag\\
&  \quad {} \quad {} \quad {} \quad {}
= 
\frac{1}{T} \, \int_0^{+\infty} \frac{dy}{\sqrt{y}} \, \frac{1}{1+y} \, 
 \ln \left( \frac{y \, R_{ij} + T}{T} \right) 
\label{i2}
\end{align}
Making the change of variable $u=1/y$ we readily see that
\[
\int_{0}^{+\infty} \frac{dy}{\sqrt{y}} \, \frac{\ln(y)}{1+y}
\, = \, 
- \int_{0}^{+\infty} \frac{du}{\sqrt{u}} \, \frac{\ln(u)}{1+u}
 \, = \, 0
\]
the combination $\{- \, i \times (\ref{i1}) + \, i \times (\ref{i2})\}$ thus
gives $i \, \pi \, \ln(R_{ij}/T)$.
In the 
third integral, we make the change of variable 
$x = y - (y-1) \, v$, the third integral thus becomes
\[
 \int_1^{+\infty} \frac{dy}{\sqrt{y}} \, (y-1)^{-1-\varepsilon} \, 
\int_0^{1} \frac{dv}{v} 
\left[  
 \left( -v \, R_{ij} + T \, (v-1) \right)^{-1-\varepsilon} 
 - 
 \left( -T \right)^{-1-\varepsilon} \, (1-v)^{-1-\varepsilon} 
\right]
\]
The $v$ integration is performed using the result of $K_2(- R_{ij}, - T)$ (cf.\ eq.\
(\ref{eqsecondontv3})). After some algebra, eq. (\ref{eqdefnl101}) thus reads:
\begin{align}
\hspace{2em}&\hspace{-2em}L_4^n(\Delta_3,0,\Delta_1^{\{i,j\}},0) \notag \\
&= \frac{1}{2} \, \Gamma(1+\varepsilon) \, 
\frac{\Gamma^2(1 - \varepsilon)}{\Gamma(1 - 2 \, \varepsilon)} \, 
\frac{1}{R_{ij} \, T} \, 
\left[ 
 \frac{1}{\varepsilon^2} \, \left( 2 \, R_{ij} \right)^{-\varepsilon} 
 + \dilog\left( \frac{T - R_{ij}}{T} \right) - \frac{\pi^2}{6} 
\right]
  \label{eqdefnl103}
\end{align}
Eq. (\ref{eqdefnl103}) is identical to eq. (\ref{eqdefnl6}); since 
$\Im(R_{ij})$ and $\Im(T)$ have the same sign  as in subsubsec. \ref{531}, 
eq. (\ref{eqdefnl103}) can be recast into a form identical to eq. 
(\ref{eqdefnl6bis}):
\begin{align}
\hspace{2em}&\hspace{-2em}L_4^n(\Delta_3,0,\Delta_1^{\{i,j\}},0) \notag \\
&= 
\frac{1}{2} \, \Gamma(1+\varepsilon) \, 
\frac{\Gamma^2(1 - \varepsilon)}{\Gamma(1 - 2 \, \varepsilon)} \, 
\frac{1}{R_{ij} \, T} \, 
\notag\\
& \quad {} \times
\left\{ 
 \frac{1}{\varepsilon^2} \, \left( 2 \, R_{ij} \right)^{-\varepsilon} 
 - \dilog \left( \frac{R_{ij}}{T} \right) 
 - \left[ 
    \ln \left( R_{ij} \right) - \ln \left( T \right)
   \right] \, 
   \ln \left( \frac{T-R_{ij}}{T} \right) 
\right\}
\label{eqdefnl103bis}
\end{align}
In summary in all four cases $L_4^n(\Delta_3,0,\Delta_1^{\{i,j\}},0)$ takes the same analytical form.
Compared with the multiplicity of forms met in the general complex mass case,
and still with the diverse cases met in subsec. \ref{subsect52}, this
simplification come from the coalescence of the pole and branch points all at 
the value 1 which is the end-point singularity causing the appearance of the
soft and collinear singularity in all four cases. 

\subsection{Consistency checks and explicit examples}\label{examplefourpoint}

In refs. \cite{Binoth:2005ff,Binoth:1999sp} the infrared structure 
of any IR divergent $N$-point one-loop integral
was shown to be carried by IR divergent three-point one-loop functions 
resulting from appropriate iterated pinchings. In the following this feature is
explicitly verified for the formulae obtained in this article for the
four-point functions, when compared with those for the three-point functions,
when the latter are formulated most conveniently according to the so-called
``indirect way'' for this purpose. 

\vspace{0.3cm}

\noindent
In the case of purely soft divergence i.e. whenever some $\Delta_2^{\{i\}} = 0$
whereas $\tD_{ijk} \ne 0$ the various cases cf. eqs. 
(\ref{eqlijkir15}), (\ref{eqlijkir16reframed}), (\ref{eqlijkir17reframed}) and 
(\ref{eqlijkir18reframed}) can be encompassed in one single formula.
For this purpose let us introduce the following notation,
where for any complex $Q$ we denote 
$Q_R \equiv \Re(Q)$ and $Q_I \equiv \Im(Q)$:
\begin{align}
\int_{_{\widetilde{(0,1)}}} du \, F(u) 
&= 
\left\{
 \begin{array}{lcl}
  \int_0^{^{- i \, S_{\!_{A}} \, \infty}} du \, F(u) + 
  \int_{_{+\infty}}^{^1} du \, F(u)
  & \mbox{if} & 0 < - \, B_I/A_I <1 \\
  & \mbox{and} & A_I \, [ A_R \, B_I - A_I \, B_R ] > 0 \\
  & & \\
  \int_{_{0}}^{^{1}} du \, F(u) & & \text{otherwise}
 \end{array}
\right.
  \label{eqdefwtilde01}
\end{align}
with $S_{A} = \mbox{sign}(A_I)$, whether
\[
F(u) \; = \; \frac{\ln(A \, u^2 + B) - \ln(A \, u_0^2 + B)}{u^2 - u_0^2}
\quad {} \mbox{with} \quad {} u_0^2 \neq - \, \frac{B}{A}
\]
or\[
  F(u) \; = \; (A \, u^2 + B)^{-1-\varepsilon}
\]
The condition 
``$ 0 < - \, B_I/A_I <1$ and $A_I \, [ A_R \, B_I - A_I \, B_R ] > 0$" 
is the condition for the discontinuity cuts of $\ln (A \, u^2 + B)$ to 
cross the real axis between $0$ and $1$ (cf. appendix~\ref{P2-cut} of \cite{paper2}). 
This enables us to rewrite $L_4^n(\Delta_3,0,\Delta_1^{\{i,j\}},\tD_{ijk})$ in a generic way in the various cases as:
\begin{align}
\hspace{2em}&\hspace{-2em}L_4^n(\Delta_3,0,\Delta_1^{\{i,j\}},\tD_{ijk}) = \frac{2^{\varepsilon}}{T} \, \Gamma(1+\varepsilon) \notag \\
& {} \times
\left\{ 
 - \frac{1}{\varepsilon} \, 
 \int_{\widetilde{(0,1)}} d u \, 
 \left( u^2 \, P_{ijk} + R_{ij} \right)^{-1-\varepsilon} - U \left( \Delta_3,\Delta_1^{\{ij\}},\tD_{ijk} \right) 
\right. 
\notag \\
&\qquad {} 
 \;\; + \int_{\widetilde{(0,1)}} \, \frac{d u }{u^2 \, P_{ijk} + R_{ij}} \, 
 \left[  
 \vphantom{\frac{d u }{u^2 \, P_{ijk} + R_{ij}}}  
  \ln \left( T \, (1-u^2) \right) - \ln \left( \frac{T \, (P_{ijk} + R_{ij})}{P_{ijk}} \right) \right.
  \notag \\
  &\qquad \qquad{} 
  - \left.
  \ln \left( u^2 \, (P_{ijk} + R_{ij} - T) + T \right) + \ln \left( \frac{(P_{ijk} + R_{ij}) \, (T - R_{ij})}{P_{ijk}} \right)
 \right] 
\notag \\
 &\qquad {} 
\left.
- \int_0^1 \, \frac{d u }{u^2 \, P_{ijk} + R_{ij}} \, \left[ \eta \left( \frac{T \, (P_{ijk} + R_{ij})}{P_{ijk}}, \frac{T - R_{ij}}{T} \right) + \ln \left( \frac{T - R_{ij}}{T} \right) \right]
\vphantom{\frac{d u }{u^2 \, P_{ijk} + R_{ij}}} 
\right\}
\label{eqnewverLijk0}
\end{align}
with:
\begin{align}
\hspace{2em}&\hspace{-2em}U(\Delta_3,\Delta_1^{\{ij\}},\tD_{ijk}) \notag \\
   &= \left\{
\begin{array}{lcl}
0 & \mbox{if} & \Im(\Delta_3) > 0 \;  \Im(\Delta_1^{\{ij\}}) > 0 \\
\int_{_{\Gamma^{+}}} \, \frac{d u }{u^2 \, P_{ijk} + R_{ij}} \, 
\ln \left( \frac{T - R_{ij}}{T} \right) & 
\mbox{if} & \Im(\Delta_3) > 0 \;  \Im(\Delta_1^{\{ij\}}) < 0 \\
\int_{_{\Gamma^{+}}} \, \frac{d u }{u^2 \, P_{ijk} + R_{ij}} \, 
\eta \left( \frac{T \, (P_{ijk} + R_{ij})}{P_{ijk}}, \frac{T - R_{ij}}{T} \right) & 
\mbox{if} & \Im(\Delta_3) < 0 \;  \Im(\Delta_1^{\{ij\}}) > 0 \\
\int_{_{\Gamma^{+}}} \, \frac{d u }{u^2 \, P_{ijk} + R_{ij}} \, \left[ 
\ln \left( \frac{T - R_{ij}}{T} \right) \right. &
\mbox{if} & \Im(\Delta_3) < 0 \;  \Im(\Delta_1^{\{ij\}}) < 0 \\
\qquad {} + \left. \eta \left( \frac{T \, (P_{ijk} + R_{ij})}{P_{ijk}}, \frac{T - R_{ij}}{T} \right) \right] & & 
\end{array}
\right.
\label{eqdeffuncu}
\end{align}
where the contour $\Gamma^{+}$ is the closed contour encircling the ``north-east" 
quadrant {\em clockwise} (cf. subsec.~\ref{P2-sectfourpointcomp} of \cite{paper2}).
Depending on the location of the cuts of $ \left( u^2 \, P_{ijk} + R_{ij} \right)^{-1-\varepsilon}$, the first term of eq. (\ref{eqnewverLijk0}) is the same as those
which appear in eqs. (\ref{eqlijsoft40}) or (\ref{eqlijsoft60}), $L_4^n(\Delta_3,0,\Delta_1^{\{i,j\}},\tD_{ijk})$ can 
thus be written as:
\begin{align}
L_4^n(\Delta_3,0,\Delta_1^{\{i,j\}},\tD_{ijk}) &= \frac{1}{T} \,  L_3^n(0,\Delta_1^{\{i,j\}},\tD_{ijk}) + \tilde{L}_4^n(\Delta_3,0,\Delta_1^{\{i,j\}},\tD_{ijk})
\label{decomp-ir}
\end{align}
with
\begin{align}
\hspace{2em}&\hspace{-2em}\tilde{L}_4^n(\Delta_3,0,\Delta_1^{\{i,j\}},\tD_{ijk}) \notag \\ 
&= 
\frac{1}{T} \, 
\left\{ 
 \vphantom{\frac{d u }{u^2 \, P_{ijk} + R_{ij}}}  
 \int_{\widetilde{(0,1)}} \, \frac{d u }{u^2 \, P_{ijk} + R_{ij}} \, 
 \Bigg[ 
  \ln \left( T \, (1-u^2) \right) - \ln \left( \frac{T \, (P_{ijk} + R_{ij})}{P_{ijk}} \right) \right.
  \notag \\
  &\qquad \qquad{} 
  - 
  \ln \left( u^2 \, (P_{ijk} + R_{ij} - T) + T \right) + \ln \left( \frac{(P_{ijk} + R_{ij}) \, (T - R_{ij})}{P_{ijk}} \right)
 \Biggr] 
\notag \\
 &\qquad \qquad {} 
- \int_0^1 \, \frac{d u }{u^2 \, P_{ijk} + R_{ij}} \, \left[ \eta \left( \frac{T \, (P_{ijk} + R_{ij})}{P_{ijk}}, \frac{T - R_{ij}}{T} \right) + \ln \left( \frac{T - R_{ij}}{T} \right) \right]
\notag \\
&\qquad \qquad {}  
  - \left. U \left( \Delta_3,\Delta_1^{\{ij\}}, \tD_{ijk} \right) 
\vphantom{\frac{d u }{u^2 \, P_{ijk} + R_{ij}}} 
\right\}
\label{eqnewvertLijk1}
\end{align}
and $L_3^n(0,\Delta_1^{\{i,j\}},\tD_{ijk})$ is given by eq.~(\ref{eqlijsoft40})
or eq.~(\ref{eqlijsoft62}) depending on the sign of the imaginary part of $\Delta_1^{\{i,j\}}$.

\vspace{0.3cm}

\noindent
For the cases where $\Delta_2^{\{i\}} = 0$ and $\tD_{ijk} = 0$, a unique 
formula for $L_4^n(\Delta_3,0,\Delta_1^{\{i,j\}},0)$ was found above whatever the signs of $\Im(\Delta_3)$ and
$\Im(\Delta_1^{\{ij\}})$. The decomposition of the form (\ref{decomp-ir}) 
stills holds, the IR divergent part of $L_4^n(\Delta_3,0,\Delta_1^{\{i,j\}},0)$ is the same as in the 
three-point case (cf.\ eqs.~(\ref{eqlijsoft8}) and~(\ref{eqlijsoft10bis})), whereas now:
\begin{align}
\tilde{L}_4^n(\Delta_3,0,\Delta_1^{\{i,j\}},0) 
&=  - \frac{1}{2 \, R_{ij} \, T} \, 
\left[ 
 \dilog\left( \frac{R_{ij}}{T} \right) + \left[ \ln(R_{ij}) - \ln(T) \right] \, 
 \ln \left( \frac{T - R_{ij}}{T} \right) 
\right]
\label{eqnewvertLijk2}
\end{align}
Coming back to $I_4^n$ and using the results of subsec.~\ref{indirway}, we have:
\begin{align}
I_4^n 
&= \sum_{i \in S_4} \frac{\bbar_i}{\detg} \, \sum_{j \in S_4 \setminus \{i\}} 
\frac{\bbj{j}{i}}{\detgj{i}} \, \frac{W\left(\detgj{i,j},\tD_{ijk}, \tD_{ijl}\right)}{T} 
\notag \\
&\quad {} 
+ \sum_{i \in S_4} \frac{\bbar_i}{\detg} \, \sum_{j \in S_4 \setminus \{i\}} 
\frac{\bbj{j}{i}}{\detgj{i}} \, \sum_{k \in S_4 \setminus \{i,j\}} 
\frac{\bbj{k}{i,j}}{\detgj{i,j}} \,  \tilde{L}_4^n(\Delta_3,0,\Delta_1^{\{i,j\}},\tD_{ijk})
\label{eqI4nb3}
\end{align}
where $l \in S_4 \setminus \{i,j,k\}$.
The first term in the r.h.s. of eq. (\ref{eqI4nb3}) is nothing but the 
combination of the three-point functions such as in refs. 
\cite{Binoth:2005ff,Binoth:1999sp} decomposed according to the so called 
``direct way" made explicit in subsec.~\ref{dirway}. Note that when 
$\tD_{ijk} = 0$, the quantity $\tilde{L}_4^n(\Delta_3,0,\Delta_1^{\{i,j\}},0)$ which does actually
not depend on $k$ factors out from the sum over $k$ which then 
yields trivially - 1. 
The functions $W(\detgj{i,j},\tD_{ijk}, \tD_{ijl})$ has been defined by eq. (\ref{eqdefwi0}). 
Its arguments have one extra subscript, tracing back the extra pinching which was involved 
compared with the three-point case. We recap here the different results concerning this function.
\begin{equation}
  W\left(\detgj{i,j},\tD_{ijk}, \tD_{ijl}\right) = \frac{2^{\varepsilon}}{\varepsilon} \, \Gamma(1+\varepsilon) \, \int^1_0 dx \, 
\left( D^{\{i,j\}(k)}(x) - i \, \lambda \right)^{-1-\varepsilon}
\label{eqdefwi01}
\end{equation}
where
\begin{equation}
D^{\{i,j\}(k)}(x) 
= 
G^{\{i,j\}(k)} \, x^2 - 2 \, V^{\{i,j\}(k)} \, x - C^{\{i,j\}(k)}
\label{eqremd11}
\end{equation}
with:
\begin{align}
G^{\{i,j\}(k)} 
&= - \cals_{ll} + 2 \, \cals_{kl} - \cals_{kk} = \detgj{i,j} 
\notag \\
V^{\{i,j\}(k)} 
&= \cals_{kl} - \cals_{kk} = \frac{1}{2} \, \left[ \detgj{i,j} - \tD_{ijk} + \tD_{ijl} \right]
\label{eqremd21}\\
C^{\{i,j\}(k)} 
&= \cals_{kk} = - \tD_{ijl}
\notag
\end{align}
Note that
$W(\detgj{i,j},\tD_{ijk}, \tD_{ijl})$ is symmetric under the exchange of $i$ and $j$.

\vspace{0.3cm}

\noindent
* If $\tD_{ijk}$ and $\tD_{ijl}$ both differ from zero, $W\left(\detgj{i,j},\tD_{ijk}, \tD_{ijl}\right)$ is given by eq.~(\ref{eqcompintlog1}).

\vspace{0.3cm}

\noindent
* If only $\tD_{ijl}$ vanishes, $W\left(\detgj{i,j},\tD_{ijk}, 0\right)$ is read from eq.~(\ref{eqdirei3n3}).

\vspace{0.3cm}

\noindent
* If $\tD_{ijk}$ and $\tD_{ijl}$ both vanish, $W\left(\detgj{i,j},0,0\right)$ is given by eq.~(\ref{eqdirei3n4}).

\vspace{0.3cm}

\noindent
For practical purpose let us stress that we do not have to compute a dedicated
formula for each IR four-point case as in ref. \cite{Ellis:2007qk} where
16 cases were distinguished. Indeed, for each contribution labelled by the 
index $i$, 
we merely distinguish two cases: either $\Delta_2^{\{i\}} \ne 0$ for which we 
use the generic formula suited to the massive case, or $\Delta_2^{\{i\}} = 0$ 
for which we use the appropriate formula suited to the IR case at hand.
The massive case is split depending on the vanishing of $\Im(\Delta_3)$ (namely if one internal mass squared has an imaginary part different from zero).
The IR case is also divided in two cases: $\tD_{ijk} = 0$ and $\tD_{ijk} \ne 0$. The latter is furthermore separated
according to the fact that one or several internal masses squared have a non vanishing imaginary part.
All these cases are depicted on the decision tree presented in fig. \ref{dectree}, for each case the appropriate formula is given.
Notice that when $\Im(\Delta_3) \ne 0$, it may appear that $\Delta_2^{\{i\}}$ and/or $\Delta_1^{\{i,j\}}$ are real, 
in this case they must be understood as having a ``$+ i \, \lambda$'' prescription.
The expense paid by the present method is a possible proliferation of 
dilogarithms, a counteraction against which would require some extra work. 
This point will be commented in some more details in the examples studied below.
\tikzstyle{post}=[->,shorten >=1pt,>=stealth,semithick]
\begin{figure}[h!]
\begin{tikzpicture}
  \node at (-4,0) [level 3] (c311) {Eq.~(\ref{eqlijkir15r})};
  \node at (-8,0) [level 2] (c312) {cases};
  \node at (-6.5,2) [level 2] (c31) {$\Im(\Delta_3) = 0$}
    edge [post] node[auto] {Yes} (c311)
    edge [post] node[auto,swap] {No} (c312);
  \node at (-2.6,2) [level 3,text width=3cm] (c32) {Eq.~(\ref{eqdefnl6bis}) for all cases};
  \node at (-4,5) [level 2] (c3) {$\tD_{ijk} = 0$}
    edge [post] node[auto,swap] {No} (c31)
    edge [post] node[auto] {Yes} (c32);
  \node at (6,2) [level 3] (c21) {Eq.~(\ref{P1-eqlijk14alternaivebbar}) of \cite{paper1}};
  \node at (0,2) [level 2] (c22) {cases};
  \node at (3,5) [level 2] (c2) {$\Im(\Delta_3) = 0$}
    edge [post] node[auto,swap] {No} (c22)
    edge [post] node[auto] {Yes} (c21);
  \node at (0,8) [level 2] (c1) {$\Delta_2^{\{i\}} = 0$}
    edge [post] node[auto] {No} (c2)
    edge [post] node[auto,swap] {Yes} (c3);
  \node at (0,10) [root] (c0) {$L_4^n(\Delta_3,\Delta_2^{\{i\}},\Delta_1^{\{i,j\}},\tD_{ijk})$}
    edge [post]  (c1);
  \begin{scope}[every node/.style={level 3}]
    \node [below of = c312,xshift=3.5cm] (c3121) {{\footnotesize $\Im(\Delta_3) > 0$, $\Im(\Delta_1^{\{i,j\}}) > 0$: eq.~(\ref{eqlijkir15c})}};
    \node [below of = c3121] (c3122) {{\footnotesize $\Im(\Delta_3) > 0$, $\Im(\Delta_1^{\{i,j\}}) < 0$: eq.~(\ref{eqlijkir16reframed})}};
    \node [below of = c3122] (c3123) {{\footnotesize $\Im(\Delta_3) < 0$, $\Im(\Delta_1^{\{i,j\}}) > 0$: eq.~(\ref{eqlijkir17reframed})}};
    \node [below of = c3123] (c3124) {{\footnotesize $\Im(\Delta_3) < 0$, $\Im(\Delta_1^{\{i,j\}}) < 0$: eq.~(\ref{eqlijkir18reframed})}};
  \end{scope}
  \begin{scope}[every node/.style={level 3}]
    \node [below of = c22,xshift=4.5cm] (c221) {{\scriptsize $\Im(\Delta_3) > 0$, $\Im(\Delta_2^{\{i\}}) > 0$, $\Im(\Delta_1^{\{i,j\}}) > 0$: eq.~(\ref{P2-eqcas1soustraite}) of \cite{paper2}}};
    \node [below of = c221] (c222) {{\scriptsize $\Im(\Delta_3) > 0$, $\Im(\Delta_2^{\{i\}}) < 0$, $\Im(\Delta_1^{\{i,j\}}) > 0$: eq.~(\ref{P2-eqcas3eclatesoustrait}) of \cite{paper2}}};
    \node [below of = c222] (c223) {{\scriptsize $\Im(\Delta_3) < 0$, $\Im(\Delta_2^{\{i\}}) > 0$, $\Im(\Delta_1^{\{i,j\}}) > 0$: eq.~(\ref{P2-eqcas4eclatesoustrait}) of \cite{paper2}}};
    \node [below of = c223] (c224) {{\scriptsize $\Im(\Delta_3) < 0$, $\Im(\Delta_2^{\{i\}}) < 0$, $\Im(\Delta_1^{\{i,j\}}) > 0$: eq.~(\ref{P2-eqcas5eclatesoustrait}) of \cite{paper2}}};
    \node [below of = c224] (c225) {{\scriptsize $\Im(\Delta_3) > 0$, $\Im(\Delta_2^{\{i\}}) > 0$, $\Im(\Delta_1^{\{i,j\}}) < 0$: eq.~(\ref{P2-eqcas2aa}) of \cite{paper2}}};
    \node [below of = c225] (c226) {{\scriptsize $\Im(\Delta_3) > 0$, $\Im(\Delta_2^{\{i\}}) < 0$, $\Im(\Delta_1^{\{i,j\}}) < 0$: eq.~(\ref{P2-eqcas2bb}) of \cite{paper2}}};
    \node [below of = c226] (c227) {{\scriptsize $\Im(\Delta_3) < 0$, $\Im(\Delta_2^{\{i\}}) > 0$, $\Im(\Delta_1^{\{i,j\}}) < 0$: eq.~(\ref{P2-eqcas2cc}) of \cite{paper2}}};
    \node [below of = c227] (c228) {{\scriptsize $\Im(\Delta_3) < 0$, $\Im(\Delta_2^{\{i\}}) < 0$, $\Im(\Delta_1^{\{i,j\}}) < 0$: eq.~(\ref{P2-eqcas2dd}) of \cite{paper2}}};
  \end{scope}
\foreach \value in {1,...,4}
  \draw[->] (c312) |- (c312\value.west);
\foreach \value in {1,...,8}
  \draw[post] (c22) |- (c22\value.west);
\end{tikzpicture}
\caption{\footnotesize Decision tree to compute $L_4^n(\Delta_3,\Delta_2^{\{i\}},\Delta_1^{\{i,j\}},\tD_{ijk})$ for a given sector labelled by $i$, $j$ and $k$. }\label{dectree}
\end{figure}
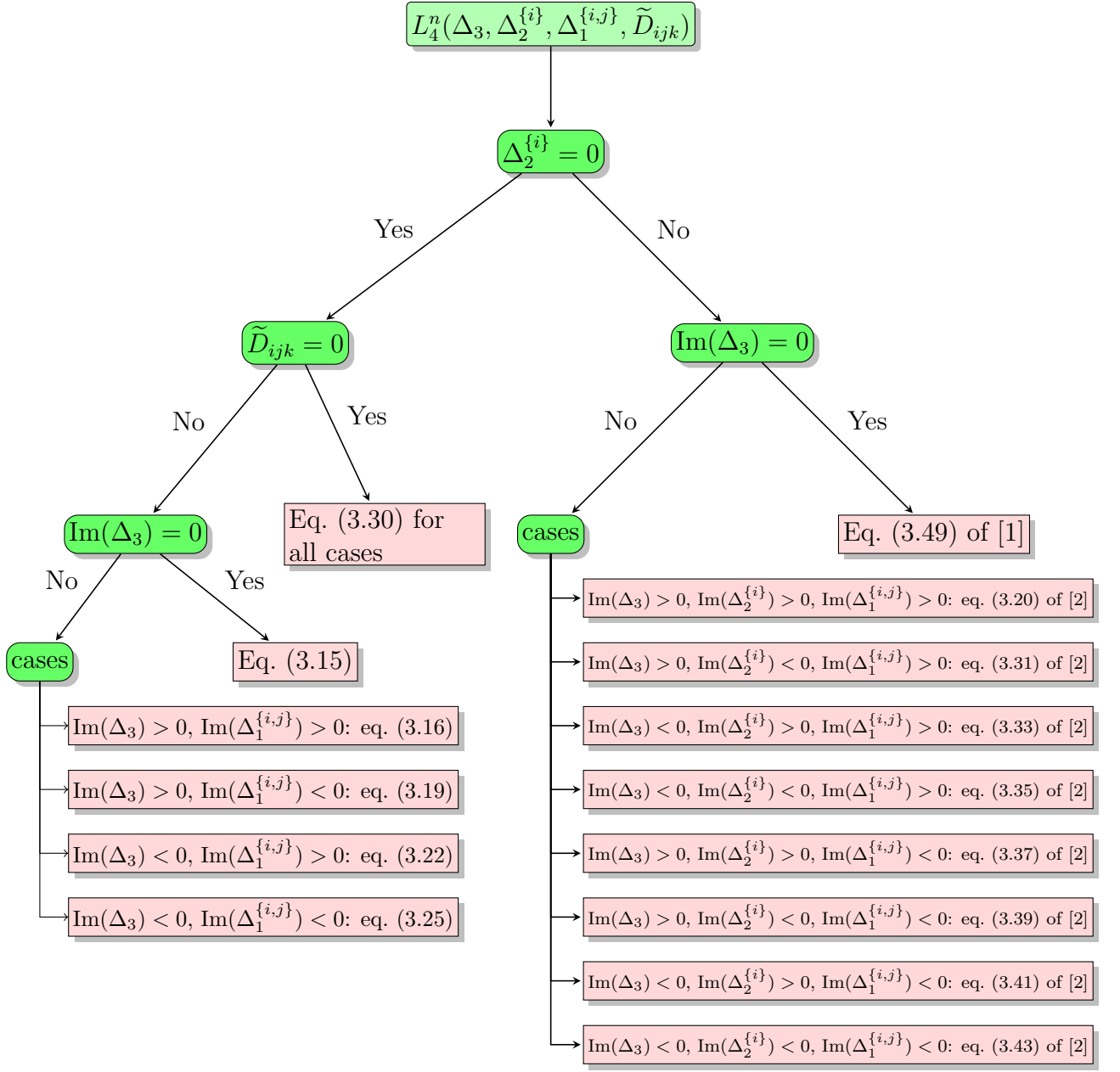

\vspace{0.3cm}

\noindent
{\bf 1) Two opposite external masses}

\vspace{0.3cm}

\noindent
In this case, all internal masses are zero and two opposite external legs
(say 1 and 3) have non lightlike four momenta (our convention is depicted in
fig. \ref{fig2}). The texture of the $\cals$ matrix is:
\begin{align}
  \cals &= \left(
  \begin{array}{cccc}
    0 & 0 & s_{23} & s_1 \\
    0 & 0 & s_3 & s_{12} \\
    s_{23} & s_3 & 0 & 0 \\
    s_1 & s_{12} & 0 & 0
  \end{array}
  \right)
  \label{eqsmatexempl1}
\end{align}
with $s_i = p_i^2$ and $s_{ij} = (p_i + p_j)^2$ (all the momenta are taken ingoing). All the contributions $i$ 
are such that $\Delta_2^{\{i\}} = 0$ and $\tD_{ijk} = 0$. In this case, 
$R_{ij} = - \Delta_1^{\{i,j\}}$ is symmetric under the exchange of $i$ and $j$. 
The four-point function is given by, cf. eq. (\ref{eqdefnl6}):
\begin{align}
I_4^n 
&= \frac{1}{2 \, \dets} \, \Gamma(1+\varepsilon) \, 
\frac{\Gamma^2(1-\varepsilon)}{\Gamma(1 - 2 \, \varepsilon)} \, 
\sum_{i \in S_4} \, \sum_{j > i} \frac{1}{\Delta_1^{\{i,j\}}} \, 
\left( 
 \frac{\bbar_i \, \bbj{j}{i}}{\detgj{i}} 
 + 
 \frac{\bbar_j \, \bbj{i}{j}}{\detgj{j}} 
\right) 
\notag \\
&\quad {} 
\times \, 
\left[ 
 \frac{1}{\varepsilon^2} \, 
 \left( - 2 \, \Delta_1^{\{i,j\}} \right)^{- \varepsilon} 
 + \dilog \left( \frac{T - R_{ij}}{T - i \, \lambda} \right) 
 - \frac{\pi^2}{6} 
\right]
  \label{eqI4nexemple1}
\end{align}
Due to the hollow texture of the reduced $\cals$ matrices, a bunch of 
$\bbj{j}{i}$
coefficients vanish. Eq. (\ref{eqI4nexemple1}) thus simplifies into:
\begin{align}
I_4^n 
&= \frac{1}{2 \, \dets} \, \Gamma(1+\varepsilon) \, 
\frac{\Gamma^2(1-\varepsilon)}{\Gamma(1 - 2 \, \varepsilon)} \, 
\sum_{i=1}^{2} \, \sum_{j=3}^{4} \frac{1}{\Delta_1^{\{i,j\}}} \, 
\left( 
 \frac{\bbar_i \, \bbj{j}{i}}{\detgj{i}} 
 + 
 \frac{\bbar_j \, \bbj{i}{j}}{\detgj{j}} 
\right) 
\notag \\
&\quad {} 
\times \, 
\left[ \frac{1}{\varepsilon^2} \, 
 \left( - 2 \, \Delta_1^{\{i,j\}} \right)^{- \varepsilon} 
 + \dilog\left( 1 - \frac{R_{ij} - i \, \lambda}{T - i \, \lambda} \right) 
 - \frac{\pi^2}{6} 
\right]
\label{eqI4nexemple2}
\end{align}
Expressing the $\bbar_{i}$ and $\bbj{j}{i}$ coefficient as well as the 
various determinants as functions of the $\cals$ matrix elements, we get:
\begin{align}
I_4^n 
&= \Gamma(1+\varepsilon) \, 
\frac{\Gamma^2(1-\varepsilon)}{\Gamma(1 - 2 \, \varepsilon)} \, \frac{2}{d} \, 
\notag\\
& \quad {}
\left\{ 
 \vphantom{\frac{s_1}{s_1}} 
 (- s_1 - i \, \lambda)^{-\varepsilon} + 
 (- s_3 - i \, \lambda)^{-\varepsilon} - 
 (- s_{12} - i \, \lambda)^{-\varepsilon} - 
 (- s_{23} - i \, \lambda)^{-\varepsilon} 
\right. 
\notag \\
&\quad {} \quad {} 
 - \dilog 
   \left( 1 - \frac{s_{12} + i \, \lambda}{d/\Sigma + i \, \lambda} \right) 
 - \dilog
   \left( 1 - \frac{s_{23} + i \, \lambda}{d/\Sigma + i \, \lambda} \right) 
\notag \\
&\quad {}\quad {}  
\left. 
 + \dilog 
   \left( 1 - \frac{s_{1} + i \, \lambda}{d/\Sigma + i \, \lambda} \right) 
 + \dilog 
   \left( 1 - \frac{s_{3} + i \, \lambda}{d/\Sigma + i \, \lambda} \right) 
\right\}
  \label{eqI4nexemple3}
\end{align}
with:
\begin{align}
  d &= s_1 \, s_3 - s_{12} \, s_{23} \label{eqdefdI4n} \\
  \Sigma &= s_1 + s_3 - s_{12} - s_{23}
  \label{eqdefSigmaI4n}
\end{align}
Eq. (\ref{eqdefSigmaI4n}) involves four dilogarithms as in the 
refs. \cite{Duplancic:2000sk,Brandhuber:2004yw}. 
The arguments of the dilogarithms in eq. (\ref{eqI4nexemple3}) vs. 
in ref. \cite{Duplancic:2000sk} are seemingly different, namely  ref. 
\cite{Duplancic:2000sk} involves
$\dilog ( 1 - (w+i \, \lambda)/(d/\Sigma) )$ whereas  
$\dilog ( 1 - (w+i \, \lambda)/(d/\Sigma + i \, \lambda) )$ appears in eq. 
(\ref{eqI4nexemple3}), where $w$ stands for $s_1$, $s_3$, $s_{12}$ or
$s_{23}$. However the $i \, \lambda$ prescriptions matter only 
when the real parts of the arguments of the dilogarithms are greater than 1 i.e. 
whenever $w \, \Sigma/d < 0$, in which case the
signs of $(d/\Sigma) - w$ and of $\Sigma/d$, which respectively control the
signs of the $i \lambda$ prescriptions in either case, are the same: the two 
results in eq. (\ref{eqI4nexemple3}) and in ref. \cite{Duplancic:2000sk} are 
actually identical.

\vspace{0.3cm}

\noindent
This is to be compared with the formula given in ref. \cite{Ellis:2007qk}. The
latter was taken from ref. \cite{Brandhuber:2004yw} which involves five 
dilogarithms instead of four. 
The authors of ref. \cite{Brandhuber:2004yw} used the so-called Mantel identity 
which entails nine dilogarithms to prove that the four-dilogarithm and
five-dilogarithm results are actually equivalent. The Mantel identity happens 
to be a corollary of the Hill identity, the former is derived by applying the 
latter three 
times to some suitable combinations of variables\footnote{See \cite{lewin}, 
chap. 1, p. 2-3 and chap. 2, p. 17-18.}.
The continuation of the Hill identity to any arbitrary two complex variables 
however requires additional combinations of $\eta$ functions handling the
mismatch between the various discontinuities of the dilogarithms involved, 
and these $\eta$ functions are often skipped in the literature\footnote{See 
  however \cite{vanOldenborgh:1989wn}.}. An even busier modification is then 
required for the Mantel identity. This drove 
of $\eta$ functions makes the analytical check of the equivalence between the 
four-dilogarithm and five-dilogarithm expressions extremely awkward in general, 
and to our understanding this drove of $\eta$ functions was not accounted in 
ref. \cite{Brandhuber:2004yw}. We did perform numerical tests accounting for 
these $\eta$ functions, which verified the equivalence for the configurations 
probed.

\vspace{0.3cm}

\noindent
{\bf 2) A simple case with one internal masses}

\vspace{0.3cm}

\noindent
With the same notations of the preceding example, the texture of the $\cals$ matrix is:
\begin{align}
  \cals &= \left(
  \begin{array}{cccc}
    0 & 0 & s_{23}-m_3^2 & 0 \\
    0 & 0 & 0 & s_{12} \\
    s_{23} - m_3^2 & 0 & - 2 \, m_3^2 & s_4 - m_3^2 \\
    0 & s_{12} & s_4 - m_3^2 & 0
  \end{array}
  \right)
  \label{eqsmatexempl2}
\end{align}
In this case, the sector $i=1$
has no soft or collinear divergence and is computed using the massive formula. The
other sectors correspond to the cases: $\Delta_2^{\{i\}} = 0$, $\tD_{ijk} \ne 0$
and $\Delta_2^{\{i\}} = 0$, $\tD_{ijk} = 0$. The four-point amplitude can be
cast in a divergent part and a finite one. The divergent part is given by:
\begin{align}
  \left( I^n_4 \right)_{div} &= \sum_{i \in S_4 \setminus \{1\}} \, \frac{\bbar_i}{\detg} \, \sum_{j \in S_4 \setminus \{i\}} \, \frac{\bbj{j}{i}}{\detgj{i}} \, \frac{W\left(\detgj{i,j},\tD_{ijk},\tD_{ijl}\right)}{T} 
  \label{eqdefdivpartI44}
\end{align}
As several $\bbj{j}{i}$ vanish due to the hollow texture of reduced $\cals$
matrices, we actually have to compute\footnote{We will use the following properties: $\detgj{i,j}$ is symmetric under the permutation $i \leftrightarrow j$ and $\tD_{ijk}$ is symmetric under any permutation of the set $\{i,j,k\}$}:
\begin{align}
  \left( I^n_4 \right)_{div} &= \frac{\bbar_2}{\dets} \, \left[ \frac{\bbj{1}{2}}{\detgj{2}} \, W\left( \detgj{1,2},\tD_{124},0 \right) + \frac{\bbj{4}{2}}{\detgj{2}} \, W\left( \detgj{2,4},\tD_{124},0 \right) \right] \notag \\
  &\quad {} + \frac{\bbar_3}{\dets} \, \frac{\bbj{1}{3}}{\detgj{3}} \, W\left( \detgj{1,3},0,0 \right) \notag \\
  &\quad {} + \frac{\bbar_4}{\dets} \, \frac{\bbj{2}{4}}{\detgj{4}} \, W\left( \detgj{2,4},\tD_{124},0 \right)
  \label{eqdefdivpartI441}
\end{align}
where $W\left( \detgj{1,2},\tD_{124},0 \right)$, $W\left( \detgj{2,4},\tD_{124},0 \right)$ are given by eq. (\ref{eqdirei3n3}) and
\linebreak $W\left( \detgj{1,3},0,0 \right)$ by eq. (\ref{eqdirei3n4}). We get:
\begin{align}
\left( I^n_4 \right)_{div} 
&= 
 \frac{1}{\varepsilon} \, \Gamma(1+\varepsilon) \,
\frac{\Gamma^2(1-\varepsilon)}{\Gamma(1 - 2 \, \varepsilon)} \, 
\frac{1}{s_{12} \, (s_{23} - m_3^2)} \, 
\notag\\
& \quad {}
\left[ 
 \;
 \frac{2}{\varepsilon} \, 
 \left(  - s_{23} + m_3^2 - i \, \lambda \right)^{-\varepsilon} 
 + 
 \frac{1}{\varepsilon} \, 
 \left( - s_{12} - i \, \lambda \right)^{-\varepsilon} 
\right. 
\notag \\ 
&\quad {} 
- \frac{1}{\varepsilon} \, \left( - s_4 + m_3^2 - i \, \lambda \right)^{-\varepsilon} - \frac{1}{2 \, \varepsilon} \left( m_3^2 - i \, \lambda \right)^{-\varepsilon} \notag \\
&\quad {} 
\left. 
 + \varepsilon \, 
 \dilog\left( \frac{s_4}{s_4 - m_3^2 + i \, \lambda} \right) 
 - 2 \, \varepsilon \, 
 \dilog \left( \frac{s_{23}}{s_{23} - m_3^2 + i \, \lambda} \right)  
 + \varepsilon \, \frac{\pi^2}{12} \; 
\right]
  \label{eqdefdivpartI442}
\end{align}
Expanding eq. (\ref{eqdefdivpartI442}) in $\varepsilon$, we recover the results 
of ref. \cite{Ellis:2007qk} (eq. (4.27)) taken from ref. 
\cite{Beenakker:2002nc} for the terms proportional to $1/\varepsilon^2$ and 
$1/\varepsilon$. Concerning the finite part, we obtain a host of terms for
which it is cumbersome to verify  analytically that they do reduce to the finite
part of eq. (4.27) of ref. \cite{Ellis:2007qk}. We verified numerically that
they do indeed.

\section{Summary and outlook}

In this article we presented an extension of a novel approach developed in companion 
articles \cite{paper1} and \cite{paper2} to the case of vanishing internal masses 
involving soft and/or collinear divergences. 
For this latter case, the method remains very similar to the massive cases: the three- and 
four-point functions are split into ``sectors'' whose coefficients are expressed 
in terms of algebraic kinematical invariants involved in reduction algorithms. 
Each ``sector'' may diverge or not when the IR regulator is sent to zero yielding 
to a simple decision tree to compute the relevant integrals. This avoids the computation of
the numerous different integrals over Feynman parameters as it is usually done 
in the literature. This extension also applies to general kinematics beyond the 
one relevant for one-loop collider processes, offering a potential application to 
the calculation of two-loop processes using one-loop (generalised) $N$-point 
functions as building blocks as discussed in the introduction of \cite{paper1}. 

\vspace{0.3cm}

\noindent
One drawback of the present method is the proliferation of dilogarithms in the expression of the four-point function
computed in closed form. This requires some extra work to be
better apprehended, in order to counteract it.
But as the method used hereby is the same as in the real mass case, up to 
slight modifications, any solution found for the latter case can be applied in the infrared divergent case.
This issue will be addressed in a
future article.

\vspace{0.3cm}

\noindent
The last goal is to provide the generalised one-loop building blocks 
entering as integrands in the computation of two-loop three- and four-point 
functions by means of an extra numerical double integration.
In this respect, let us mention that the expansion around $\varepsilon = 0$ of the results 
given in this article has been truncated in order to keep only the divergent and the constant terms.
This is sufficient for any one-loop computation but may be not enough for two-loop applications of the method. 
This article contains already a lot of results, so the expansion around $\varepsilon = 0$ at the necessary orders is postponed to a future work.

\section*{In memoriam}

Various ideas and techniques used in this work were initiated by Prof. Shimizu
after a visit to LAPTh. He explained us his ideas about the numerical 
computation of scalar two-loop three- and four-point functions, he shared his 
notes partly in English, partly in Japanese with us and he encouraged us to 
push this project forward. J.Ph. G. would like to thank Shimizu-sensei for 
giving him a taste of the Japanese culture and for his kindness.

\section*{Acknowledgements}

We would like to thank P. Aurenche for his support along this project and for a careful reading of the manuscript.


\appendix

\section{Two basic integrals \label{appendJ}} 

In what follows $A$ and $B$ are assumed dimensionless and complex valued, the
signs of their real parts is unknown, and the signs of their imaginary parts 
may or may not be the same either. 

\subsection{First kind}

The computation of the three- and four-point functions in a space-time of arbitrary dimensions, 
involves the following extension of the case treated in appendix \ref{P1-appendJ} of ref.\ \cite{paper1}:
\begin{equation}
K(\nu) = \int^{\infty}_0 \frac{d \xi}{(\xi^{\nu}+A) \, (\xi^{\nu}+B)} 
\label{eqdefk1ext}
\end{equation} 
After partial fraction decomposition the r.h.s. of eq. (\ref{eqdefk1ext}) becomes:
\begin{equation}
K(\nu) = \frac{1}{B-A} \, 
\int^{\infty}_0 d \xi \, 
\left[ \frac{1}{\xi^{\nu}+A} - \frac{1}{\xi^{\nu}+B} \right]
\label{eqdefk2ext}
\end{equation}
Let us assume that $\nu > 1$ such that
the r.h.s.\ of eq.\ (\ref{eqdefk1ext}) can be split into a difference of two 
convergent integrals at infinity, which can be separately computed using
appendix \ref{P1-ap2} of ref. \cite{paper1} with $\mu =1$. $K(\nu)$ thus reads:
\begin{equation}
K(\nu) 
= 
\frac{1}{B-A} \, 
\frac{1}{\nu} \, B \left(1 - \frac{1}{\nu}, \frac{1}{\nu} \right) \,
\left[ A^{\frac{1}{\nu} - 1} - B^{\frac{1}{\nu} - 1} \right]
\label{eqdefk3ext}
\end{equation}
regardless of the signs of $\Im(A)$ vs. $\Im(B)$.
In the case of the three-point function, $\nu = 1/(1-\varepsilon)$ which is slightly less 
that $1$ for $\varepsilon < 0$, the r.h.s.\ of eq.~(\ref{eqdefk3ext}) can be analytically 
continued in $\nu$ as long as $\nu \neq -1/n$ or $\nu \neq 1/(n+1)$ with $n$ an arbitrary 
positive integer. So the result of eq.~(\ref{eqdefk3ext}) can be used in the case where 
the dimension of the space-time is shifted by a small positive amount from $n=4$ to $n=4 - 2 \, \varepsilon$.
We also note that the limit $\nu \to 1$ of $K(\nu)$ leads to
eq. (\ref{P1-eqdefk4}) of ref. \cite{paper1}. 

\vspace{0.3cm}

\noindent
A practical case met in sec.\ \ref{3point_ir} is $A = - i \, \lambda$ and $B$ remaining an arbitrary complex number with an imaginary part different from $0$.
In the limit $\lambda \to 0^{+}$, eq.~(\ref{eqdefk3ext}) becomes: 
\begin{align}
\int^{+\infty}_0   
\frac{d \xi}{(\xi^{\nu} - i \, \lambda)\, (\xi^{\nu} + B)} 
&= - \frac{1}{\varepsilon} \, (1-\varepsilon) \, \Gamma(1+\varepsilon) \, \Gamma(1-\varepsilon) 
\, B^{-1-\varepsilon}
\label{eqmodifk}
\end{align}
This is a well-known fact that the two limits $\lambda \to 0^{+}$ and $\varepsilon \to 0$ do not commute.

\subsection{Second kind}

The most general case for the integral
\begin{equation}
  J(\nu) 
  = 
  \int^{+\infty}_0  
   \frac{d \xi}{\left(\xi^{\nu}+A \right) \, \sqrt{\xi^{\nu}+B}}
  \label{eqdefj1}
\end{equation}
has been treated in appendix \ref{P2-appendJ} of ref.\ \cite{paper2}, let us just recap the results.
Two cases can be distinguished according to the signs of the imaginary parts of $A$ and $B$.

\vspace{0.3cm}

\noindent
{\bf 1)} $\Im(A)$ and $\Im(B)$ of the same sign
\begin{equation}
J(\nu) 
= 
\frac{1}{\nu} \, 
B \left( \frac{3}{2}- \frac{1}{\nu},\frac{1}{\nu} \right) \, 
\int^1_0 dz \, \left( (1-z^2) \, A + z^2 \, B \right)^{-3/2 + 1/\nu}
\label{eqdeffuncj2}
\end{equation}

\vspace{0.3cm}

\noindent
{\bf 2)} $\Im(A)$ and $\Im(B)$ of opposite signs
\begin{align}
J(\nu) 
&= - \, \frac{1}{\nu} \, 
B \left( \frac{3}{2}-\frac{1}{\nu},\frac{1}{\nu} \right) \notag \\ 
&\quad {} \times  
\left[ 
 e^{- i \, S_B \, \pi/\nu} \, 
 \int^{+\infty}_0  dz \, 
 \left( B \, z^2  - (1+z^2) \, A \right)^{-3/2+1/\nu} 
\right. 
\notag \\ 
&
\;\;\;\;\;\;\;\;\;\;\;
+ 
\left. 
 \int^{+\infty}_{1}   dz \, 
 \left( B \, z^2 + (1-z^2) \, A \right)^{-3/2+1/\nu} 
\right]
\label{eqdeffuncj7}
\end{align}
with $S_B = \sign\left( \Im\left( B \right) \right)$.

\vspace{0.3cm}

\noindent
Let us remind that the two cases 1) vs. 2) can be reunified by seeing 
eq. (\ref{eqdeffuncj7}) as an analytic continuation in $A$ of eq. 
(\ref{eqdeffuncj2}) which possibly requires a deformation of the contour 
$[0,1]$ originally drawn along the real axis in eq. (\ref{eqdeffuncj2}), cf.\ appendix \ref{P2-appendJ} of \cite{paper2} for more details.

\section{The function $J(x_1,x_2)$}\label{calculJx1x2}

This appendix computes the function 
\[
J(x_1,x_2) 
= \int^1_0 dx \, 
\frac{\ln\left( (x-x_1) \, (x-x_2) \right)}{(x - x_1) \, (x - x_2)}
\]
Using partial fraction decomposition, $J(x_1,x_2)$ can be written as:
\begin{align}
&J(x_1,x_2)
\notag \\
&= \frac{1}{x_1-x_2} 
\left[
\quad {} \quad {}
 \int^1_0 dx \, \frac{\ln(x-x_1)}{x-x_1} 
  \quad {} \quad {} \quad {} - \quad {} \quad {}  \quad {} 
 \int^1_0 dx \, \frac{\ln(x-x_2)}{x-x_2} 
\right. 
\notag \\
&\quad {} \quad {} \quad {} 
+ \int^1_0 dx \, \frac{\ln(x-x_2) - \ln(x_1-x_2)}{x-x_1} 
\; - \; 
\int^1_0 dx \, \frac{\ln(x-x_1) - \ln(x_2-x_1)}{x-x_2} 
\notag \\
&\quad {} \quad {} \quad {} \quad {} \quad {} 
+ 
\left. 
 \ln(x_1-x_2) \, \int^1_0 dx \, \frac{d x}{x-x_1} \;\; - \;\;
 \ln(x_2-x_1) \, \int^1_0 dx \, \frac{d x}{x-x_2} 
\quad {}
 \right]
\label{eqcompj2}
\end{align}
With the help of appendix~\ref{P1-appF} of \cite{paper1} (cf.\ also appendix B of ref. \cite{tHooft:1978jhc}), we immediately get:
\begin{align}
\int^1_0 dx \, \frac{\ln(x-x_1) - \ln(x_2-x_1)}{x-x_2} 
&= R^{\prime}(x_1,x_2) \notag \\
& = \dilog \left( \frac{x_2}{x_2-x_1} \right) 
- \dilog \left( \frac{x_2-1}{x_2-x_1} \right) 
\notag \\
&\quad {} 
+ \eta \left(-x_1, \frac{1}{x_2-x_1} \right) \, 
\ln \left( \frac{x_2}{x_2-x_1} \right) 
\notag \\
&\quad {} 
- \eta \left( 1-x_1, \frac{1}{x_2-x_1} \right) \, 
\ln \left( \frac{x_2-1}{x_2-x_1} \right)
\label{eqappbth1}
\end{align}
where $R^{\prime}(x_1,x_2)$ is 
given by eq.~(\ref{P1-eqdefrprime4}) of \cite{paper1}\footnote{The subtlety discussed in appendix~\ref{P1-appF} of \cite{paper1} 
does not show up here because $x_1$ and $x_2$ have imaginary parts of opposite signs.}.
Since $x_1$ and $x_2$ have imaginary parts of opposite signs
all the $\eta$ functions vanish in eq.~(\ref{eqappbth1}).
Then using the Landen identity: 
\begin{equation}
\dilog(z) + \dilog \left( \frac{z}{z-1} \right) 
= - \, \frac{1}{2} \, \ln^2(1-z)
\label{eqlanden}
\end{equation}
we can rewrite eq. (\ref{eqappbth1}) as:
\begin{align}
&\int^1_0 dx \, 
\frac{\ln(x-x_1) - \ln(x_2-x_1)}{x-x_2} 
\notag\\
& \quad {} \quad {} \quad {} \quad {} \quad {} \quad {} 
 =
\dilog \left( \frac{x_2-1}{x_1-1} \right) 
 - \dilog \left( \frac{x_2}{x_1} \right) 
\notag\\
& \quad {} \quad {} \quad {} \quad {} \quad {} \quad {} \quad {}  
+ \frac{1}{2} 
\left[ 
 \ln^2 \left( x_1-1 \right)  \vphantom{\frac{x_1-1}{x_1}} 
 - \ln^2 \left( x_1 \right) - 2 \, \ln(x_1-x_2) \, 
 \ln \left( \frac{x_1-1}{x_1} \right) 
\right]
\label{eqappbth2}
\end{align}
Substituting into eq. (\ref{eqcompj2}) and easily computing the remaining
$x$ integrals, we get:
\begin{align}
&J(x_1,x_2) 
\notag\\
&= \frac{1}{x_1-x_2} 
\left\{ 
 \;\;
\frac{1}{2} \, \left[ \ln^2(1-x_1) - \ln^2(-x_1) \right] 
 - 
 \frac{1}{2} \, \left[ \ln^2(1-x_2) - \ln^2(-x_2) \right] 
 \right. 
\notag \\
&\quad {} \quad {} \quad {} \quad {} \quad {} \quad {}
 + \dilog \left( \frac{x_1-1}{x_2-1} \right) 
 - \dilog \left( \frac{x_1}{x_2} \right) 
 - \dilog \left( \frac{x_2-1}{x_1-1} \right) 
 + \dilog \left( \frac{x_2}{x_1} \right) 
\notag\\
&
\quad {} \quad {} \quad {} \quad {} \quad {} \quad {}
+ 
\frac{1}{2} 
\left[ 
 \ln^2 \left( x_2-1 \right) \vphantom{\frac{x_2-1}{x_2}} 
 - 
 \ln^2 \left( x_2 \right) 
 - 
 2\, \ln(x_2-x_1) \, \ln \left( \frac{x_2-1}{x_2} \right) 
\right] 
\notag \\
& \quad {} \quad {} \quad {} \quad {} \quad {} \quad {}
 - 
\frac{1}{2} 
\left[ 
 \ln^2 \left( x_1-1 \right) \vphantom{\frac{x_1-1}{x_1}} 
 - 
 \ln^2 \left( x_1 \right) 
 - 
 2\, \ln(x_1-x_2) \, \ln \left( \frac{x_1-1}{x_1} \right) 
\right] 
\notag \\
&\quad {} \quad {} \quad {} \quad {} \quad {} \quad {}
+ 
\left. 
 \ln(x_1-x_2) \, \ln \left( \frac{x_1-1}{x_1} \right) 
 - 
 \ln(x_2-x_1) \, \ln \left( \frac{x_2-1}{x_2} \right) 
\;\;
\right\}
\label{eqcompj3}
\end{align}
The following identity:
\[
\ln^2(1-x_1) - \ln^2(-x_1) - \ln^2(x_1-1)+\ln^2(x_1) 
= - 2 \, i \, \pi \, S(x_1) \, \ln \left( \frac{x_1-1}{x_1} \right)
\]
where $S(x_1)$ is the sign of the imaginary part of $x_1$,
allows to simplify  
eq. (\ref{eqcompj3}) into:
\begin{align}
&J(x_1,x_2) 
= \frac{1}{x_1-x_2} 
\left\{ 
 \dilog \left( \frac{x_1-1}{x_2-1} \right) 
 - 
 \dilog \left( \frac{x_1}{x_2} \right) 
 - 
 \dilog \left( \frac{x_2-1}{x_1-1} \right) 
 + 
 \dilog \left( \frac{x_2}{x_1} \right) 
\right. 
\notag \\
&\quad {} \quad {} \quad {} \quad {} \quad {}  \quad {} 
\quad {}  \quad {} \quad {}  \quad {}
 + \left[ 2\, \ln(x_1-x_2) - i \, \pi \, S(x_1) \right] \, 
 \ln \left( \frac{x_1-1}{x_1} \right) 
\notag \\
&\quad {} \quad {} \quad {} \quad {} \quad {} \quad {} 
\quad {} \quad {} \quad {}  \quad {}\left.
 - \left[ 2\, \ln(x_2-x_1) - i \, \pi \, S(x_2) \right] \, 
 \ln \left( \frac{x_2-1}{x_2} \right) 
\right\}
\label{eqcompj4}
\end{align}
Then, we use the identity relating $\dilog (z)$ and $\dilog(1/z)$ \cite{abramowitz}, 
and also the following relation:
\begin{equation}
  2 \, \ln \left( x_1 - x_2 \right) - i \, \pi \, S(x_1) = 2 \, \ln \left( x_2 - x_1 \right) - i \, \pi \, S(x_2) = \ln \left( - (x_1 - x_2)^2 \right)
  \label{eqrelsimp1}
\end{equation}
whose validity relies on the fact that $x_1$ and $x_2$ have 
imaginary parts of opposites signs. This yields:
\begin{align}
&J(x_1,x_2) 
\notag\\
&= \frac{1}{x_1-x_2} 
\left\{ 
\;\;
 2 \, \dilog \left( \frac{x_1-1}{x_2-1} \right) - 
 2 \, \dilog \left( \frac{x_1}{x_2} \right)
 + \frac{1}{2} \, \ln^2 \left( - \frac{x_1-1}{x_2-1} \right) 
 - \frac{1}{2} \, \ln^2 \left( - \frac{x_1}{x_2} \right) 
\right.
\notag \\
&\quad {} \quad {} \quad {} \quad {} \quad {} \quad {} \quad {}  
+ 
\left. 
 \ln \left( - (x_1-x_2)^2 \right) \, 
 \left[ 
  \ln \left( \frac{x_1-1}{x_1} \right) - \ln \left( \frac{x_2-1}{x_2} \right) 
 \right] 
\;\;
\right\}
\label{eqcompj5}
\end{align}

\section{Herbarium of utilitarian integrals \label{herba}}

This appendix collects a bunch of integrals appearing in the three- and four-point cases to make the reading easier.

\vspace{0.3cm}

\noindent
The first series appears under the following forms and can be computed in terms of the Euler Beta function:
\begin{align}
\int^{+\infty}_0 dz \left(1+z^2 \right)^{-1-\varepsilon} & = \frac{1}{2} \, \int^{+\infty}_0 \frac{dy}{\sqrt{y}} \, (1+y)^{-1-\varepsilon} 
 = \frac{1}{2} \, B \left( \frac{1}{2},\frac{1}{2} + \varepsilon \right) 
\label{firstinty0} \\
\int^{+\infty}_1 dz \left(z^2-1 \right)^{-1-\varepsilon} & = \frac{1}{2} \, \int^{+\infty}_1 \frac{dy}{\sqrt{y}} \, (y-1)^{-1-\varepsilon} 
 =
 \frac{1}{2} \,B \left( \frac{1}{2} + \varepsilon, - \, \varepsilon \right) 
\label{secondinty0} \\
\int^1_0 dz \left(1-z^2 \right)^{-1-\varepsilon} & =  \frac{1}{2} \,\int^{1}_0 \frac{dy}{\sqrt{y}} \, (1-y)^{-1-\varepsilon} 
\quad {} 
 = \frac{1}{2} \, B \left( \frac{1}{2}, - \, \varepsilon \right) 
\label{thirdinty0}
\end{align}
Using the duplication formula for the Gamma functions \cite{abramowitz}, The $z$ integrals computed in closed form read:
\begin{align}
\int^{+\infty}_0 dz \left(1+z^2 \right)^{-1-\varepsilon} 
&= \frac{\tan(\pi \, \varepsilon)}{2\varepsilon} \, 
\frac{\Gamma^2(1-\varepsilon)}{\Gamma(1- 2\, \varepsilon)} \, 
2^{-2 \, \varepsilon} \, 
\label{firstinty} \\
\int^{+\infty}_1 dz \left(z^2-1 \right)^{-1-\varepsilon} 
&= - \frac{1}{2 \, \varepsilon} \, \frac{1}{\cos(\pi \, \varepsilon)}
\frac{\Gamma^{2}(1-\varepsilon)}{\Gamma(1-2 \, \varepsilon)} \, 
2^{- 2 \, \varepsilon}
\label{secondinty} \\
\int^1_0 dz \left(1-z^2 \right)^{-1-\varepsilon} 
&= 
- \frac{1}{2 \, \varepsilon} \, 2^{-2 \, \varepsilon} \, 
\frac{\Gamma(1-\varepsilon)^2}{\Gamma(1-2 \, \varepsilon)}
\label{thirdinty}
\end{align}

\vspace{0.3cm}

\noindent
For the four-point functions in the real mass case or in the complex mass case when $\sign(\Im(R_{ij})) = \sign(\Im(T))$ , the following integral needs to be evaluated:
\begin{align}
  K_2(R_{ij},T) 
&= \int^1_0 \frac{d v}{v} \, 
\left[ 
 \frac{1}{[ v \, R_{ij} + (1-v) \, T \, - i \, \lambda ]^{1+\varepsilon}}  
 - 
 \frac{1}{[(1-v) \, T \, - i \, \lambda]^{1+\varepsilon}} 
\right]
\label{eqdefK2}
\end{align}
Note that in both cases, $\Im(v \, R_{ij} + (1-v) \, T \, - i \, \lambda)$ has a constant sign when $v$ spans $[0,1]$.
After a partial fraction decomposition w.r.t. the variable $v$,  $K_2(R_{ij},T)$ 
can be written as :
\begin{align}
K_2(R_{ij},T) 
&= 
\frac{1}{T} \, \int^{1}_{0} dv \, 
\Biggl\{ 
- (R_{ij}-T) \, 
\left[ R_{ij} \, v + T \, (1-v) - i \, \lambda \right]^{-1-\varepsilon} 
 - 
 (T - i \, \lambda)^{-\varepsilon}  \, (1-v)^{-1-\varepsilon} 
\notag \\
&\quad {}\quad {}\quad {} \quad {} \quad {}\quad {}
  +
 \frac{1}{v} \, 
 \left[ 
  \left( R_{ij} \,v + T \, (1-v) - i \, \lambda \right)^{- \varepsilon} 
  - (T - i \, \lambda)^{- \varepsilon} 
 \right] 
\notag \\
&\quad {}\quad {}\quad {} \quad {}\quad {}\quad {}
 + \frac{(T - i \, \lambda)^{-\varepsilon}}{v} \, 
 \left[ 1 - (1-v)^{- \varepsilon} \right] 
\Biggr\}
\label{eqsecondontv}
\end{align}
The first two terms of eq. (\ref{eqsecondontv}) which yield a $1/\varepsilon$
pole are integrated in closed form, whereas 
the last three terms of eq. (\ref{eqsecondontv}) which are not divergent can be
expanded around $\varepsilon=0$ up to order $\varepsilon$: 
\begin{align}
K_2(R_{ij},T) 
&= \frac{1}{T} \, 
\Biggl[
 \;\;\;\; \frac{1}{\varepsilon} \, (R_{ij} - i \, \lambda)^{-\varepsilon} 
\notag \\
&\quad {}  \quad {}\quad {}
 - \; \varepsilon \, \int^{1}_{0} \frac{dv}{v}  \, 
 \left[ 
  \ln \left( R_{ij} \,v + T \, (1-v) - i \, \lambda \right) 
  - \ln \left(T - i \, \lambda \right) 
 \right] 
\notag \\
&\quad {}  \quad {}\quad {}
 + \; \varepsilon \, (T - i \, \lambda)^{-\varepsilon} \, 
 \int^1_0 \frac{dv}{v} \, \ln(1-v) 
\; 
\Biggr]
\label{eqsecondontv1}
\end{align}
Since $\sign(\Im(R_{ij} \, v + T \, (1-v) - i \, \lambda)) 
= \sign(\Im(T - i \, \lambda))$ when $v \in [0,1]$ the logarithms in the first 
integral can be combined together. The last integration is performed explicitly
and we end with :
\begin{align}
K_2(R_{ij},T) 
&= \frac{1}{T} \, 
\Biggl[ 
 \frac{1}{\varepsilon} \,  (R_{ij} - i \, \lambda)^{-\varepsilon} 
 + \varepsilon \, \dilog \left( \frac{T - R_{ij}}{T - i \, \lambda} \right) 
 - \varepsilon \, \frac{\pi^2}{6} \Biggr]
\label{eqsecondontv3}
\end{align}

\vspace{0.3cm}

\noindent
With respect to the preceding case, two new integrals show up when $\sign(\Im(T)) \ne \sign(\Im(R_{ij}))$:
\begin{align}
K_3(A,B) 
&= \int^{+\infty}_0 \frac{dv}{v} \, 
\left[ (A \, v + B)^{-1-\varepsilon} - (B \, (1+v))^{-1-\varepsilon} \right]
\label{eqdefk3text} \\
K_4(A^{\prime},B^{\prime}) 
&= \int^{+\infty}_1 \frac{dv}{v} \, 
\left[ (A^{\prime} \, v + B^{\prime})^{-1-\varepsilon} - (B^{\prime} \, (1-v))^{-1-\varepsilon} \right]
\label{eqdefk4text}
\end{align}
where $\sign(\Im(A \, v + B))$ (resp. $\sign(\Im(A^{\prime} \, v + B^{\prime}))$) keeps a constant sign when $v$ spans $[0, + \infty[$ (resp. $[1, + \infty[$).
To compute $K_3(A,B)$, we expand the r.h.s. of eq. (\ref{eqdefk3text}) around
$\varepsilon=0$ and we get:
\begin{align}
K_3(A,B) 
&= \frac{1}{B} \, 
\ln \left( \frac{B}{A} \right) \, \left[ 1 - \varepsilon \, \ln(B) \right]
\label{eqresulK3}
\end{align}
To compute $K_4(A^{\prime},B^{\prime})$, we expand  $(A^{\prime} \, v + B^{\prime})^{-1-\varepsilon}$ in eq. 
(\ref{eqdefk4text}) around $\varepsilon=0$, keeping in mind that 
$\sign(\Im(A^{\prime}+B^{\prime})) = \sign(\Im(A^{\prime}))$, and we get:
\begin{align}
K_4(A^{\prime},B^{\prime})
&=
\frac{1}{B^{\prime}}
\left\{
 - \, \frac{1}{\varepsilon} (-B^{\prime})^{\varepsilon} 
 + \left( \ln(A^{\prime}+B^{\prime}) - \ln(A^{\prime}) \right) 
\right.
\notag\\
& \quad {}\quad {}\quad {}
 + \left. \varepsilon \, 
 \left[
  \dilog \left( - \, \frac{B^{\prime}}{A^{\prime}} \right) - \frac{\pi^2}{6} 
  -  
  \frac{1}{2} 
  \left( \ln^2(A^{\prime}+B^{\prime}) - \ln^2(A^{\prime}) \right)
 \right]
\right\}
\label{eqresulK4proto}
\end{align}
With the additional assumption that $\sign(\Im(A^{\prime})) = - \sign(\Im(B^{\prime}))$ and after some algebra $K_4(A^{\prime},B^{\prime})$ can be recast in the 
alternative more useful form: 
\begin{align}
K_4(A^{\prime},B^{\prime})
&=
\frac{1}{B^{\prime}}
\left\{
 - \, \frac{1}{\varepsilon} (A^{\prime}+B^{\prime})^{\varepsilon}
 + \left( \ln(-B^{\prime}) - \ln(A^{\prime}) \right)
\right.
\notag\\
& \quad {}\quad {}\quad {}
 + \left. \varepsilon \, 
 \left[
  \dilog \left( \frac{A^{\prime}+B^{\prime}}{B^{\prime}} \right) + 
  \ln \left( A^{\prime}+B^{\prime} \right) \, 
  \ln \left( - \, \frac{A^{\prime}}{B^{\prime}} \right) 
 \right]
\right\}
\label{eqresulK4}
\end{align}
Notice that this additional assumption is always fulfilled in the cases met.

\section{Detailed comparisons with the ``direct way" 
( cf. sec. \ref{3point_ir})}\label{direcway3pIR}

Our present goal is to check that the ``indirect way'' leads to the same results for
infrared divergent three-point functions as the ``direct way''. By performing
explicitly the sum over the $j$ index in eq. (\ref{eqdef3n3}), we recover the
results of section (\ref{3point_ir}). 
This part is not necessary for the understanding of the method proposed in this article and can be skipped in a first reading.

\subsubsection*{Real mass case}

Let us start by the real mass case. We successively revisit
the examples examined in subsection (\ref{exp_exemp_ir}).

\vspace{0.3cm}

\noindent
{\bf 1. Occurrence of a soft divergence}\\  
\noindent
Let us recap the texture of the $\cals$ matrix (cf.\ eq.~(\ref{eqcalssoft})):
\begin{equation}
  \cals =
  \left(
  \begin{array}{ccc}
    0 & 0     & 0 \\
    0 & -2 \, m_2^2 & s_3 - m_2^2 - m_3^2 \\
    0 & s_3 - m_2^2 - m_3^2 & - 2 \, m_3^2
  \end{array}
  \right)
  \label{eqcalssoft_ver}
\end{equation}
thus $\det(\cals) = 0$. Let us single out row and column 1. We readily see that
the vector $V^{(1)}$ vanishes and so do the coefficients $\bbar_2$ and 
$\bbar_3$; thus $\bbar_1 =  \detg$ since the coefficients $\bbar_i$ fulfil 
$\sum_{i \in S_3} \bbar_i = \detg$. The three-point function is thus given by (cf.\ eq.~(\ref{eqdef3n3})):
\begin{align}
I_3^n 
&= 
\frac{\bbj{2}{1}}{\detgj{1}} \, 
L_{3}^{n} \left( 0,\Delta_{1}^{\{1\}},\tD_{12} \right)
+  \frac{\bbj{3}{1}}{\detgj{1}} \, 
L_{3}^{n} \left( 0,\Delta_{1}^{\{1\}},\tD_{13} \right)
\label{eq_verif_ir1}
\end{align}
The only relevant reduced $\cals$ matrix is:
\begin{equation}
  \cals^{\{1\}} =
  \left(
  \begin{array}{cc}
    -2 \, m_2^2 & s_3 - m_2^2 - m_3^2 \\
    s_3 - m_2^2 - m_3^2 & - 2 \, m_3^2
  \end{array}
  \right)
  \label{eqcalssoft_ver1}
\end{equation}
whose determinant $\det (\cals^{\{1\}}) = - {\cal K}(s_3,m_2^2,m_3^2)$ 
involves the K\"all\'en function given by eq.~(\ref{eqkallenfunc}).
The associated Gram ``matrix" degenerates into a single scalar:
\begin{equation}
  G^{\{1\}(2)} = \detgj{1} = \left( 2 \, s_3 \right)
  \label{eqgrammat_ver1}
\end{equation}
One easily reads the $\bbj{j}{1}$ coefficients from the reduced Gram matrix $G^{\{1\}(2)}$ and the vector $V^{\{1\}(2)}$ (cf.\ the group of eqs.~(\ref{P1-Rij}) of ref.~\cite{paper1}):
\begin{align}
  \bbj{2}{1} &= m_2^2 - m_3^2 - s_3, \quad \bbj{3}{1} = m_3^2 - m_2^2 - s_3 \label{bbar21}
\end{align}
Since $\tD_{12} = 2 \, m_3^2$ and $\tD_{13} = 2 \, m_2^2$, the quantities
$L_3^n(0,\Delta_1^{\{1\}},\tD_{12})$ and $L_3^n(0,\Delta_1^{\{1\}},\tD_{13})$ are given by eq. (\ref{eqlijsoft41}). The
$\tD_{12} +
\Delta_1^{\{1\}}$, $\tD_{13} + \Delta_1^{\{1\}}$ and $\Delta_1^{\{1\}}$ terms are given by eqs.~(\ref{P1-eqtruc1}) and (\ref{P1-eqtruc3}) of \cite{paper1}:
\begin{align}
\tD_{12} + \Delta_1^{\{1\}} 
&= \frac{(s_3 + m_3^2 - m_2^2)^2}{2 \, s_3} 
\label{eqdelta1pd12_ver} \\
\tD_{13} + \Delta_1^{\{1\}} 
&= \frac{(s_3 + m_2^2 - m_3^2)^2}{2 \, s_3} 
\label{eqdelta1pd13_ver} \\
\Delta_1^{\{1\}} 
&= \frac{{\cal K}(s_3,m_2^2,m_3^2)}{2 \, s_3} 
\label{eqdelta1_ver}
\end{align}
The roots of the denominator of eq. (\ref{eqlijsoft41}) are such that:
\begin{align}
(\bar{z}^{12})^2 
&= \frac{\Delta_1^{\{1\}} + i \, \lambda}{\tD_{12} + \Delta_1^{\{1\}}} 
= 
\frac{{\cal K}(s_3,m_2^2,m_3^2) + i \, \lambda \, \sigma_s}
{(s_3 + m_3^2 - m_2^2)^2} 
\label{eqroot_ver12} \\
(\bar{z}^{13})^2 
&= 
\frac{\Delta_1^{\{1\}} + i \, \lambda}{\tD_{13} + \Delta_1^{\{1\}}} 
= \frac{{\cal K}(s_3,m_2^2,m_3^2) + i \, \lambda \, \sigma_s}
{(s_3 + m_2^2 - m_3^2)^2} 
\label{eqroot_ver13}
\end{align}
with $\sigma_s = \sign(s_3)$. We specify:
\begin{align}
\bar{z}^{12} 
= \frac{\sqrt{{\cal K}(s_3,m_2^2,m_3^2) + i \, \lambda \, \sigma_s}}
{s_3 + m_3^2 - m_2^2} 
& \quad {} , \quad {}
\bar{z}^{13} 
= \frac{\sqrt{{\cal K}(s_3,m_2^2,m_3^2) + i \, \lambda \, \sigma_s}}
{s_3 + m_2^2 - m_3^2} 
\label{eqroot_ver13r}
\end{align}
Introducing two new quantities: $\tx_1$ and $\tx_2$ which are the two roots of the equation $D^{\{1\}\, (2)}(x) = 0$ appearing in the ``direct way'':
\begin{align}
\tx_1 
&= 
\frac{s_3 + m_2^2 - m_3^2 + \sqrt{{\cal K}(s_3,m_2^2,m_3^2) 
+ i \, \lambda \, \sigma_s}}{2 \, s_3} 
\label{eqxtilde1def} \\
\tx_2 
&= \frac{s_3 + m_2^2 - m_3^2 - \sqrt{{\cal K}(s_3,m_2^2,m_3^2) 
+ i \, \lambda \, \sigma_s}}{2 \, s_3} 
\label{eqxtilde2def}
\end{align}
Notice that the quantities $1-\tx_1$ and $1-\tx_2$ are the 
roots of the equation $D^{\{1\} \, (3)}(x) = 0$. 
The quantities $\bar{z}^{12}$, one of the roots of the equation $(\tD_{12} + \Delta_1^{\{1\}}) \, z^2 - \Delta_1^{\{1\}} -i \, \lambda = 0$, 
and $\bar{z}^{13}$, a root of the equation $(\tD_{13} + \Delta_1^{\{1\}}) \, z^2 - \Delta_1^{\{1\}} -i \, \lambda = 0$,
can be related
to the roots $\tx_1$ and $\tx_2$ by the following relations:
\begin{align}
\bar{z}^{12} 
= \frac{\tx_1 - \tx_2}{2 - \tx_1 - \tx_2} 
&\quad {} , \quad {} 
\bar{z}^{13}
= \frac{\tx_1 - \tx_2}{\tx_1 + \tx_2} \label{eqroot_ver13r1}
\end{align}
Putting things together, the $I_3^n$ can be written:
\begin{align}
I_3^n 
&= 
- \frac{2^{\varepsilon} \, \Gamma(1 + \varepsilon)}
{2 \, \sqrt{{\cal K}(s_3,m_2^2,m_3^2) + i \, \lambda \, \sigma_s}} \, 
\left\{
 - \frac{1}{\varepsilon} \, 
 \left[ 
  \ln \left( \frac{\bar{z}^{12}-1}{\bar{z}^{12}+1} \right) + 
  \ln \left( \frac{\bar{z}^{13}-1}{\bar{z}^{13}+1} \right) 
 \right] 
\right. 
\notag \\
&\quad 
+ 
\left. 
 \bar{H}_{0,1}
 \left( \tD_{12}+\Delta_1^{\{1\}},-\Delta_1^{\{1\}}- i \, \lambda \right)  
 + 
 \bar{H}_{0,1}
 \left( \tD_{13}+\Delta_1^{\{1\}},-\Delta_1^{\{1\}}- i \, \lambda \right) 
 \vphantom{\frac{\bar{z}^{12}-1}{\bar{z}^{12}+1}} \right\}
\label{eq_verif_ir2}
\end{align}
with:
\begin{equation}
\bar{H}_{0,1}
 \left( A,B \right) 
= 
2 \, A \, \sqrt{ - \frac{B}{A} } \, 
H_{0,1} 
\left( A,B \right) 
\label{eqdefhbarij}
\end{equation}
which is effectively the content between the curly brackets of
eq. (\ref{eqdefh05}) in appendix \ref{appF}.
Expressing all the arguments of logarithms and dilogarithms in eq. 
(\ref{eqdefh05}) in terms of $\tx_1$ and $\tx_2$, we get:
\begin{align}
I_3^n 
&= 
- \, \frac{2^{\varepsilon} \, \Gamma(1 + \varepsilon)}
{(\tx_1 - \tx_2) \, \detgj{1}} \, 
\notag\\
& \quad {} \times
\left\{ 
 - \, \frac{1}{\varepsilon} \, 
 \left[ 
  \ln \left( - \, \frac{1 - \tx_1}{1 - \tx_2} \right) 
  + 
  \ln \left( - \, \frac{\tx_2}{\tx_1} \right) 
 \right] 
\right. 
\notag \\
&\qquad \quad {}
+ \ln \left( - \, \frac{1 - \tx_1}{1- \tx_2} \right) \, 
\left[ 
 \; \ln 
 \left( 
  \frac{(2 - \tx_1 - \tx_2)^2}{4} \, \detgj{1} + i \, \lambda \, \sigma_s
 \right) 
\right. 
\notag \\
&\qquad \qquad \quad \quad \quad \quad \quad \quad \quad {}
\left.
 + 
 \frac{1}{2}
 \left( 
  \ln \left( \frac{4 \, (1-\tx_1) \, (1-\tx_2)}{(2-\tx_1-\tx_2)^2} \right) 
  + 
  \ln \left( - \, \frac{4 \, (\tx_1 - \tx_2)^2}{(2 - \tx_1 - \tx_2)^2} \right) 
 \right) 
\right] 
\notag \\ 
&\qquad \quad {} 
+ 
\ln \left( - \, \frac{\tx_2}{\tx_1} \right) \, 
\left[ 
 \ln 
 \left( 
  \frac{(\tx_1 + \tx_2)^2}{4} \, \detgj{1} + i \, \lambda \, \sigma_s
 \right) 
\right. 
\notag \\
&\qquad \qquad \quad \quad \quad \quad \quad \quad {}
 \left. 
  + \frac{1}{2} \, 
  \left( \ln \left( \frac{4 \, \tx_1 \, \tx_2}{(\tx_1+\tx_2)^2} \right) 
  + \ln \left( - \frac{4 \, (\tx_1 - \tx_2)^2}{(\tx_1 + \tx_2)^2} \right) 
 \right) 
\right] 
\notag \\ 
&\qquad \quad {} +
\left. 
\dilog \left( \frac{1-\tx_2}{\tx_1 - \tx_2} \right) 
- \dilog \left( \frac{\tx_1 - 1}{\tx_1 - \tx_2} \right) 
 + \dilog \left( \frac{\tx_1}{\tx_1 - \tx_2} \right) 
 - \dilog \left( - \, \frac{\tx_2}{\tx_1 - \tx_2} \right) 
\right\}
\label{eq_verif_ir3}
\end{align}
As $\Im(\tx_1)$ and $\Im(\tx_2)$ have
opposite signs, eq. (\ref{eq_verif_ir3}) can be rearranged as:
\begin{align}
I_3^n 
&= 
- \, 
\frac{2^{\varepsilon} \, \Gamma(1 + \varepsilon)}{(\tx_1 - \tx_2) \, \detgj{1}} 
\notag\\
& \quad {} \times
\left\{
 - \, \frac{1}{\varepsilon} \, 
 \left[ 
  \ln \left( \frac{\tx_1 - 1}{\tx_1} \right) 
  - 
  \ln \left( \frac{\tx_2 - 1}{\tx_2} \right) 
 \right] 
\right. 
\notag \\
&\qquad \quad {} 
 + 
 \left[ 
  \ln \left( \detgj{1} + i \, \lambda \, \sigma_s \right) 
  + 
  \frac{1}{2} \, \ln \left( - \, (\tx_1 - \tx_2)^2 \right) 
 \right] \, 
 \left[ 
  \ln \left( \frac{\tx_1 - 1}{\tx_1} \right) 
  - 
  \ln \left( \frac{\tx_2 - 1}{\tx_2} \right) 
 \right] 
\notag \\
&\qquad \quad {} 
 + \frac{1}{2} \, \ln \left( - \, \frac{1 - \tx_1}{1 - \tx_2} \right) \, 
 \ln \left( (1-\tx_1) \, (1-\tx_2) \right) 
 + \frac{1}{2} \, \ln \left( - \, \frac{\tx_2}{\tx_1} \right) \, 
 \ln ( \tx_1 \, \tx_2 ) 
\notag \\
&\qquad \quad {} 
 - \frac{1}{2} \, \ln^2 \left( \frac{\tx_1 - 1}{\tx_1 - \tx_2} \right) 
 + \frac{1}{2} \, \ln^2 \left( \frac{1 - \tx_2}{\tx_1 - \tx_2} \right) 
 - \frac{1}{2} \, \ln^2 \left( \frac{- \, \tx_2}{\tx_1 - \tx_2} \right) 
 + \frac{1}{2} \, \ln^2 \left( \frac{\tx_1}{\tx_1 - \tx_2} \right) 
\notag \\
&\qquad \quad {} + 
\left. 
 2 \, \dilog \left( \frac{\tx_1 - 1}{\tx_2 - 1} \right) 
 + \frac{1}{2} \, \ln^2 \left( \frac{1 - \tx_2}{\tx_1 - 1} \right) 
 - 2 \, \dilog \left( \frac{\tx_1}{\tx_2} \right) 
 - \frac{1}{2} \, \ln^2 \left( - \,\frac{\tx_2}{\tx_1} \right) \right\}
  \label{eq_verif_ir4}
\end{align}
Using the relation
between $\ln(z)$ and $\ln(-z)$, after some algebra the quantity
\begin{align*}
E &= \;\;\; 
\frac{1}{2} \, \ln \left( - \, \frac{1 - \tx_1}{1 - \tx_2} \right) \, 
\ln \left( (1-\tx_1) \, (1-\tx_2) \right) 
+ \frac{1}{2} \, \ln \left( - \, \frac{\tx_2}{\tx_1} \right) \, 
\ln ( \tx_1 \, \tx_2 ) 
\notag \\
&\quad 
- \frac{1}{2} \, \ln^2 \left( \frac{\tx_1 - 1}{\tx_1 - \tx_2} \right)
+ \frac{1}{2} \, \ln^2 \left( \frac{1 - \tx_2}{\tx_1 - \tx_2} \right) 
- \frac{1}{2} \, \ln^2 \left( \frac{- \, \tx_2}{\tx_1 - \tx_2} \right) 
+ \frac{1}{2} \, \ln^2 \left( \frac{\tx_1}{\tx_1 - \tx_2} \right) 
\notag 
\end{align*}
can be rewritten: 
\begin{align}
E 
&= 
\frac{1}{2} \, 
\left[ 2 \, \ln (\tx_1 - \tx_2) - i \, \pi \, S(\tx_1) \right] \,
\left[ 
  \ln \left( \frac{\tx_1 - 1}{\tx_1} \right) 
  - 
  \ln \left( \frac{\tx_2 - 1}{\tx_2} \right) 
\right] 
\notag \\
&= 
\frac{1}{2} \, \ln \left( - \, (\tx_1 - \tx_2)^2 \right) \, 
\left[ 
  \ln \left( \frac{\tx_1 - 1}{\tx_1} \right) 
  - 
  \ln \left( \frac{\tx_2 - 1}{\tx_2} \right) 
\right]
  \label{eqaux_vere1}
\end{align}
with $S(\tx_1) = \sign(\Im(\tx_1))$. Substituting eq. (\ref{eqaux_vere1}) into 
eq. (\ref{eq_verif_ir4}), we end up with:
\begin{align}
I_3^n 
&= 
- 
\frac{2^{\varepsilon} \, \Gamma(1 + \varepsilon)}
{(\tx_1 - \tx_2) \, \detgj{1}} \, 
\notag \\
&\quad {} \times
\left\{ 
 - \frac{1}{\varepsilon} \, 
 \left[ 
  \ln \left( \frac{\tx_1 - 1}{\tx_1} \right) 
  - 
  \ln \left( \frac{\tx_2 - 1}{\tx_2} \right) 
 \right] 
\right. 
\notag \\
&\qquad \quad {} 
+ 
\left[ 
 \ln \left( \detgj{1} + i \, \lambda \, \sigma_s \right) + 
 \ln \left( - (\tx_1 - \tx_2)^2 \right) 
\right] \, 
\left[
  \ln \left( \frac{\tx_1 - 1}{\tx_1} \right) 
 - 
 \ln \left( \frac{\tx_2 - 1}{\tx_2} \right) 
\right]
\notag \\
&\qquad \quad {} 
+ 2 \, \dilog \left( \frac{\tx_1 - 1}{\tx_2 - 1} \right) 
+ \frac{1}{2} \, \ln^2 \left( \frac{1 - \tx_2}{\tx_1 - 1} \right) 
- 
\left. 
 2 \, \dilog \left( \frac{\tx_1}{\tx_2} \right) - 
 \frac{1}{2} \, \ln^2 \left( -\frac{\tx_2}{\tx_1} \right) 
\right\} 
\notag \\
 &= 
- \frac{2^{\varepsilon} \, \Gamma(1 + \varepsilon)}{\varepsilon \, \detgj{1}} 
\, 
\left\{ 
 \left[ 1 - \varepsilon \, \ln \left( \detgj{1} + i \, \lambda \, \sigma_s \right) \right]
 \, K(\tx_1,\tx_2) - \varepsilon \, J(\tx_1,\tx_2) 
\right\} 
\label{eq_verif_ir5}
\end{align}
where $K(\tx_1,\tx_2)$ is given by eq.~(\ref{eqcompk11}) and $J(\tx_1,\tx_2)$
 by eq.~(\ref{eqcompj5}).
Last we note that the prescription  ``$+ i \, \lambda \, \sigma_s$" in 
eq.(\ref{eq_verif_ir5}) can be replaced by ``$- i \, \lambda$" as it matters
only when $\detgj{1} < 0$.
Thus eq. (\ref{eq_verif_ir5}) is nothing but eq. (\ref{eqcompintlog1}):
the indirect and direct ways lead to the same result indeed.  

\vspace{0.3cm}

\noindent
{\bf 2. Occurrence of a collinear divergence }\\
\noindent
We recap the texture of the $\cals$ matrix in this case (cf.\ eq.~(\ref{eqcalscoll})):
\begin{equation}
  \cals =
  \left(
  \begin{array}{ccc}
    0 & 0     & s_1 - m_3^2 \\
    0 & 0 & s_3 - m_3^2 \\
    s_1 - m_3^2 & s_3 - m_3^2 & - 2 \, m_3^2
  \end{array}
  \right)
  \label{eqcalscoll_ver}
\end{equation}
Obviously, we have $\det(\cals) = 0$. As explained in section
\ref{3point_ir}, (eqs. (\ref{eqdefg3coll}) and (\ref{eqdefv3coll})), the
coefficient $\bbar_3$ vanishes whereas $\bbar_2$ and $\bbar_1$ are different
from zero. So the three-point function is given by:
\begin{align}
I_3^n 
&= \;\;\;
\frac{\bbar_1}{\detg} \, 
\left[ 
 \frac{\bbj{2}{1}}{\detgj{1}} \, 
 L_{3}^{n} \left( 0,\Delta_{1}^{\{1\}},\tD_{12} \right) +  
 \frac{\bbj{3}{1}}{\detgj{1}} \, 
 L_{3}^{n} \left( 0,\Delta_{1}^{\{1\}},\tD_{13} \right)
\right] 
\nonumber \\
&\quad 
+ \frac{\bbar_2}{\detg} \, 
\left[ 
 \frac{\bbj{1}{2}}{\detgj{2}} \, 
 L_{3}^{n} \left( 0,\Delta_{1}^{\{2\}},\tD_{21} \right) +  
 \frac{\bbj{3}{2}}{\detgj{2}} \, 
 L_{3}^{n} \left( 0,\Delta_{1}^{\{2\}},\tD_{23} \right)
\right]
  \label{eq_verif_irc1}
\end{align}
As $\tD_{12} = \tD_{21} = 2 \, m_3^2$ and $\tD_{13} = \tD_{23} = 0$, 
$L_{3}^{n} ( 0,\Delta_{1}^{\{1\}},\tD_{12} )$ and 
$L_{3}^{n} ( 0,\Delta_{1}^{\{2\}},\tD_{12} )$
are given by eq. (\ref{eqlijsoft41}) whereas 
$L_{3}^{n} ( 0,\Delta_{1}^{\{1\}},\tD_{13} )$ and 
$L_{3}^{n} ( 0,\Delta_{1}^{\{2\}},\tD_{23} )$
are given by (\ref{eqlijsoft8}). 

\vspace{0.3cm}

\noindent
Let us first focus on the first term of the r.h.s. 
of eq. (\ref{eq_verif_irc1}). The relevant reduced $\cals$ matrix is:
\begin{equation}
  \cals^{\{1\}} =
  \left(
  \begin{array}{cc}
    0 & s_3 - m_3^2 \\
    s_3 - m_3^2 & - 2 \, m_3^2
  \end{array}
  \right)
  \label{eqcalscol_ver1}
\end{equation}
whose determinant is: $\detsj{1} = - (s_3- m_3^2)^2$. The $1 \times 1$ 
associated Gram matrix is: $G^{\{1\}(2)} = ( 2 \, s_3)$ and the reduced 
$\bar{b}$ coefficients are given by:
\begin{align}
\frac{\bbj{2}{1}}{\detgj{1}} 
= - \, \frac{s_3 + m_3^2}{2 \, s_3}
& \quad {} , \quad {} 
\frac{\bbj{3}{1}}{\detgj{1}} 
= - \, \frac{s_3 - m_3^2}{2 \, s_3} \label{eqbbar_vercol2}
\end{align}
whereas:
\begin{align}
\Delta_1^{\{1\}} 
= \frac{(s_3 - m_3^2)^2}{2 \, s_3} 
& \quad {} , \quad {}
\tD_{12} + \Delta_1^{\{1\}} = \frac{(s_3 + m_3^2)^2}{2 \, s_3} 
 \label{eqdtildepdelta1_vercol}
\end{align}
The square of the root of the polynomial 
$(\tD_{12} + \Delta_1^{\{1\}}) \, z^2 - \Delta_1^{\{1\}} - i \, \lambda$ is:
\begin{equation}
\bar{z}^2 = \frac{(s_3 - m_3^2)^2}{(s_3 + m_3^2)^2} + i \, \lambda \, \sigma_s
\label{eqsqroot_vercol}
\end{equation}
with $\sigma_s = \sign(s_3)$. Whereas
\begin{equation}
\sqrt{\bar{z}^2} 
= \left| \frac{s_3 - m_3^2}{s_3 + m_3^2} \right| + i \, \lambda \, \sigma_s
\nonumber
\end{equation}
a more handy choice for further manipulations is instead: 
\begin{equation}
\bar{z} =  
\frac{s_3 - m_3^2}{s_3 + m_3^2} + i \, \lambda \, \sigma_s \, \sigma_r
\label{eqroot_vercol2}
\end{equation}
where $\sigma_r = \sign( (s_3 - m_3^2)/(s_3 + m_3^2) )$. Let us note:
\begin{align}
\Sigma_3^n(s_3) 
&= \frac{\bbj{2}{1}}{\detgj{1}} \, 
L_{3}^{n} \left( 0,\Delta_{1}^{\{1\}},\tD_{12} \right)  
+  \frac{\bbj{3}{1}}{\detgj{1}} \, 
L_{3}^{n} \left( 0,\Delta_{1}^{\{1\}},\tD_{13} \right)  
\label{eq_verif_irc20}
\end{align}
we get:
\begin{align}
\Sigma_3^n(s_3) 
&= \frac{2^{\varepsilon} \, \Gamma(1 + \varepsilon)}{2 \, (m_3^2 - s_3)} 
\notag\\
&\quad {} \times
\left[ 
 - \, \frac{1}{\varepsilon} \, 
 \ln \left( -\frac{m_3^2}{s_3} + i \, \lambda \, \sigma_s \, \sigma_r \right) 
 + \bar{H}_{0,1} 
 \left( \tD_{12} + \Delta_1^{\{1\}},-\Delta_1^{\{1\}} - i \, \lambda \right) 
\right. 
\notag \\
&\qquad \quad {} - 
\left. 
 \frac{1}{\varepsilon^2} \, 
 \frac{\Gamma(1 - \varepsilon)^2}{\Gamma(1 - 2 \, \varepsilon)} \, 
 \left( - \frac{2 \, (s_3 - m_3^2)^2}{s_3} - i \, \lambda \right)^{-\varepsilon} 
\right]
\label{eq_verif_irc2}
\end{align}
Using the definition of $\bar{H}_{0,1}(X,Y)$ (cf.\ eq.~(\ref{eqdefhbarij})) and eq. (\ref{eqdefh05}), we have:
\begin{align}
& \bar{H}_{0,1}(\tD_{12} + \Delta_1^{\{1\}},-\Delta_1^{\{1\}} - i \, \lambda) 
\notag\\
&= 
\dilog 
 \left( \frac{s_3}{s_3 - m_3^2} - i \, \lambda \sigma_s \, \sigma_r \right) 
 - \dilog 
 \left( 
  - \, \frac{m_3^2}{s_3 - m_3^2} + i \, \lambda \, \sigma_s \, \sigma_r 
 \right)
\notag\\
&\quad {}
+
\ln 
\left( - \frac{m_3^2}{s_3} + i \, \lambda \, \sigma_s \, \sigma_r \right) 
\left[ 
 \ln 
 \left(  \frac{(s_3 + m_3^2)^2}{2 \, s_3} + i \, \lambda \, \sigma_s \right) 
 + \frac{1}{2} \, 
  \ln 
  \left( 
   \frac{4 \, s_3 \, m_3^2}{(s_3 + m_3^2)^2} - i \, \lambda \sigma_s 
  \right) 
\right. 
\notag \\
&\qquad \qquad \qquad \qquad \quad \quad \quad \quad {} + 
\left. 
  \frac{1}{2} \, 
  \ln 
  \left( 
   -\, \frac{4 \, (s_3 - m_3^2)^2}{(s_3 + m_3^2)^2} - i \, \lambda \, \sigma_s 
  \right) 
\right] 
\label{eqresubarh_vercol1}
\end{align}
which can be rewritten as:
\begin{align}
&\bar{H}_{0,1}
\left( \tD_{12} + \Delta_1^{\{1\}},-\Delta_1^{\{1\}} - i \, \lambda \right) 
\notag\\
&= 
\ln \left( - \frac{m_3^2}{s_3} + i \, \lambda \, \sigma_s \, \sigma_r \right) 
\, 
\left[ 
 -\ln \left(  2 \, s_3 - i \, \lambda \, \sigma_s \right)  
 + \frac{1}{2} \, 
  \ln \left( 4 \, s_3 \, m_3^2 - i \, \lambda \sigma_s \right) 
\right.
\notag \\
&\quad \quad \quad \quad \quad \quad \quad \quad
\quad \quad \quad \quad {} +
\left. 
  \frac{1}{2} \, 
 \ln \left( - \, 4 \, (s_3 - m_3^2)^2 - i \, \lambda \, \sigma_s \right)  
\right] 
\notag \\
&\quad {}
+ 2 \, 
\dilog 
\left( \frac{s_3}{s_3 - m_3^2} - i \, \lambda \sigma_s \, \sigma_r \right) 
- \frac{\pi^2}{6} 
\notag \\
&\quad {} 
+ 
\ln 
\left( 
 - \, \frac{m_3^2}{s_3 - m_3^2} + i \, \lambda \, \sigma_s \, \sigma_r 
\right) \, 
\ln \left( \frac{s_3}{s_3 - m_3^2} - i \, \lambda \sigma_s \, \sigma_r \right)
\label{eqresubarh_vercol2}
\end{align}
Putting eq. (\ref{eqresubarh_vercol2}) into eq. (\ref{eq_verif_irc2}), the
$\ln(2)$ drop out and we end with:
\begin{align}
\Sigma_3^n(s_3) 
&= \frac{\Gamma(1 + \varepsilon)}{2 \, (m_3^2 - s_3)} \, 
\notag\\
&\quad {} \times
\left\{ 
 - \, \frac{1}{\varepsilon^2} + \frac{1}{\varepsilon} \, 
 \left[ 
  \ln \left( - \frac{(s_3 - m_3^2)^2}{s_3} - i \, \lambda \right) 
  - 
  \ln\left( -\frac{m_3^2}{s_3} + i \, \lambda \, \sigma_s \, \sigma_r \right) 
 \right] 
\right. 
\notag \\
&\qquad \quad {}
 - \frac{1}{2} \, 
 \ln^2 \left( - \frac{(s_3 - m_3^2)^2}{s_3} - i \, \lambda \right) 
 + 2 \, 
 \dilog 
 \left( \frac{s_3}{s_3 - m_3^2} - i \, \lambda \sigma_s \, \sigma_r \right) 
\notag \\
&\qquad \quad {}
 + \ln \left( - \frac{m_3^2}{s_3} + i \, \lambda \, \sigma_s \, \sigma_r \right) 
 \, 
\left[ 
 \vphantom{\frac{1}{2}}-\ln \left(  s_3 - i \, \lambda \, \sigma_s \right) 
\right. 
\notag \\
&\qquad \quad {}
 \left. 
  + \frac{1}{2} \, 
   \ln \left( s_3 \, m_3^2 - i \, \lambda \sigma_s \right) 
  +  \frac{1}{2} \,
   \ln \left( - (s_3 - m_3^2)^2 - i \, \lambda \, \sigma_s \right) 
 \right] 
\notag \\
&\qquad \quad {} +
\left. 
  \ln 
 \left( 
  - \, \frac{m_3^2}{s_3 - m_3^2} + i \, \lambda \, \sigma_s \, \sigma_r 
 \right) \, 
 \ln \left( \frac{s_3}{s_3 - m_3^2} - i \, \lambda \sigma_s \, \sigma_r \right) 
\right\}
\label{eq_verif_irc3}
\end{align}
$\Sigma_3^n(s_3)$ can be shown to be equal to the following quantity:
\begin{align}
\Upsilon_3^n(s_3) 
&= \frac{\Gamma(1 + \varepsilon)}{2 \, (m_3^2 - s_3)} \, 
\left\{
 - \, \frac{1}{\varepsilon^2} + \frac{1}{\varepsilon} \, 
 \left[ 
  2 \, \ln \left( - s_3 + m_3^2 - i \, \lambda \right) 
  - 
  \ln \left( m_3^2 - i \, \lambda \right) 
 \right] 
\right. 
\notag \\
&\quad 
\left.
 - \ln^2 \left( - s_3 + m_3^2 - i \, \lambda \right) 
 + \frac{1}{2} \, \ln^2 \left( m_3^2 - i \, \lambda \right) 
 + 2 \, \dilog \left( \frac{s_3}{s_3 - m_3^2 + i \, \lambda} \right) 
\right\}
\label{eq_verif_irc4}
\end{align}
To show that, 
one has to distinguish the following cases: 1) $s_3 < - \, m_3^2$, 
2) $- \, m_3^2 < s_3 < 0$, 3) $0 < s_3 < m_3^2$ and 4) $m_3^2 < s_3$. 
For each of them, some tedious algebra performed on the r.h.s. of eqs. 
(\ref{eq_verif_irc3}) shows that the two results are 
equal.
Let us discuss in detail only the case $s_3 > m_3^2$, for which 
$\sigma_s = \sigma_r = +$. The argument of the dilogarithm in eq. 
(\ref{eq_verif_irc3}) has a real part greater than $1$, hence:
\begin{equation}
\dilog 
\left( \frac{s_3}{s_3 - m_3^2} - i \, \lambda \sigma_s \, \sigma_r \right)
= 
\dilog 
\left( \frac{s_3}{s_3 - m_3^2} - i \, \lambda \right) 
= 
\dilog \left( \frac{s_3}{s_3 - m_3^2 + i \, \lambda} \right)
\nonumber
\end{equation}
The logarithms in eq. (\ref{eq_verif_irc3}) can be modified in such a way 
that their arguments are ratios of positive quantities, $\Sigma_3^n(s_3)$ 
thus reads:
\begin{align}
\Sigma_3^n(s_3) 
&= \frac{\Gamma(1 + \varepsilon)}{2 \, (m_3^2 - s_3)} \, 
\notag\\
& \quad {}
\left\{
 - \, \frac{1}{\varepsilon^2} + \frac{1}{\varepsilon} \, 
 \left[ 
  \ln \left( \frac{(s_3 - m_3^2)^2}{s_3} \right) 
  - 
  \ln \left( \frac{m_3^2}{s_3} \right) - 2 \, i \, \pi 
 \right] 
\right. 
\notag \\
&\quad 
 - \frac{1}{2} \, 
 \left( \ln \left( \frac{(s_3 - m_3^2)^2}{s_3} \right) - i \, \pi \right)^2 
 + 2 \, \dilog \left( \frac{s_3}{s_3 - m_3^2 + i \, \lambda} \right) 
\notag \\
&\quad  
 + \left( \ln \left( \frac{m_3^2}{s_3} \right) + i \, \pi \right) \, 
\notag \\
&\quad \quad 
 \times 
 \left[ 
  \vphantom{\frac{1}{2}} -\ln \left(  s_3 \right) 
  + \frac{1}{2} \, 
  \left( 
   \ln \left( s_3 \, m_3^2  \right) + \ln \left( (s_3 - m_3^2)^2 \right) 
   - i \, \pi 
  \right) 
 \right] 
\notag \\
&\quad 
\left.
 + 
 \left( \ln \left(  \frac{m_3^2}{s_3 - m_3^2} \right) + i \, \pi \right) \, 
 \ln \left( \frac{s_3}{s_3 - m_3^2} \right)
\right\}
\label{eq_verif_irc5}
\end{align}
Splitting logarithms of ratios, we get:
\begin{align}
\Sigma_3^n(s_3)
&= 
\frac{\Gamma(1 + \varepsilon)}{2 \, (m_3^2 - s_3)} \, 
\left\{
 - \, \frac{1}{\varepsilon^2} + \frac{1}{\varepsilon} \, 
 \left[ 
  2 \, \left( \ln \left( s_3 - m_3^2 \right) - i \, \pi \right) 
  - \ln\left( m_3^2 \right)  
 \right] 
\right. 
\notag \\
&\quad {}\quad {}\quad {}\quad {}\quad {}\quad {}\quad {}
 -  
 \left( 
  \ln^2 \left( s_3 - m_3^2 \right)
  - 2 \, i \, \pi \, \ln \left( s_3 - m_3^2  \right)  - \pi^2 
 \right)
\notag \\
&\quad{} \quad {}\quad {}\quad {}\quad {}\quad {}\quad {}
\left.
 + \frac{1}{2} \, \ln^2(m_3^2) 
 + 2 \, \dilog \left( \frac{s_3}{s_3 - m_3^2 + i \, \lambda} \right)
\right\}
\label{eq_verif_irc6}
\end{align}
In eq.~(\ref{eq_verif_irc6}), for the case at hand, we recognise eq. (\ref{eq_verif_irc4}).
Similar handling can be performed in the other three cases so as to reach the
same conclusion. We thus conclude:
\begin{align}
\Sigma_3^n(s_3)
& = \Upsilon_3(s_3)
\notag \\
&= 
\frac{\Gamma(1 + \varepsilon)}{(m_3^2 - s_3)} \, 
\left\{ 
 - \, \frac{1}{\varepsilon^2} \, 
 \left( -s_3 + m_3^2 - i \, \lambda \right)^{-\varepsilon} 
 + \frac{1}{2 \, \varepsilon^2} \, 
 \left( m_3^2 - i \, \lambda \right)^{-\varepsilon} 
\right. 
\notag \\
&\quad {} \quad {} \quad {} \quad {} \quad {} \quad {}  \quad {}  \quad {}
\left. 
 + \dilog \left( \frac{s_3}{s_3 - m_3^2 + i \, \lambda} \right) 
\right\}
  \label{eq_verif_irc7}
\end{align}
So long for the first term of eq. (\ref{eq_verif_irc1}).
The second term can be obtained from the first one by replacing $s_3$ by $s_1$.
The coefficients $\bbar_1$ and $\bbar_2$ as well as $\detg$ are easily extracted 
from the $\cals$ matrix cf. eq. (\ref{eqcalscoll_ver}):
\begin{align}
\bbar_1 \; = \; (s_3 - m_3^2) \, (s_1 - s_3) &, \quad {}
\bbar_2 \; = \; (s_1 - m_3^2) \, (s_3 - s_1) 
\notag\\
\detg &= - (s_1 - s_3)^2
\label{eqdefbb1bb2detg}
  \end{align}
Thus we finally get:
\begin{align}
I_3^n 
&= \frac{\bbar_1}{\detg} \, \Sigma_3^n(s_3) + \frac{\bbar_2}{\detg} \, 
\Sigma_3^n(s_1) 
\notag \\
&= \frac{\Gamma(1 + \varepsilon)}{ (s_1 - s_3)} \, 
\left\{ 
 - \frac{1}{\varepsilon^2} \, 
 \left[ 
  \left( -s_3 + m_3^2 - i \, \lambda \right)^{-\varepsilon} 
  - \left( -s_1 + m_3^2 - i \, \lambda \right)^{-\varepsilon} 
 \right] 
\right. 
\notag \\
 &\quad {} \quad {} \quad {}  \quad {}  \quad {}  \quad {} \quad {} 
 + \left. 
 \dilog \left( \frac{s_3}{s_3 - m_3^2 + i \, \lambda} \right) 
 - \dilog \left( \frac{s_1}{s_1 - m_3^2 + i \, \lambda} \right) 
\right\}
 \label{eq_verif_irc8}
\notag
\end{align}
which coincides with eq. (\ref{eqdirei3n8}):
the direct and indirect ways lead to the same result.

\vspace{0.3cm}

\noindent
{\bf 3. Concomitant occurrence of a soft and a collinear divergences} \\
\noindent
Here again, we start to recap the texture of the $\cals$ matrix\footnote{Contrary to 
  the example $3$ in subsec.~\ref{exp_exemp_ir}, we choose to set $m_3^2 = 0$ because, 
for the purpose of this appendix, a non vanishing mass does not bring anything new with 
respect to the previous case.}:
\begin{equation}
  \cals =
  \left(
  \begin{array}{ccc}
    0 & 0     & 0 \\
    0 & 0 & s_3 \\
    0 & s_3 & 0
  \end{array}
  \right)
  \label{eqcalscollsof_ver}
\end{equation}
Obviously, $\det(\cals) = 0$. As in example ``1. Occurrence of a soft divergence'', the coefficients $\bbar_2$ 
and $\bbar_3$ vanish whereas $\bbar_1 = \detg$, and, as 
$\tD_{12} = \tD_{13} = 0$, $L(0,\Delta_1^{\{1\}},0)$ is given by
eq. (\ref{eqlijsoft8}), thus:
\begin{align}
 I_3^n &= 
 \frac{\bbj{2}{1}}{\detgj{1}} \, 
 L_{3}^{n} \left( 0,\Delta_{1}^{\{1\}},0 \right) 
 + 
 \frac{\bbj{3}{1}}{\detgj{1}} \, 
 L_{3}^{n} \left( 0,\Delta_{1}^{\{1\}},0 \right)
 \label{eq_verif_ircs1}
\end{align}
The relevant reduced $\cals$ matrix is:
\begin{equation}
  \cals^{\{1\}} =
  \left(
  \begin{array}{cc}
    0 & s_3  \\
    s_3 & 0
  \end{array}
  \right)
  \label{eqcalscols_ver1}
\end{equation}
whose determinant is $\detsj{1} = - s_3^2$. The $1 \times 1$ associated Gram
``matrix" is $G^{\{1\}(2)} = ( 2 \, s_3)$ and the reduced $\bar{b}$ 
coefficients are given by:
\begin{align}
\frac{\bbj{2}{1}}{\detgj{1}} 
&= 
\frac{\bbj{3}{1}}{\detgj{1}} 
\; = \; 
- \, \frac{1}{2} 
\label{eqbbar_vercols2}
\end{align}
whereas:
\begin{align}
  \Delta_1^{\{1\}} &= \frac{s_3}{2} \label{eqdelat1_vercols}
\end{align}
We thus obtain:
\begin{align}
I_3^n
&= - L_3^n\left( 0,\Delta_1^{\{1\}},0 \right) \notag \\
&= -\frac{1}{\varepsilon^2} \, \Gamma(1+\varepsilon) \, 
\frac{\Gamma^2(1-\varepsilon)}{\Gamma(1-2 \, \varepsilon)} \, 
\left( 
 - \,s_3 - i \, \lambda  
\right)^{-1-\varepsilon} 
\notag \\
&= W\left( \detgj{1}, 0, 0 \right)
\notag
\end{align}
(cf.\ eqs.~(\ref{eqdirei3n1}), (\ref{eqdefwi0}) and (\ref{eqdirei3n4})), so the direct and
indirect ways lead to the same results.

\subsubsection*{Complex mass case}

We now treat the complex mass case. As discussed in section
(\ref{3point_ir}), the only relevant case is the collinear case where the
non vanishing internal mass, say $m_3^2$, is complex\footnote{We follow the convention of appendix~\ref{P2-ImofdetS} of ref.\ \cite{paper2}}: 
$m_3^2 = m_R^2 + i \, m_I^2$ 
with $m_R^2$ and $m_I^2$ real and $m_R^2 > 0$, $m_I^2 < 0$\footnote{
  One could be tempted, in this subsec. to recover the real mass case results by setting 
  $m^2_I = - \lambda$. Doing that could lead to wrong formulae because, when deriving the 
  complex mass case, we have already assumed that $|m^2_I| \gg \lambda$ and so dropped some $i \, \lambda$ terms. 
}. The $\cals$ matrix is given
by eq. (\ref{eqcalscoll_ver}) and $I_3^n$ by eq. (\ref{eq_verif_irc1}).
As in the real mass case, let us focus on the first line of eq.~(\ref{eq_verif_irc1}), the second line can be obtained by changing $s_3$ in $s_1$.
To
compute $L(0,\Delta_1^{\{1\}},\tD_{12})$ and $L(0,\Delta_1^{\{1\}},\tD_{13})$, 
we have to determine which formulae to use, depending on the sign of
$\Im(\Delta_1^{\{1\}})$. The quantity $\Delta_1^{\{1\}}$ is given by eq. (\ref{eqdtildepdelta1_vercol}) which reads:
\begin{align}
\Delta_1^{\{1\}} 
&= \frac{1}{2 \, s_3} \, 
\left( (s_3 - m_R^2)^2 - m_I^4 - 2 \, i \, m_I^2 \, (s_3 - m_R^2) \right)
\label{eqdelta1_vercolc}
\end{align}
When $(s_3 - m_R^2)/s_3 > 0$ i.e. either $s_3 > m_R^2$ or $s_3 < 0$, 
$L(0,\Delta_1^{\{1\}},\tD_{12})$ 
is given by eq. (\ref{eqlijsoft41}) as in the real mass case, 
and when $(s_3 - m_R^2)/s_3 < 0$ i.e. $0 < s_3 < m_R^2$,
$L(0,\Delta_1^{\{1\}},\tD_{12})$ 
is given by eq. (\ref{eqlijsoft61}). 
As discussed previously, there is no such dichotomy for 
$L(0,\Delta_1^{\{1\}},\tD_{13})$ 
which is given by eq. (\ref{eqlijsoft8}).

\vspace{0.3cm}

\noindent
Let us first consider $s_3 > m_R^2$ or $s_3 < 0$. We define
$\bar{z} = (s_3 - m_3^2)/(s_3 + m_3^2)$ reminiscent of eq. 
(\ref{eqroot_vercol2}) and $\Sigma_3^n(s_3)$ is given by eq.
(\ref{eq_verif_irc3}), in which the infinitesimal imaginary parts 
$\propto \lambda$ are dropped out except for arguments of logarithms 
which do not depend on $m_3^2$, namely:
\begin{align}
\Sigma_3^n(s_3) 
&= \frac{\Gamma(1 + \varepsilon)}{2 \, (m_3^2 - s_3)} 
\left\{ 
 - \, \frac{1}{\varepsilon^2} + \frac{1}{\varepsilon} \, 
 \left[ 
  \ln \left( - \, \frac{(s_3 - m_3^2)^2}{s_3}  \right) 
  - 
  \ln \left( - \, \frac{m_3^2}{s_3} \right) 
 \right] 
\right. 
\notag \\
&\quad {} \quad {}  \quad {} \quad {} \quad {} \quad {}\quad {}
 - \frac{1}{2} \, \ln^2 \left( - \, \frac{(s_3 - m_3^2)^2}{s_3} \right) 
 + 2 \, \dilog \left( \frac{s_3}{s_3 - m_3^2} \right) 
\notag \\
&\quad {} \quad {}   \quad {} \quad {}  \quad {} \quad {}\quad {}
 + \ln \left( - \,  \frac{m_3^2}{s_3} \right) 
\notag \\
&\quad {} \quad {}   \quad {} \quad {}  \quad {} \quad {} \quad {} 
 \times \left[ 
  \vphantom{\frac{1}{2}}- \ln \left(  s_3 - i \, \lambda \, \sigma_s \right) 
  + \frac{1}{2} \, 
  \left[ 
  \ln \left( s_3 \, m_3^2 \right) + \ln \left( - \, (s_3 - m_3^2)^2  \right) 
  \right] 
 \right] 
\notag \\
 &\quad {} \quad {}  \quad {} \quad {} \quad {} \quad {}\quad {}
\left. 
 + \ln \left(  - \, \frac{m_3^2}{s_3 - m_3^2} \right) \, 
 \ln \left( \frac{s_3}{s_3 - m_3^2} \right) 
\right\}
\label{eq_verif_ircc3}
\end{align}
Logarithms of products and ratios in eq. (\ref{eq_verif_ircc3}) are further 
split. For this purpose we use eq. (\ref{eqdeflnzlnmz}) as well as:
\[
\ln \left( (s_3 - m_3^2)^2 \right) 
= 
2 \, \ln(s_3 - m_3^2) 
- 2 \, i \, \pi \, \theta \left( - s_3 + m_R^2  \right)
\]
and, for any real $a$ and complex $b$:
\begin{align}
\ln \left( \frac{b}{a} \right) 
&= - \ln \left( a + i \, \lambda \, S(b) \right) + \ln(b) 
\notag\\
\ln(a \, b) 
&= \;\;\;\, \ln \left( a - i \, \lambda \, S(b) \right) + \ln(b)
\notag
\end{align}
We also use the notation $S(z) = \sign\left( \Im(z) \right)$. Let us note that 
$S\left( (s_3 - m_3^2)^2 \right) = \sign(s_3 - m_R^2) \equiv \sigma_p$ and that
$\ln(  s_3 - i \, \lambda \, \sigma_s)$ is equivalent to 
$\ln(  s_3 + i \, \lambda)$. 
Then, to compactify eq.~(\ref{eq_verif_ircc3}), we take advantage of the following relations:
\begin{align}
  \ln\left( - s_3 \pm i \, \lambda \right) &= \ln\left( |s_3| \right) \pm i \, \pi \, \theta(s_3) \notag \\
  \ln\left( s_3 \pm i \, \lambda \right) &= \ln\left( |s_3| \right) \pm i \, \pi \, \theta(- s_3) \notag \\
  \theta(\pm s_3) &= \frac{(1 \pm \sigma_s)}{2} \notag \\
  \theta(- s_3 + m_R^2) &= \frac{1- \sigma_p}{2}
  \label{eqnewrelovers3}
\end{align}
Eq. (\ref{eq_verif_ircc3}) can be rewritten:
\begin{align}
\Sigma_3^n(s_3) 
&= \frac{\Gamma(1 + \varepsilon)}{2 \, (m_3^2 - s_3)} \, 
\left\{ 
 - \frac{1}{\varepsilon^2} + \frac{1}{\varepsilon} \, 
 \left[ 
  2 \, \left( \ln \left( s_3 - m_3^2 \right) - i \, \pi \right) 
  - \ln \left( m_3^2 \right) 
 \right] 
\right.
\notag \\
 &\qquad \qquad \qquad \quad {}
 - \frac{1}{2} \, 
 \left[ 
  2 \, \ln \left( s_3 - m_3^2 \right) 
  - i \, \pi \, \left( 1 - \frac{\sigma_p}{2} + \frac{\sigma_p \, \sigma_s}{2} \right)
\right]^2 
\notag \\
 &\qquad \qquad \qquad \quad {}
 + 2 \, \dilog \left( \frac{s_3}{s_3 - m_3^2} \right) 
 + \left( \ln \left( m_3^2 \right) + i \, \pi \, \frac{1+\sigma_s}{2}
\right)
\notag \\
 &\qquad \qquad \qquad \quad {}
 \times
\left[ 
 \frac{1}{2} \, \ln \left( m_3^2 \right) 
 + \ln \left( s_3 - m_3^2 \right) - i \, \pi \, (3 - \sigma_s)
\right] 
\notag \\
 &\qquad \qquad \qquad \quad {}
 + 
 \left( 
  \ln \left( m_3^2 \right) - \ln \left( s_3 - m_3^2 \right) + i \, \pi 
 \right) \, 
 \left( 
 i \, \pi \, \frac{1-\sigma_s}{2} - \ln \left( s_3 - m_3^2 \right) 
 \right) \notag \\
 &\qquad \qquad \qquad \quad {}
 + \left. i \, \frac{\pi}{2} \, \ln\left( |s_3| \right) \, (1-\sigma_s) \, (1+\sigma_p)
 \vphantom{\frac{1}{\varepsilon^2}}
\right\}
\label{eq_verif_ircc4}
\end{align}
Let us treat the case where $s_3 > m_R^2$. We have $\sigma_p = +$ and 
$\sigma_s=+$. By expanding eq. (\ref{eq_verif_ircc4}), we get:
\begin{align}
\Sigma_3^n(s_3) 
&= \frac{\Gamma(1 + \varepsilon)}{2 \, (m_3^2 - s_3)} \, 
\left\{ 
 - \frac{1}{\varepsilon^2} + \frac{1}{\varepsilon} \, 
 \left[ 
  2 \, \left( \ln \left( s_3 - m_3^2 \right) -  i \, \pi  \right)
  - \ln\left( m_3^2\right) 
 \right] 
\right. 
\notag \\
&\quad {} \quad {}\quad {} \quad {} \quad {}\quad {} \quad {}
 -  
 \left( 
  \ln^2 \left( s_3 - m_3^2 \right)
  - 2 \, i \, \pi \, \ln \left( s_3 - m^2 \right)  - \pi^2 
 \right)
\notag \\
&\quad {} \quad {}\quad {} \quad {} \quad {}\quad {} \quad {}
\left. 
 + \frac{1}{2} \, \ln \left( m_3^2 \right) 
 + 2 \, \dilog \left( \frac{s_3}{s_3 - m_3^2} \right) 
\right\}
\label{eq_verif_ircc5}
\end{align}
As $\ln(s_3 - m_3^2) - i \, \pi = \ln(-s_3 + m_3^2)$, we recover 
$\Upsilon_3^n(s_3)$ given by eq. (\ref{eq_verif_irc4}). 
The same exercise can be easily done for the case $s_3 < 0$ leading to 
the same conclusion.

\vspace{0.3cm}

\noindent
In the case where $0 < s_3 < m_R^2$, 
$L(0,\Delta_1^{\{1\}},\tD_{12})$
is given by eq. (\ref{eqlijsoft61}) i.e.:
\begin{align}
L(0,\Delta_1^{\{1\}},\tD_{12})
&= 
\frac{2^{\varepsilon} \, \Gamma(1+\varepsilon)}
{2 \, (\tD_{12} + \Delta_1^{\{1\}}) \, \bar{z}} \, 
\left[ 
 \frac{1}{\varepsilon} \, 
 \left( 
  i \, \pi \, S(i \, \bar{z} )
  + \ln \left( \frac{1+\bar{z}}{1-\bar{z}} \right) 
 \right) 
\right. 
\notag \\
&\qquad \qquad \qquad \qquad \quad {}
 - \bar{H}_{0,\infty}(- \tD_{12} - \Delta_1^{\{1\}}, - \Delta_1^{\{1\}}) 
\notag \\
&\qquad \qquad \qquad \qquad \quad {}
 - \left. 
 \bar{H}_{1,\infty}(\tD_{12} + \Delta_1^{\{1\}}, - \Delta_1^{\{1\}}) 
\vphantom{\frac{1}{\varepsilon}} \right]
\label{eqdefnewL12}
\end{align}
where $\bar{z} = \sqrt{\Delta_1^{\{1\}}/(\tD_{12}+\Delta_1^{\{1\}})}$. We have chosen for the root of the equation 
$(\tD_{12} + \Delta_1^{\{1\}})\, z^2 + \Delta_1^{\{1\}} = 0$, 
$\tilde{z} = i \, \bar{z}$. The functions $\bar{H}_{1,\infty}(x,y)$ 
is given by the expression in curly brackets of eq.
(\ref{eqdefh27}) and $\bar{H}_{0,\infty}(x,y)$ by the expression in square
brackets in eq. (\ref{eqdefh23}) of appendix \ref{appF} i.e.:
\begin{align}
\bar{H}_{0,\infty}(- \tD_{12} - \Delta_1^{\{1\}}, - \Delta_1^{\{1\}}) 
&= i \, \pi \, S(i \, \bar{z}) \, 
\left[ 
 2 \, \ln \left( 2 \, i \, \bar{z} \right) 
 + \ln \left( - \tD_{12} - \Delta_1^{\{1\}} \right) 
\right. 
\notag \\
&\qquad \qquad \qquad {}
 + \left.
 \eta \left( -\tD_{12} - \Delta_1^{\{1\}}, \bar{z}^2 \right) 
\right] 
+ \pi^2 
\label{eqbarh0inf1} \\
\bar{H}_{1,\infty}(\tD_{12} + \Delta_1^{\{1\}}, - \Delta_1^{\{1\}}) 
&= 
\ln \left( \frac{1 + \bar{z}}{1 - \bar{z}} \right) \, 
\left[ \vphantom{\frac{\tD_{12}}{\tD_{12} + \Delta_1^{\{1\}}}} 
 \ln \left( \tD_{12} + \Delta_1^{\{1\}} \right)  
 + \frac{1}{2} \, \ln \left( 1 - \bar{z}^2 \right) 
 + \ln \left( 2 \, \bar{z} \right) 
\right. 
\notag \\
&\qquad \qquad \qquad \quad {}
+ \left. 
 \eta 
 \left( 
  \tD_{12} + \Delta_1^{\{1\}}, \frac{\tD_{12}}{\tD_{12} + \Delta_1^{\{1\}}} 
 \right) 
\right] 
 + \frac{\pi^2}{2} 
\notag \\
&\quad {}
- i \, \pi \, S(\bar{z}) \, \ln \left( \frac{\bar{z} + 1}{ 2 \, \bar{z}} \right) 
 - \dilog \left( \frac{\bar{z} + 1}{ 2 \, \bar{z}} \right) 
 + \dilog \left( \frac{\bar{z} - 1}{ 2 \, \bar{z}} \right)
\label{eqbarh1inf1}
\end{align}
The quantities  $\tD_{12}+\Delta_1^{\{1\}}$ and $\Delta_1^{\{1\}}$ 
can be expressed in terms of $s_3$ and $m_3^2$ with the group of eqs.~(\ref{eqdtildepdelta1_vercol}). The expression obtained contains
the ratio $(s_3 - m_3^2)/(s_3 + m_3^2)$ given by:
\begin{equation}
\frac{s_3 - m_3^2}{s_3 + m_3^2} 
= 
\frac{s_3^2 - m_R^4 - m_I^4 - 2 \, i \, m_I^2 \, s_3}{(s_3 + m_R^2)^2 + m_I^4}
\label{eqratios3pmm2}
\end{equation}
which has a negative real part and a positive imaginary part for 
$0 < s_3 < m_R^2$. This implies that:
\begin{align}
  S \left( i \, \bar{z} \right) &= -1, \quad
  S \left( \bar{z} \right) = +1
  \label{eqvariouss1}
\end{align}
In addition, since $\Im( (s_3 - m_3^2)^2 ) = - 2 \, m_I^2 \, (s_3 - m_R^2) < 0$
and $\Im( (s_3 + m_3^2)^2 ) = 2 \, m_I^2 \, (s_3 + m_R^2) < 0$, one can show
that:
\[
\eta 
\left( 
 \frac{(s_3+m_3^2)^2}{2 \, s_3}, \frac{4 \, m_3^2 \, s_3}{(s_3+m_3^2)^2} 
\right) 
= \eta 
\left( 
 - \frac{(s_3+m_3^2)^2}{2 \, s_3}, \frac{(s_3-m_3^2)^2}{(s_3+m_3^2)^2} 
\right) 
= 0
\]
Putting all things together, $L(0,\Delta_1^{\{1\}},\tD_{12})$ becomes:
\begin{align}
\hspace{2em}&\hspace{-2em}L(0,\Delta_1^{\{1\}},\tD_{12}) \notag \\ 
&= 
\frac{2^{\varepsilon} \, \Gamma(1+\varepsilon) \, s_3}
{(s_3 + m_3^2) \, (s_3 - m_3^2)} \, 
\left\{ 
 \frac{1}{\varepsilon} \, 
 \left[ \ln \left( \frac{s_3}{m_3^2} \right) - i \, \pi  \right] - \frac{10 \, \pi^2}{6}
 + i \, \pi \, \ln \left( \frac{s_3}{s_3 - m_3^2} \right) 
\right. 
\notag \\
&\qquad \qquad \qquad \qquad \qquad \quad 
 + i \, \pi \, 
 \left[
  2 \, \ln \left( 2 \, i \, \frac{s_3-m_3^2}{s_3+m_3^2} \right) 
  + \ln \left( - \frac{(s_2-m_3^2)^2}{2 \, s_3} \right) 
 \right] 
\notag \\
&\qquad \qquad \qquad \qquad \qquad \quad 
 + 2 \, \dilog \left( \frac{s_3}{s_3-m_3^2} \right) 
 + \ln \left( -\frac{m_3^2}{s_3-m_3^2} \right) \, 
   \ln \left( \frac{s_3}{s_3 - m_3^2} \right) 
\notag \\
&\qquad \qquad \qquad \qquad \qquad \quad 
 - \ln \left( \frac{s_3}{m_3^2} \right) \, 
 \left[ 
  \ln \left( \frac{(s_3+m_3^2)^2}{2 \, s_3} \right)  
   + \frac{1}{2} \, \ln \left( \frac{4 \, m_3^2 \, s_3}{(s_3+m_3^2)^2} \right) 
 \right. 
\notag \\
&\qquad \qquad \qquad \qquad \qquad \qquad \qquad \qquad \quad
+ \left. 
 \left.
  \ln \left( 2 \, \frac{s_3 - m_3^2}{s_3 + m_3^2} \right) 
 \right] 
\right\}
\label{eqdefnewL1210}
\end{align}
Substituting eq. (\ref{eqdefnewL1210}) into eq. (\ref{eq_verif_irc20}) 
with the explicit values for $\bbj{2}{1}$, $\bbj{3}{1}$ and $\detgj{1}$ 
and using eq.(\ref{eqlijsoft8}) for $L(0,\Delta_1^{\{1\}},\tD_{13})$, we get:
\begin{align}
\Sigma_3^n(s_3) 
&= \frac{\Gamma(1+\varepsilon)}{2 \, (m_3^2 - s_3)} \, 
\left\{ 
 - \frac{1}{\varepsilon^2} + \frac{1}{\varepsilon} \, 
 \left[ 
  \ln \left( \frac{s_3}{m_3^2} \right) 
  + 
  \ln \left( - \frac{(s_3 - m_3^2)^2}{s_3} \right) - i \, \pi 
 \right] 
\right. 
\notag \\
&\qquad \qquad \qquad \quad 
 - \frac{1}{2} \, \ln^2 \left( - \frac{(s_3 - m_3^2)^2}{s_3} \right) - \frac{3 \, \pi^2}{2}  
 + i \, \pi \, \ln \left( \frac{s_3}{s_3 - m_3^2} \right) 
\notag \\
&\qquad \qquad \qquad \quad 
 + i \, \pi \, 
 \left[ 
  2 \, \ln \left( i \, \frac{s_3-m_3^2}{s_3+m_3^2} \right) 
  + \ln \left( - \frac{(s_2-m_3^2)^2}{s_3} \right) 
 \right] 
 - \ln \left( \frac{s_3}{m_3^2} \right)
\notag \\
&\qquad \qquad \qquad \quad 
\times
 \left[ 
  \ln \left( \frac{(s_3+m_3^2)^2}{s_3} \right)  
  + \frac{1}{2} \, \ln \left( \frac{m_3^2 \, s_3}{(s_3+m_3^2)^2} \right) 
  + \ln \left( \frac{s_3 - m_3^2}{s_3 + m_3^2} \right) 
 \right] 
\notag \\
&\qquad \qquad \qquad \quad 
\left.
 + 2 \, \dilog \left( \frac{s_3}{s_3-m_3^2} \right) 
 + \ln \left( -\frac{m_3^2}{s_3-m_3^2} \right) \, 
   \ln \left( \frac{s_3}{s_3 - m_3^2} \right) 
\right\}
\label{eq_verif_ircc2}
\end{align}
Keeping in mind that $0 < s_3 < m_R^2$, we split
the logarithms and expand the terms to end with:
\begin{align}
\Sigma_3^n(s_3) 
&= \frac{\Gamma(1+\varepsilon)}{2 \, (m_3^2 - s_3)} \, 
\left\{
  - \frac{1}{\varepsilon^2} + \frac{1}{\varepsilon} \, 
 \left[ 
  2 \, \ln \left( s_3 - m_3^2 \right) - 2 \, i \, \pi - \ln \left( m_3^2 \right)
 \right] 
\right. 
\notag \\
&\qquad \qquad \qquad \quad 
 - \left( \ln^2 \left( s_3 - m_3^2 \right) - \pi^2 
 - 2 \, i \, \pi \, \ln \left( s_3-m_3^2 \right) \right) 
\notag \\
&\qquad \qquad \qquad \quad 
 + \left.
  \frac{1}{2} \, \ln^2 \left( m_3^2 \right) 
  + 2 \, \dilog \left( \frac{s_3}{s_3-m_3^2} \right)  
\right\}
\label{eq_verif_ircc3final}
\end{align}
and using $\ln(s_3 - m_3^2) = \ln(-s_3 + m_3^2) + i \, \pi$, we again 
recover
eq. (\ref{eq_verif_irc4}). 
We note that the same formula holds both for
$\Im(\Delta_{1}^{\{1\}})>0$ i.e. either $s_3<0$ or $s_3> m_R^2$, and for 
$\Im(\Delta_{1}^{\{1\}})<0$ i.e. $0< s_3< m_R^2$. This is because in the last 
case, the integration contour $\int_{0}^{+ i \infty} + \int_{+\infty}^{1}$ can 
actually be deformed into $\int_{0}^{1}$, i.e. eq. (\ref{eqlijsoft61}) can be
deformed into eq. (\ref{eqlijsoft41}) by means of the Cauchy theorem. Indeed, 
when $0< s_3< m_R^2$,
$\Im(z^2 \, (\tD_{12} + \Delta_{1}^{\{1\}}) - \Delta_{1}^{\{1\}})$ never
vanishes as $z$ spans the real interval $[0,1]$ hence the cut of 
$\ln(z^2 \, (\tD_{12} + \Delta_{1}^{\{1\}}) - \Delta_{1}^{\{1\}})$ in the
half plane $\{\Re(z) >0\}$ entirely lies inside the ``south-east'' quadrant 
$\{\Re(z) >0, \Im(z) <0\}$. 

\vspace{0.3cm}

\noindent
$\Sigma_3^n(s_1)$ is read from eq.~(\ref{eq_verif_ircc3final})
by replacing $s_3$ by $s_1$ and the coefficients $\bbar_1$ and $\bbar_2$ as well as $\detg$ are
still given by eq.~(\ref{eqdefbb1bb2detg}). So, in the complex mass case the same result
is obtained as in the real mass case for $I_3^n$ and this leads to the conclusion that
for the case of complex masses, the ``direct way'' and the ``indirect way'' also coincide.

\section{Basic integrals in terms of dilogarithms and
logarithms}\label{appF}

In presence of vanishing internal masses, specific integrals of the
following type
\[
H
=  
\int^a_b du \, 
\frac{\ln ( A \, u^2 + B)}{A \, u^2 + B}
\]
for the contour $[0,1]$ as 
well as the two other contours $[0,+\infty[$ and $[1,+\infty[$ are involved. 

\vspace{0.3cm}

\noindent
We hereby compute all the above types of integrals successively.
The presentation is ordered according to the integration contours $(a,b)$
considered.
We last provide an extra load of back-up integrals.
This appendix often makes use of the identity
\begin{align}
\ln(z) &= \ln(-z) + i \, \pi \, S(z) 
\label{eqdeflnzlnmz}
\end{align}
where $S(z)$ is given by eq.~(\ref{eqdeffuncS0}).

\subsection{$H$-type integrals for the IR case}

\subsubsection{First kind}
\[
  H_{0,1}(A,B) = \int^1_0 du \, \frac{\ln(A \, u^2 + B)}{A \, u^2 + B} 
\]
The cases of real masses ($A$ real) and of complex masses ($\Im(A) \ne 0$) are 
treated all at once considering $A$ and $B$ both complex yet such that 
sign$(\Im(A \, u^2 + B)$ is kept constant when $u$ spans the range $[0,1]$, 
as is always the case for all our needs (cf. sections \ref{3point_ir} and \ref{sectfourpointir}). 
We write:
\begin{equation}
H_{0,1}(A,B) 
= 
\frac{1}{A} \, \int^1_0 du \, \frac{C_A + \ln(u^2 - \baru^2)}{u^2-\baru^2}
\label{eqdefh01}
\end{equation}
where
\begin{align}
C_A &= \ln(A - i \, \lambda \, S(-\baru^2)) 
\quad \text{if} \; \Im(A)=0 
\label{eqdefca1a}\\
C_A &= \ln(A) + \eta(A,-\baru^2) 
\quad \text{otherwise}
\label{eqdefca1b}
\end{align}
and $\baru^2 = - B/A$.
The $\eta$ function is given by eq.~(\ref{P1-eqdefeta01}) of ref.\ \cite{paper1}.
The term $\ln(u^2-\baru^2)$ can be split without $\eta$ function since 
$\baru$ and $-\baru$ have imaginary parts of opposite signs. Performing a 
partial fraction decomposition, we get:
\begin{align}
H_{0,1}(A,B) 
&= \frac{1}{2 \, A \, \baru} 
\left[ 
 C_A \, \int^1_0 du \, \left( \frac{1}{u-\baru} - \frac{1}{u+\baru} \right) 
\right.
\notag \\
&\quad {}\quad {} \quad {} \quad {}\quad {}
+ 
 \int^1_0 du \, \frac{\ln(u-\baru)}{u-\baru} 
 -  
 \int^1_0 du \, \frac{\ln(u-\baru)}{u+\baru}
\notag \\
&\quad {}\quad {} \quad {} \quad {}\quad {}
 \left.
 + 
 \int^1_0 du \, \frac{\ln(u+\baru)}{u-\baru} - 
 \int^1_0 du \, \frac{\ln(u+\baru)}{u+\baru} 
\right]
\label{eqdefh02}
\end{align}
We can rearrange the terms of the r.h.s. of
eq. (\ref{eqdefh02}) in the following way:
\begin{align}
H_{0,1}(A,B) 
&= \frac{1}{2 \, A \, \baru} 
\left\{ 
 C_A \, 
 \left[ 
  \ln \left( \frac{\baru-1}{\baru} \right) - 
  \ln \left( \frac{\baru+1}{\baru} \right) 
 \right] 
\right. 
\notag \\
&\quad {}\quad {}\quad {}\quad {} + 
 \frac{1}{2} \, 
 \left[ 
  \ln^2(1-\baru) - \ln^2(-\baru) -\ln^2(1+\baru) + \ln^2(\baru) 
 \right] 
\notag \\
&\quad {} \quad {}\quad {}\quad {}+ 
 \int^1_0 du \, \frac{\ln(u+\baru) - \ln(2 \, \baru)}{u-\baru} 
 \;\;\; + \;\;\;
 \int^1_0 du \, \frac{\ln(2 \, \baru)}{u-\baru}
\notag \\
&\quad {}\quad {}\quad {}\quad {} -
\left. 
 \int^1_0 du \, \frac{\ln(u-\baru) - \ln(-2 \, \baru)}{u+\baru} - 
 \int^1_0 du \, \frac{\ln(-2 \, \baru)}{u+\baru} 
\;\;
\right\}
\label{eqdefh03}
\end{align}
Using eq. (\ref{eqdeflnzlnmz}) we can write eq.
(\ref{eqdefh03}) in the following way:
\begin{align}
H_{0,1}(A,B) 
&= \frac{1}{2 \, A \, \baru} 
\left\{ 
 \ln \left( \frac{\baru-1}{\baru+1} \right) \, 
 \left[ 
  C_A + 
  \frac{1}{2} \, \left[ \ln(\baru-1) + \ln(\baru+1) \right] + 
  i \, \pi \, S(-\baru) + \ln(2 \, \baru) 
 \right] 
\right. 
\notag \\
&\quad {} \quad {}\quad {}\quad {}+ 
\left. 
  R^{\prime}(-\baru,\baru)
 \vphantom{\ln \left( \frac{\Lambda - \baru}{1 - \baru} \right)}
\right\}
\label{eqdefh04}
\end{align}
where the function $R^{\prime}$ has been defined in eq. (\ref{P1-eqdeffctr10}) of \cite{paper1}\footnote{The subtlety discussed in \cite{paper1} does not appear in this case.}.
Using eq. (\ref{P1-eqdefrprime4}) of \cite{paper1} with
$y=-\baru$ and $z=\baru$ and rearranging the term in square brackets, we get:
\begin{align}
H_{0,1}(A,B) 
&= \frac{1}{2 \, A \, \baru} 
\left\{
 \ln \left( \frac{\baru-1}{\baru+1} \right) \, 
 \left[ 
  C_A + 
 \frac{1}{2} \, 
  \left[ 
   \ln \left( 1 - \baru^2 \right) + \ln \left( - 4 \, \baru^2 \right) 
  \right] 
 \right]
\right. 
\notag \\
&\quad {} \quad {}\quad {}\quad {} + 
\left. 
 \dilog \left( \frac{\baru+1}{2 \, \baru} \right) - 
 \dilog \left( \frac{\baru-1}{2 \, \baru} \right)
\right\}
\label{eqdefh05}
\end{align}

\subsubsection{Second kind}

With complex masses we need also to compute:
\begin{equation}
H_{1,\infty}(A,B) = \int^{\infty}_1 du \, 
\frac{\ln(A \, u^2 + B)}{A \, u^2 + B} 
\label{eqdefh11}
\end{equation}
where $A$ and $B$ are complex yet such that $\sign(\Im(A \, u^2 + B))$ is kept 
constant while $u$ spans $[1,+\infty[$. The quantity $H_{1,\infty}(A,B)$ can 
be written as:
\begin{equation}
H_{1,\infty}(A,B) 
= 
\frac{1}{A} \, \int^{\infty}_1 du \, 
\frac{C^{\prime}_A + \ln(u^2 - \baru^2)}{u^2 - \baru^2} 
\label{eqdefh12}
\end{equation}
where $\baru^2 = - B/A$ and $C^{\prime}_A$ is given by:
\begin{align}
C^{\prime}_A &= \ln(A) + \eta(A,1-\baru^2) 
\label{eqdefca1c}
\end{align}
We perform a partial fraction decomposition and, writing 
$H_{1,\infty}(A,B)$ as a sum 
of terms which are individually divergent when $u \rightarrow \infty$, 
we face a
situation similar to the one met in subsec. \ref{P2-g2} of \cite{paper2}. We proceed likewise,
introducing a large $u$-cut-off $\Lambda$ and write $H_{1,\infty}(A,B)$ as:
\begin{align}
H_{1,\infty}(A,B) 
&= \frac{1}{2 \, A \, \baru} \lim_{\Lambda \rightarrow + \infty} 
{\cal H}_{1,\infty}^{\Lambda}(A,B)
\notag
\end{align}
where
\begin{align}
{\cal H}_{1,\infty}^{\Lambda}(A,B)
&=
\left\{
  C^{\prime}_A \, 
 \int^{\Lambda}_1 du \, 
 \left[ \frac{1}{u-\baru} - \frac{1}{u+\baru} \right]
\right. 
\notag \\
& \quad {}  \quad {} + 
 \int^{\Lambda}_1 du \, \frac{\ln(u-\baru)}{u-\baru} 
 - 
 \int^{\Lambda}_1 du \, \frac{\ln(u+\baru)}{u+\baru} 
\notag \\
& \quad {} \quad {} 
+ 
 \int^{\Lambda}_1 du \, \frac{\ln(u+\baru)- \ln(2 \, \baru)}{u-\baru}
\;\;\;+\;\;\;
 \ln ( 2 \, \baru) \, \int^{\Lambda}_1 \frac{du}{u - \baru}
\notag \\
&\quad {} \quad {}
\left. 
 - 
 \int^{\Lambda}_1 du \, \frac{\ln(u-\baru) - \ln(- 2 \, \baru)}{u+\baru} 
 -  
 \; \ln ( - 2 \, \baru) \, \int^{\Lambda}_1 \frac{du}{u + \baru}
\right\}
\label{eqdefh24}
\end{align}
We express ${\cal H}_{1,\infty}^{\Lambda}(A,B)$ in term of the function 
$R^{\Lambda}(y,z) $ defined by eq. (\ref{P2-eqdefrlamda1}) of \cite{paper2}:
\begin{align}
{\cal H}_{1,\infty}^{\Lambda}(A,B)
&= 
\left\{
  \left[ C^{\prime}_A + \ln ( 2 \, \baru) \right] \,
 \ln \left( \frac{\Lambda - \baru}{1 - \baru} \right) - 
 \left[ C^{\prime}_A + \ln ( - 2 \, \baru) \right] \,\ln \left( \frac{\Lambda + \baru}{1 + \baru} \right)
\right. 
\notag \\
&\quad {} + 
\frac{1}{2} \, \left[ \ln^2 \left( \Lambda - \baru \right) - \ln^2 \left( 1 - \baru \right) \right] -
\frac{1}{2} \, \left[ \ln^2 \left( \Lambda + \baru \right) - \ln^2 \left( 1 + \baru \right) \right]
\notag \\
&\quad {} + 
\left. 
R^{\Lambda}(-\baru,\baru) - R^{\Lambda}(\baru,-\baru) 
\vphantom{\ln \left( \frac{\Lambda - \baru}{1 - \baru} \right)}
\right\}
  \label{eqdefh25}
\end{align}
Using eq. (\ref{P2-eqcalcrlambda3}) of \cite{paper2} for the $R^{\Lambda}$ terms we take the 
limit $\Lambda \rightarrow \infty$. The terms proportional to 
$\ln^2(\Lambda)$ and those proportional to $\ln (\Lambda)$ drop out 
and we get:
\begin{align}
H_{1,\infty}(A,B) 
&= \frac{1}{2 \, A \, \baru}
\left\{ \vphantom{\dilog \left( \frac{\baru + 1}{2 \, \baru} \right)}
 \left[ C^{\prime}_A + \ln ( - 2 \, \baru) \right] \,\ln \left( 1 + \baru \right) -
  \left[ C^{\prime}_A + \ln ( 2 \, \baru) \right] \, \ln \left( 1 - \baru \right)
\right. 
\notag \\
&\quad {} \quad {}\quad {}\quad {}+ 
\frac{1}{2} \, 
\left[ 
 \ln^2 \left( 1 + \baru \right) - \ln^2 \left( 1 - \baru \right) + 
 \ln^2 \left( 2 \, \baru \right) - \ln^2 \left( - 2 \, \baru \right) 
\right]
\notag \\
&\quad {}\quad {}\quad {} \quad {}  
\left. 
 - \dilog \left( \frac{\baru + 1}{2 \, \baru} \right)
 + \dilog \left( \frac{\baru - 1}{2 \, \baru} \right) 
  \label{eqdefh26}
\right\}
\end{align}
Noting that:
\[
 \ln \left( \frac{1 + \baru}{1 - \baru} \right) 
= \ln (1 + \baru) - \ln (1 - \baru)
\]
eq. (\ref{eqdefh26}) becomes after some algebra:
\begin{align}
H_{1,\infty}(A,B) 
&= \frac{1}{2 \, A \, \baru}
\left\{
  \ln \left( \frac{1 + \baru}{1 - \baru} \right) \, 
  \left[ C^{\prime}_A + \frac{1}{2} \, \ln ( 1 - \baru^2) + \ln ( 2 \, \baru) \right] +
  \frac{\pi^2}{2}
\right. 
\notag \\
&\quad {} \quad {}\quad {} \quad {} - 
\left. 
i \, \pi \, S(\baru) \, \ln \left( \frac{\baru+1}{2 \, \baru} \right) 
 - 
 \dilog \left( \frac{\baru + 1}{2 \, \baru} \right)
 +
 \dilog \left( \frac{\baru - 1}{2 \, \baru} \right) 
\right\}  
\label{eqdefh27}
\end{align}
We remark that the same combination of dilogarithms - up to a sign - 
appears in eq. (\ref{eqdefh27}) and in $H_{0,1}(A,B)$ so that we can rewrite 
$H_{1,\infty}$ as:
\begin{align}
H_{1,\infty}(A,B) 
&=  - \, H_{0,1}(A,B) + 
\frac{1}{2 \, A \, \baru}
\left\{ \vphantom{\frac{1+\baru}{1-\baru}}
  i \, \pi \, S(\baru) \, \left[ 2 \, \ln ( 2 \, \baru) + C_A \right] + \pi^2 \right.
  \notag \\
  &\qquad {} +
  \left. \left[ \eta(A,-\baru^2) - \eta(A,1-\baru^2) \right] \, \ln \left( \frac{1+\baru}{1-\baru} \right)
\right\}
 \label{eqdefh28}
\end{align}

\subsubsection{Third kind}

With complex masses a third kind of integrals has also to be considered:
\begin{equation}
  H_{0,\infty}(A,B) = \int^{\infty}_0 du \, \frac{\ln(A \, u^2 + B)}{A \, u^2 + B} 
\label{eqdefh21}
\end{equation}
where $A$ and $B$ are complex yet such that $\sign(\Im(A \, u^2 + B))$ is kept 
constant while $u$ spans $[0,+\infty[$. The quantity $H_{0,\infty}(A,B)$ can 
be split as:
\begin{equation}
 H_{0,\infty}(A,B) = H_{0,1}(A,B) + H_{1,\infty}(A,B)
\label{eqdefh22}
\end{equation}
From eq. (\ref{eqdefh28}) and reminding that the assumption on the sign of $\Im(A \, u^2 + B)$ implies that $\eta(A,-\baru^2) = \eta(A,1-\baru^2)$
, we immediately get:
\begin{align}
  H_{0,\infty}(A,B) 
&=  \frac{1}{2 \, A \, \baru}
\Bigl[
  i \, \pi \, S(\baru) \, \left[ 2 \, \ln ( 2 \, \baru) + C_A \right] + \pi^2
\Bigr]
\label{eqdefh23}
\end{align}
As happened for $K^{C}_{0,\infty}(A,B)$ (cf.\ appendix \ref{P2-appF} of \cite{paper2}), $H_{0,\infty}(A,B)$ contains only logarithmic 
terms.

\subsection{An extra load of back-up integrals}

We also need the following load of simpler integrals:
\begin{align}
 W_1(u_0^2) &= \int^1_0 du \, \frac{\ln(1-u^2)}{u^2 - u_0^2} \\
 W_2(u_0^2) &= \int^1_0 du \, \frac{\ln(u)}{u^2 - u_0^2} \\
 W_3(u_0^2) &= \int^{\infty}_1 du \, \frac{\ln(u^2-1)}{u^2 - u_0^2} \\
 W_4(u_0^2) &= \int^{\infty}_0 du \, \frac{\ln(u)}{u^2 - u_0^2} \\
 W_5(u_0^2) &= \int^{\infty}_0 du \, \frac{\ln(u^2+1)}{u^2 + u_0^2} 
\end{align}
For all these integrals, $u_0^2$ is assumed to be a complex number, this is indeed the 
case because these integrals appear in the computation of the four-point function in 
the IR case where $u_0^2$ is either a complex number with an imaginary part $\propto \lambda$ 
or a genuine complex number.

\vspace{0.3cm}

\noindent
One might compute these integrals using specific values for $A$ and $B$ 
in $K^R_{0,1}(A,B)$, $K^C_{0,1}(A,B)$, $K^C_{1,\infty}(A,B)$ and $K^C_{0,\infty}(A,B)$ 
given in appendices~\ref{P1-appF} of ref. \cite{paper1} and~\ref{P2-appF} of ref. \cite{paper2}, 
however these integrals are simple enough to be computed 
directly (we verified that the results can be retrieved using the K-type 
integrals after some transformation on the dilogarithms).
We give here the result of these integrals without any details:
\begin{align}
W_1(u_0^2) 
&= \frac{1}{2 \, u_0} \, 
\left[ 
 \dilog \left( \frac{2}{1+u_0} \right) - 
 \dilog \left( \frac{2}{1-u_0} \right) - 
 2 \, \ln(2) \, \ln \left( \frac{u_0+1}{u_0-1} \right) 
\right] 
\label{eqresw1} \\
W_2(u_0^2) 
&= \frac{1}{2 \, u_0} \, 
\left[ 
 \dilog \left( \frac{1}{u_0} \right) - 
 \dilog \left( - \, \frac{1}{u_0} \right) 
\right] 
\label{eqresw3} \\
W_3(u_0^2) 
&= - W_1(u_0^2) + \frac{1}{2 \, u_0} \, i \, S(u_0) \, \pi \ln \left( 1 - u_0^2 \right)
\label{eqresw2} \\
W_4(u_0^2) 
&= \frac{1}{4 \, u_0} \, i \, S(u_0) \, \pi \, \ln(-u^2_0) 
\label{eqresw4} \\
W_5(u_0^2)
&= \frac{\pi}{u_0} \, \ln(1+u_0) 
\label{eqresw5}
\end{align}
with $u_0 = \sqrt{u_0^2}$.

\section{Change of contour prescription for the pole in the IR four-point 
integral}\label{ir-lambda}

The appendix legitimates the replacement
\begin{equation}\label{subst1}
\int_{0}^{1} du \,
\frac{\left[  u^2 \, P_{ijk} + R_{ij} - i \, \lambda \right]^{-\varepsilon}}
{u^2 \, P_{ijk} + R_{ij} + i \, \lambda}   
\to 
\int_{0}^{1} du \,
\frac{\left[  u^2 \, P_{ijk} + R_{ij} - i \, \lambda \right]^{-\varepsilon}}
{u^2 \, P_{ijk} + R_{ij} - i \, \lambda}   
\end{equation}
when $\lambda \to 0^{+}$ for fixed $0 < - \varepsilon \ll 1$, whenever 
$P_{ijk}$ and $R_{ij}$ are both real with $0 < - \, R_{ij}/P_{ijk} < 1$. 
Intuitively this replacement is based on the fact that
for any function $f(u)$ analytic along $[0,1]$ and $0 < u_{0} < 1$,
\[
\int_{0}^{1} du \,
\frac{f(u)}{u - u_{0} - i \, \lambda} 
-   
\int_{0}^{1} du \,
\frac{f(u)}{u - u_{0} + i \, \lambda}
\to
2 i \, \pi \, f(u_{0})
\;\;
\mbox{when} \;\; \lambda \to 0^{+}
\]
which vanishes if $u_{0}$ is a zero of $f(u)$. However the situation is
made trickier when $f(u) = (u-u_{0})^{-\varepsilon}$ 
thus has a branch point at $u = u_{0}$ and a cut running along part of the
interval of integration. To make the above argument apply one could think of 
shifting the branch point and cut by a contour prescription 
$- i \, a \lambda$ with $a>1$ so as to pass whether above or below
the pole while remaining on the same side of the cut. 
We would then get a residue 
value $\propto \lambda^{\; -\varepsilon}$ vanishing in 
the limit $\lambda \to 0^{+}$ keeping $\varepsilon < 0$ fixed. 
The ``hierarchised lambdalogy" underpinning this disentanglement of pole from 
branch point may look awkward to the rigorous reader, let us thus back up 
this hand waving argument more rigorously as follows.

\vspace{0.3cm}

\noindent
We first perform a partial fraction decomposition of the pole term:
\begin{eqnarray}
\lefteqn{\frac{1}{u^2 \, P_{ijk} + R_{ij} + i \, s \, \lambda}}
\nonumber\\  
& = &
\frac{1}{P_{ijk}} \, \frac{1}{2 \, \sqrt{- R_{ij}/P_{ijk}}} \,
\left[
\frac{1}{u - (u_{0} - i \, s \lambda^{\prime})}   
-
\frac{1}{u + (u_{0} - i \, s \lambda^{\prime})}   
\right]
\label{subst2}
\end{eqnarray}
where $s = \pm$, $u_{0} = \sqrt{- R_{ij}/P_{ijk}}$ is assumed in $]0,1[$ and 
$\lambda^{\prime} = \lambda/(2 u_{0} P_{ijk})$\footnote{We keep track of this 
multiplicative change to control various normalisations in the reasoning so as
to check the independence of the conclusion w.r.t. any assumption of 
``hierarchised lambdalogy".}. We focus on the pole at $+ u_{0}$ in
decomposition (\ref{subst2}): since the pole at $-u_{0}$ involved in the second 
term lies outside the integration region the contour prescription for this pole
is irrelevant and so is the corresponding pole term in the discussion.
We then study the legitimacy of the replacement
\begin{equation}\label{subst3}
\int_{0}^{1} du \,
\frac{\left[ u^2 \, P_{ijk} + R_{ij} - i \, a \lambda \right]^{-\varepsilon}}
{u - u_{0} + i \, \lambda^{\prime}
}   
\to 
\int_{0}^{1} du \,
\frac{\left[ u^2 \, P_{ijk} + R_{ij} - i \, a \lambda\right]^{-\varepsilon}}
{u - u_{0} - i \, \lambda^{\prime}}   
\end{equation}
($a$ being positive yet kept arbitrary). 
We consider ``$\delta \equiv$ l.h.s. minus r.h.s." of eq. (\ref{subst3}), 
which can be written:
\[
\delta =
\int_{0}^{1} \frac{du \, ( 2 \, i \, \lambda^{\prime})}
{(u - u_{0})^{2} + \lambda^{\prime \, 2}} 
 \left[ 
  P_{ijk} \, 
  \left( (u - u_{0})(u + u_{0}) - i \, (2 u_{0} a) \, \lambda^{\prime} \right) 
 \right]^{-\varepsilon}  
\]
we make the change of variable $(u - u_{0}) = |\lambda^{\prime}| \, v$ and
$\delta$ reads:
\begin{equation}\label{subst4}
\delta 
=  
2 \, i \, \sigma \, 
\left( \frac{\lambda}{2 u_{0}} \right) ^{- \varepsilon}
\int_{-u_{0}/|\lambda^{\prime}| }^{(1- u_{0})/|\lambda^{\prime}|} 
\frac{dv}{v^2+1} 
 \left[ 
   \sigma v \left( \sigma \lambda^{\prime} v \, + 2 u_{0} \right) 
   - i (2 u_{0} a)
 \right]^{-\varepsilon} 
\end{equation}
(where $\sigma = \mbox{sign}(\lambda^{\prime}) = \mbox{sign}(P_{ijk})$).
For any $b > 1$ and $|\lambda^{\prime}|$ small enough we have, for all real 
$v$:
\[
\left| 
 \sigma v ( \sigma \lambda^{\prime} v \, + 2 u_{0}) - i (2 u_{0} a) 
\right| 
< \left[ v^4 + (2u_{0})^2 b \, v^2 + (2 u_{0} a)^2 \right] ^{1/2}
\] 
and the integral 
\[
\int_{- \infty}^{+\infty} dv \, 
\frac{[ v^4 + (2u_{0})^2 b \, v^2 + (2 u_{0} a)^2]^{- \varepsilon/2}}{v^2+1}
\]
is convergent when $-\varepsilon>0$ is small enough. 
It provides an ``integrable 
hat" for the application of Lebesgue's theorem of dominated convergence.
When $\lambda \to 0^{+}$ keeping $-\varepsilon>0$ 
fixed and small enough, the integral in eq. (\ref{subst4}) has the limit
\[
{\cal L} = (2 u_{0})^{- \varepsilon}
\int_{-\infty}^{+\infty} 
\frac{dv}{v^2+1} \left( \sigma v - i a \right)^{-\varepsilon} 
\]
which is finite regardless of $a > 0$ and $\varepsilon$ small enough
(its  actual value, readily computable using the residue theorem, 
is irrelevant for the conclusion). 
We thus see that $\delta \sim {\cal O}(\lambda^{- \varepsilon})$ as 
anticipated. q.e.d.

\bibliographystyle{unsrt}
\bibliography{../biblio,../publi}

\end{fmffile}

\end{document}